\documentclass[twocolumn,aps,floatfix,superscriptaddress]{revtex4}
\usepackage{array}
\usepackage{amsmath}
\usepackage{amssymb}
\usepackage{graphicx}
\usepackage{color}

\newcommand{\bea}{\begin{eqnarray}}
\newcommand{\eea}{\end{eqnarray}}
\newcommand{\bse}{\begin{subequations}}
\newcommand{\ese}{\end{subequations}}
\newcommand{\bma}{${\rm BaMn_2As_2}$}
\newcommand{\cma}{${\rm CaMn_2As_2}$}
\newcommand{\sma}{${\rm SrMn_2As_2}$}
\newcommand{\tcs}{${\rm ThCr_2Si_2}$}

\newcommand{\cca} {CaCo$_{2-y}$As$_2$}

\newcommand{\sca} {${\rm SrCo_2As_2}$}
\newcommand{\sna} {${\rm SrNi_2As_2}$}
\newcommand{\bca} {${\rm BaCo_2As_2}$}
\newcommand{\eca} {${\rm EuCo_2As_2}$}
\newcommand{\ecp} {${\rm EuCo_2P_2}$}
\newcommand{\csca}{Ca$_{1-x}$Sr$_x$Co$_{2-y}$As$_2$}
\newcommand{\sccp}{Sr$_{1-x}$Ca$_x$Co$_2$P$_2$}

\newcommand{\scna}{Sr(Co$_{1-x}$Ni$_x$)$_2$As$_2$}
\newcommand{\bkfa}{$\rm (Ba_{0.6}K_{0.4})Fe_2As_2$}

\begin{document}

\title{Non-Fermi-liquid behaviors associated with a magnetic quantum-critical point in Sr(Co$_{1-x}$Ni$_x$)$_2$As$_2$ single crystals}

\author{N. S. Sangeetha}
\affiliation{Ames Laboratory, Iowa State University, Ames, Iowa 50011, USA}
\author{L.-L. Wang}
\affiliation{Ames Laboratory, Iowa State University, Ames, Iowa 50011, USA}
\author{A. V. Smirnov}
\affiliation{Ames Laboratory, Iowa State University, Ames, Iowa 50011, USA}
\author{V. Smetana}
\affiliation{Department of Materials and Environmental Chemistry, Stockholm University, Svante Arrhenius v\"ag 16 C, 106 91 Stockholm, Sweden}
\author{A.-V. Mudring}
\affiliation{Department of Materials and Environmental Chemistry, Stockholm University, Svante Arrhenius v\"ag 16 C, 106 91 Stockholm, Sweden}
\author{D. D. Johnson}
\affiliation{Department of Materials Science \& Engineering, Iowa State University, Ames, IA 50011, USA}
\author{M. A. Tanatar}
\affiliation{Ames Laboratory, Iowa State University, Ames, Iowa 50011, USA}
\affiliation{Department of Physics and Astronomy, Iowa State University, Ames, Iowa 50011, USA}
\author{R.~Prozorov}
\affiliation{Ames Laboratory, Iowa State University, Ames, Iowa 50011, USA}
\affiliation{Department of Physics and Astronomy, Iowa State University, Ames, Iowa 50011, USA}
\author{D. C. Johnston}
\affiliation{Ames Laboratory, Iowa State University, Ames, Iowa 50011, USA}
\affiliation{Department of Physics and Astronomy, Iowa State University, Ames, Iowa 50011, USA}

\date{\today}

\begin{abstract}
Electron-doped \scna\ single crystals with compositions $x=0$ to~0.9 were grown out of self-flux and \sna\ single crystals out of Bi flux.  The crystals were characterized using single-crystal x-ray diffraction (XRD) at room temperature, and magnetic susceptibility $\chi(H,T)$, isothermal magnetization $M(H,T)$, heat capacity $C_{\rm p}(H,T)$, and electrical resistivity $\rho(H,T)$ measurements versus applied magnetic field~$H$ and temperature~$T$\@. The XRD studies show that the system undergoes a continuous structural crossover from the uncollapsed-tetragonal (ucT)  structure to the collapsed tetragonal (cT) structure with increasing Ni doping. The $\chi(T)$ data show that \sca\ exhibits an antiferromagnetic (AFM) ground state almost immediately upon Ni doping on the Co site. {\it Ab-initio} electronic-structure calculations for $x=0$ and $x=0.15$ indicate that a flat band with a peak in the density of states just above the Fermi energy is responsible for this initial magnetic-ordering behavior on Ni doping. The AFM ordering is observed in the range \mbox{$0.013\leq x\leq 0.25$} with the ordered moments aligned in the $ab$~plane and with a maximum ordering temperature \mbox{$T_{\rm N} = 26.5$~K} at~$x=0.10$. The Curie-Weiss-like $T$ dependence of~$\chi$ in the paramagnetic (PM) state indicates dominant ferromagnetic (FM) interactions.  The behavior of the anisotropic susceptibilities below $T_{\rm N}$ suggest a planar helical magnetic ground state with a composition-dependent pitch based on a local-moment molecular-field-theory model, with FM interactions in the $ab$~plane and weaker AFM interactions along the helix $c$~axis.   However, the small ordered (saturation) moments $\sim0.1~\mu_{\rm B}$ per transition metal atom, where $\mu_{\rm B}$ is the Bohr magneton, and the values of the Rhodes-Wohlfarth ratio indicate that the magnetism is itinerant. The high-field $M(H)$ isotherms and the low-field $\chi^{-1}(T>T_{\rm N})$ data were successfully analyzed within the framework of Takahashi's theory of FM spin fluctuations. The $C_{\rm p}(T)$ at low~$T$ exhibits Fermi-liquid behavior for $0\leq x \leq 0.15$ whereas an evolution to a logarithmic non-Fermi-liquid (NFL) behavior is found for $x=0.2$ to~0.3. The logarithmic dependence is suppressed in an applied magnetic field.  The low-$T$ $\rho(H=0,T)$ data  show a $T^2$ dependence for $0 \leq x \leq 0.20$ and a power-law dependence $\rho(H=0,T) = \rho_0+AT^n$ with $n<2$ for $x = 0.20$ and 0.30. The exponent~$n$ shows a notable field dependence, suggesting both doping- and magnetic-field-tuned quantum critical phenomena.  These low-$T$ NFL behaviors observed in the $C_{\rm p}$ and $\rho$ measurements are most evident near the quantum-critical concentration $x \approx 0.3$ at which a $T=0$ composition-induced transition from the AFM phase to the PM phase occurs.
 
\end{abstract}

\maketitle

\section{Introduction}
The discovery of high-$T{\rm_c}$ superconductivity (SC) in iron-based pnictides and chalcogenides motivated many studies of their correlated lattice, electronic, magnetic, and superconducting properties  \cite{{Johnston2010},{Stewart2011},{Scalapino2012}, {Dagotto2013},{Fernandes2014},{Hosono2015},{Dai2015}, {Inosov2016},{Si2016}}. Among such compounds, the class of $AT_2X_2$ (abbreviated as 122) ternary compounds ($A=$ alkaline earth and lanthanides), which have the body-centered tetragonal \tcs-type crystal structure with space group $I4/mmm$, have been well studied. By electron or hole doping via chemical substitution which modify the  lattice, electronic, and/or magnetic degrees of freedom, the $A{\rm Fe_2As_2}$ ($A=$ Sr, Ca, Ba, and Eu) compounds were found to exhibit a variety of ground states including different crystallographic structures, magnetically-ordered states or SC\@. This in turn prompted the search for novel physical properties with different transition-metal-based 122-type compounds such as described in Refs.~\cite{{An2009},{Singh2009}, {Singh2009b}, {Johnston2011}, {Antal2012}, {Calder2014}, {Zhang2016}, {Sangeetha2016a}, {Das2017}, {Sangeetha2017}}.

The Co-based pnictides $A{\rm Co_2As_2}$ ($A=$ Ca, Sr, Ba, and Eu) have a variety of magnetic properties. For example \eca\ and \ecp\ exhibit Eu (4$f$) local-moment AFM magnetic ordering with the uncollapsed tetragonal structure (ucT) at ambient pressure \cite{{Sangeetha2016},{Sangeetha2018}}. These compounds show changes from the ucT to the collapsed tetragonal (cT) structure under pressure, resulting in a change from Eu (4$f$) local-moment ordering to Co (3$d$) itinerant magnetic ordering \cite{{Huhnt1997},{Chefki1998},{Bishop2010}}.  On the other hand, \cca\ reveals a different magnetic ordering which has a cT structure and undergoes an itinerant A-type AFM ordering where the Co moments are FM aligned along the tetragonal $c$~axis within a Co layer in the $ab$~plane and are AFM aligned between planes along the $c$ axis. Thus the magnetism is dominated by strong $ab$-plane FM interactions compared to weak AFM interlayer interactions along the $c$~axis~\cite{Anand2014}. Inelastic neutron studies revealed that \cca\ exhibits perfect magnetic frustration, rarely seen in itinerant magnets, between nearest and next-nearest neighbor magnetic interactions.  This means that stripe-type AFM and itinerant FM ground states compete which is unique among quasi-two-dimensional AFMs \cite{Sapkota2017}. In contrast to metallic \cca , metallic \sca\ and \bca\ have the ucT structure with no long-range magnetic ordering and exhibit Stoner-enhanced paramagnetism \cite{An2009, Pandey2013, Anand2014b}. Furthermore, inelastic neutron scattering measurements on \sca\ revealed strong stripe-type AFM correlations at high energies as seen in iron based superconductors, a key factor for inducing SC at high temperatures~\cite{{Jayasekara2015},{Jayasekara2013}}, whereas NMR and inelastic neutron scattering measurements reveal strong FM correlations in addition to AFM correlations~\cite{Wiecki2015, Li2019a, Li2019}, suggesting a reason that $\rm{SrCo_2As_2}$ does not exhibit SC\@.

The growth of pseudoternary crystals having different properties can also reveal additional interesting physics.  For example, crystals of the mixed system \csca\ exhibit a continuous composition-induced crossover in the itinerant magnetic properties between the above-noted $A$-type AFM~I phase ($0\leq x \leq 0.2$) with $c$-axis moment alignment to an AFM II phase ($x=0.40$ and 0.45) with the ordered moments aligned in the $ab$ plane, followed by a paramagnetic (PM) region for $x \geq 0.52$~\cite{Ying2013, Sangeetha2017a}.

The ground state of the 122-type pseudoternary series \sccp\ changes from a nearly-FM Fermi liquid to AFM, then to FM like, and finally back to AFM upon Ca substitution \cite{Jia2009}.  These changes in the magnetism are accompanied by a continuous structural crossover from the ucT ($x=0$) to the cT structure ($x=1$) that is correlated with the variation in the interlayer P-P bonding distance. It was inferred that the valence state of the transition metal and the P-P interlayer distance are correlated and that the charge redistribution in the ${\rm Co_2P_2}$ layers during the ucT to cT transition changes the magnetic ground state. Thus the system SrCo$_2$(Ge$_{1-x}$P$_x$)$_2$  develops weak itinerant FM during the course of  P-P dimer breaking and a quantum critical point (QCP) is observed at the onset of the FM phase although both end members ${\rm SrCo_2P_2}$ (ucT) and ${\rm SrCo_2Ge_2}$ (cT) are PM metals.  Here, the FM  and QCP are not induced by a simple electron doping effect giving rise to a valence change of the Co atoms, but rather by the breaking of the dimer. When the dimer is fully broken, the FM state is not present and the PM state appears.

On the other hand, CaCo$_2$(Ge$_{1-x}$P$_x$)$_2$ compounds show nonmagnetic to AFM transitions without FM in the intermediate composition region while retaining fully intact X-X dimers (X = Ge/P) over the  whole $x$ doping range~\cite{Jia2011}. In contrast, throughout the BaCo$_2$(Ge$_{1-x}$P$_x$)$_2$ system only a nonmagnetic ground state is observed in which the X-X dimer is fully broken (there is no X-X bond) because of the large size of the Ba$^{+2}$ ions~\cite{Jia2011}.

The above studies on mixed Co122 systems illustrate the importance of pnictogen X-X interlayer bonding on the  electronic states of Co-X layers. Unlike local-moment magnetism, itinerant magnetism originates from the properties of band electrons near the Fermi surface. As a result, the type of magnetic order is determined by the wave vector {\bf q} at which the the wave-vector dependent susceptibility $\chi({\bf q})$ has a peak.  For FM ordering,  ${\bf q}=0$ and the presence or absence of FM long-range ordering is determined by the value of the density of states (DOS) at the Fermi energy~$E_{\rm F}$. In $A{\rm Co_2As_2}$ ($A$ =  Ba, Sr, Ca), a sharp peak in the DOS at $E_{\rm F}$ arises from a Co $d_{x^2-y^2}$ flat band~\cite{Mao2018}. A sufficiently large value leads to FM ordering according to Stoner theory. For $\rm CaCo_2As_2$, the peak in the DOS is very close to $E_{\rm F}$ whereas it is about 35 meV above the Fermi level in \bca\ and \sca~\cite{Mao2018}, suggesting that $\rm CaCo_2As_2$ would have much stronger low-energy FM spin fluctuations than \bca\ and \sca. This is evidently the reason why only $\rm CaCo_2As_2$ has an A-type AFM transition, where the ordered moments are aligned ferromagnetically within the Co layers and antiferromagnetically between layers, with the AFM interactions much weaker than the FM interactions, whereas \bca\ and \sca\ exhibit Pauli paramagnetism down to 2~K~\cite{Mao2018}. Moreover, $\rm (Ba\ and\ Sr)Co_2As_2$ have ucT structures whereas $\rm CaCo_2As_2$ has a cT structure.

Previous reports revealed that electron-doped Sr$_{1-x}$La$_x{\rm Co_2As_2}$ exhibits a sudden change in the magnetic bahavior from PM to FM with only 2.5\% of La doping~\cite{Shen2018, Shen2019}, indicating that FM order is preferred over A-type AFM or stripe-type AFM order for this type of doping.  A report about the electron doping effects on \scna\ polycrystalline samples \cite{Ohta2015} revealed AFM ordering for $0<x<0.3$.

Herein, we report a study of the influence of electron doping in \scna\ single crystals on the crystallographic, magnetic, thermal, and electronic transport properties. We find that the incorporation of only 1.3\% of Ni at the Co sites is sufficient to induce a transition from PM to AFM ground states. This is consistent with our {\it ab initio} electronic structure calculations which show a peak in the density of states slightly above the Fermi energy due to a flat band.  In addition, the ground-state magnetic phase diagram of \scna\ and the magnetic properties within the frameworks of local-moment magnetism using molecular-field theory and of Takahashi's spin fluctuation theory for weak itinerant ferromagnets are also discussed.  The most important result of our studies is the observation of non-Fermi-liquid signatures in the heat capacity and electrical resistivity for compositions in the vicinity of the quantum-critical point at $x\approx0.30$ between the magnetically-ordered and paramagnetic phases at $T\to0$.

The experimental details are given in Sec.~\ref{Sec:ExpDetails}. The crystallographic data and composition analysis are presented in Sec.~\ref{Sec:crystalstudy}. The $M (H)$ and $\chi(T)$ data are presented in Sec.~\ref{Sec:magresult} and these data are analyzed within a local moment magnetism model in Sec.~\ref{Sec:LocMomMag}.  A more appropriate analysis of the magnetic properties in terms of an itinerant picture is presented in Sec.~\ref{Sec:itinerantMag}, together with Arrott ($M^2$ vs $H/M$) and Takahashi's $M^4$ vs $H/M$ isotherm plots. The inverse susceptibility in the paramagnetic state above $T_{\rm N}$ is also discussed within the framework of Takahashi's spin-fluctuation theory.  The implications of our electronic structure calculations for the magnetism are also included in this section.

The heat capacity capacity $C_{\rm_p}(T)$ data are presented and discussed in Sec.~\ref{Sec:heatcapacity}, where non-Fermi-liquid behavior is found at low temperatures. A comparison of the experimental density of states versus~$x$ obtained from the $C_{\rm_p}(T)$ measurements with the electronic-structure prediction is also given in this section.  Electrical resistivity measurements are presented in Sec.~\ref{Sec:Res} where we discuss non-Fermi-liquid behavior induced by doping and magnetic field.  An overall discussion of the non-Fermi-liquid behaviors in \scna\ is given in Sec.~\ref{Sec:NFLdiscuss}.

A summary of the paper is given in Sec.~\ref{Sec:Sum}. Eight tables of fitted parameters are given in the Appendix.

\section{\label{Sec:ExpDetails} Experimental and Theoretical Details}

Single crystals of  \scna\ with compositions ($x$ = 0, 0.013, 0.04, 0.06, 0.1, 0.15, 0.2, 0.25, 0.3, 0.4, 0.5, 0.6, 0.7, 0.8, 0.9) were grown using self flux. The starting materials were high-purity elemental Sr(99.999\%), Co (99.998\%), Ni (99.999\%)  and As (99.99999\%) from Alfa Aesar. A mixture of the elements in a Sr:Co:Ni:As 1:4($1-x$):4$x$:4 molar ratio was placed in an alumina crucible that was sealed under $\approx$ 1/4 atm high purity argon in a silica tube. The sealed samples were preheated at 600~$^{\circ}$C for 5~h, and then heated to 1300~$^{\circ}$C at the rate of 50~$^{\circ}$C/h and held there for 15~h. Then the furnace was slowly cooled at the rate of 6~$^{\circ}$C/h to 1180~$^{\circ}$C. The single crystals were separated from the self flux by decanting the flux with a centrifuge at that temperature. Several 2 to~4~mm size shiny platelike single crystals were obtained from each growth. 

Single crystals of \sna\ were grown in Bi flux. The sample to flux in a 1:10 molar ratio was placed in an alumina crucible that was sealed under argon gas in a silica tube. A sealed sample was preheated at 600~$^{\circ}$C for 6~h. Then the mixture was placed in a box furnace and heated to 1050~$^{\circ}$C at a rate of 50~$^{\circ}$C/h, held there for 20~h, and then cooled to 700~$^{\circ}$C at a rate of 2~$^{\circ}$C/h and then to 400~$^{\circ}$C at a rate of 5~$^{\circ}$C/h. At this temperature the molten Bi flux was decanted using a centrifuge. Shiny platelike crystals with plate-surface dimension 2 to 5~mm ($\approx 15$ mg) were obtained. 

The phase purity and chemical composition of the \scna\ crystals were checked by energy dispersive x-ray (EDS) semiquantitative chemical analysis using an EDS attachment to a JEOL scanning-electron microscope (SEM). SEM scans were taken on  cleaved surfaces of the crystals which verified the single-phase nature of the crystals. The compositions of each side of a platelike crystal were measured at six or seven positions on each face, which revealed good homogeneity in each crystal. The average compositions and error bars were obtained from these data.

Single-crystal x-ray diffraction (XRD) measurements were performed at room temperature on a Bruker D8 Venture diffractometer operating at tube voltage of 50~kV and a current of 1~mA equipped with a Photon 100 CMOS detector, a flat graphite monochromator and a Mo~K$\alpha$ I$\mu$S microfocus source ($\lambda = 0.71073$~\AA). The preliminary quality testing has been performed on a set of 32 frames. The raw frame data were collected using the Bruker APEX3 software package~\cite{APEX2015}. The frames were integrated with the Bruker SAINT program~\cite{SAINT2015}  using a narrow-frame algorithm integration and the data were corrected for absorption effects using the multi-scan method (SADABS)~\cite{Krause2015} within the APEX3 package. The occupancies of the Sr and As sites were set to unity and that of the Co/Ni site was refined.  No vacancies on the Co/Ni site were found for any of the crystals.  The atomic displacement parameters were refined anisotropically. Initial models of the crystal structures were first obtained with the program SHELXT-2014~\cite{Sheldrick2015A} and refined using the program SHELXL-2014~\cite{Sheldrick2015C} within the APEX3 software package.

The results of the EDS and XRD composition analyses are given in Table~\ref{CrystalData} below.  The same crystals were used to perform the physical-property measurements reported below.

Magnetization data were obtained using a Quantum Design, Inc., magnetic-properties measurement system (MPMS) and a vibrating sample magnetometer (VSM) in a Quantum Design, Inc., physical-properties measurement system (PPMS) in magnetic fields up to 14~T where 1~T~$\equiv10^4$~Oe. Heat capacity $C_{\rm p}(T)$ measurements were performed in a PPMS system using a relaxation method.

Electrical resistivity measurements were performed with current flow along the tetragonal plane. Samples were cleaved in the shape of a bar with typical size $2\times 0.5 \times0.1~{\rm mm}^3$. The contacts were made by soldering Ag wires using Sn \cite{Tanatar2010, patent}. The contact resistance was similarly low as in BaFe$_2$As$_2$-based superconductors. Four-probe resistivity measurements were performed in a Quantum Design, Inc., PPMS setup. Magnetic fields up to 9~T were applied along the tetragonal $c$~axis.

Density-functional theory~\cite{Hohenberg1964, Kohn1965} (DFT) with local-density approximation~\cite{Ceperley1980, Perdew1981} (LDA) for exchange-correlation functional were used to solve the KKR-CPA equations using a spherical harmonic basis set within an atomic sphere approximation for scattering sites~\cite{Johnson1986, Johnson1990}. A conventional unit cell and ($10\times10\times4$) Monkhorst-Pack~\cite{Monkhorst1976} $k$-point mesh was used in the calculations based on the experimental structural parameters.

\section{\label{Sec:crystalstudy} Crystal structures}

\begin{figure}
\includegraphics[width=1.8in]{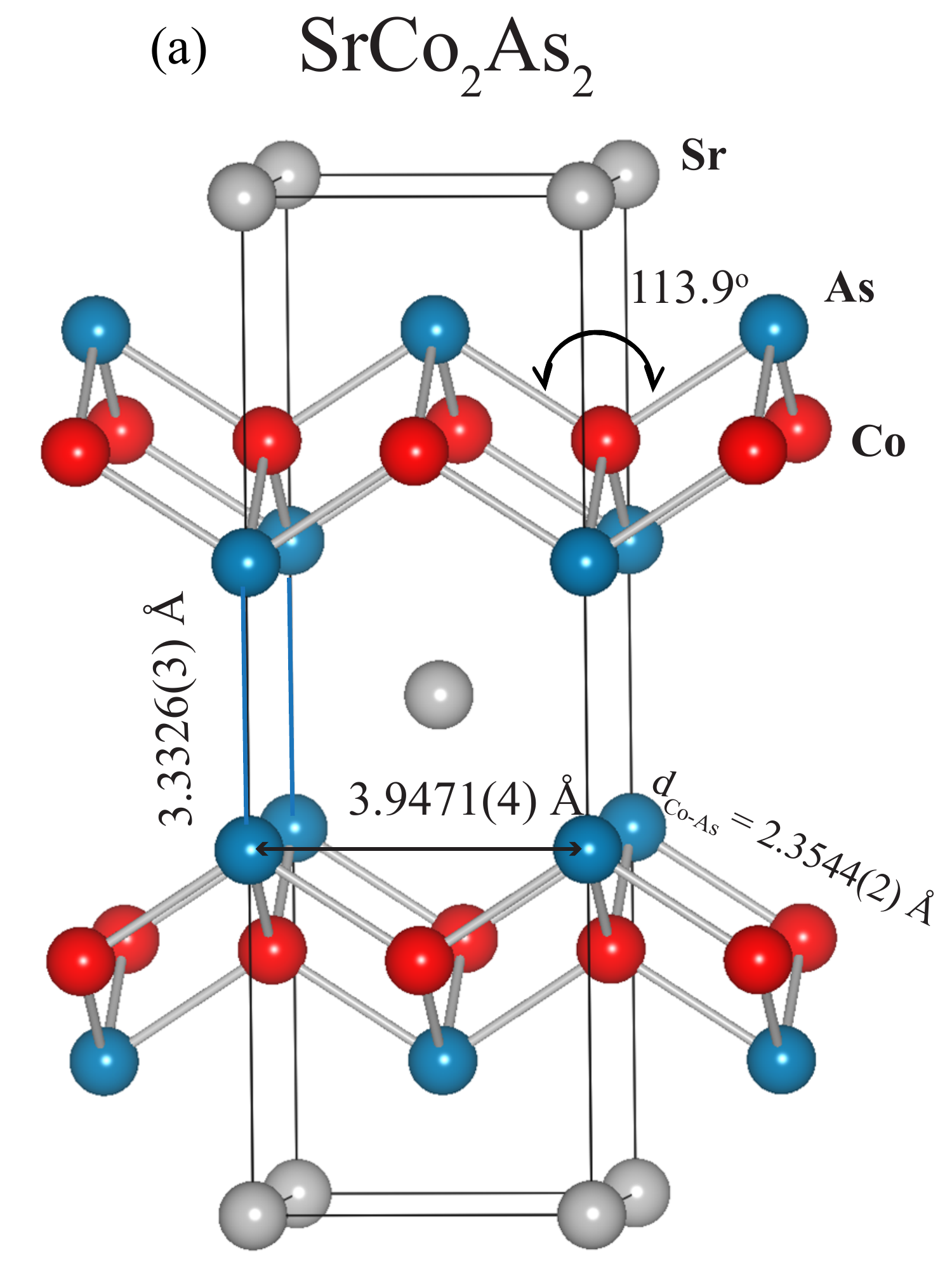}\vspace{0.1in}
\includegraphics[width=2.3in]{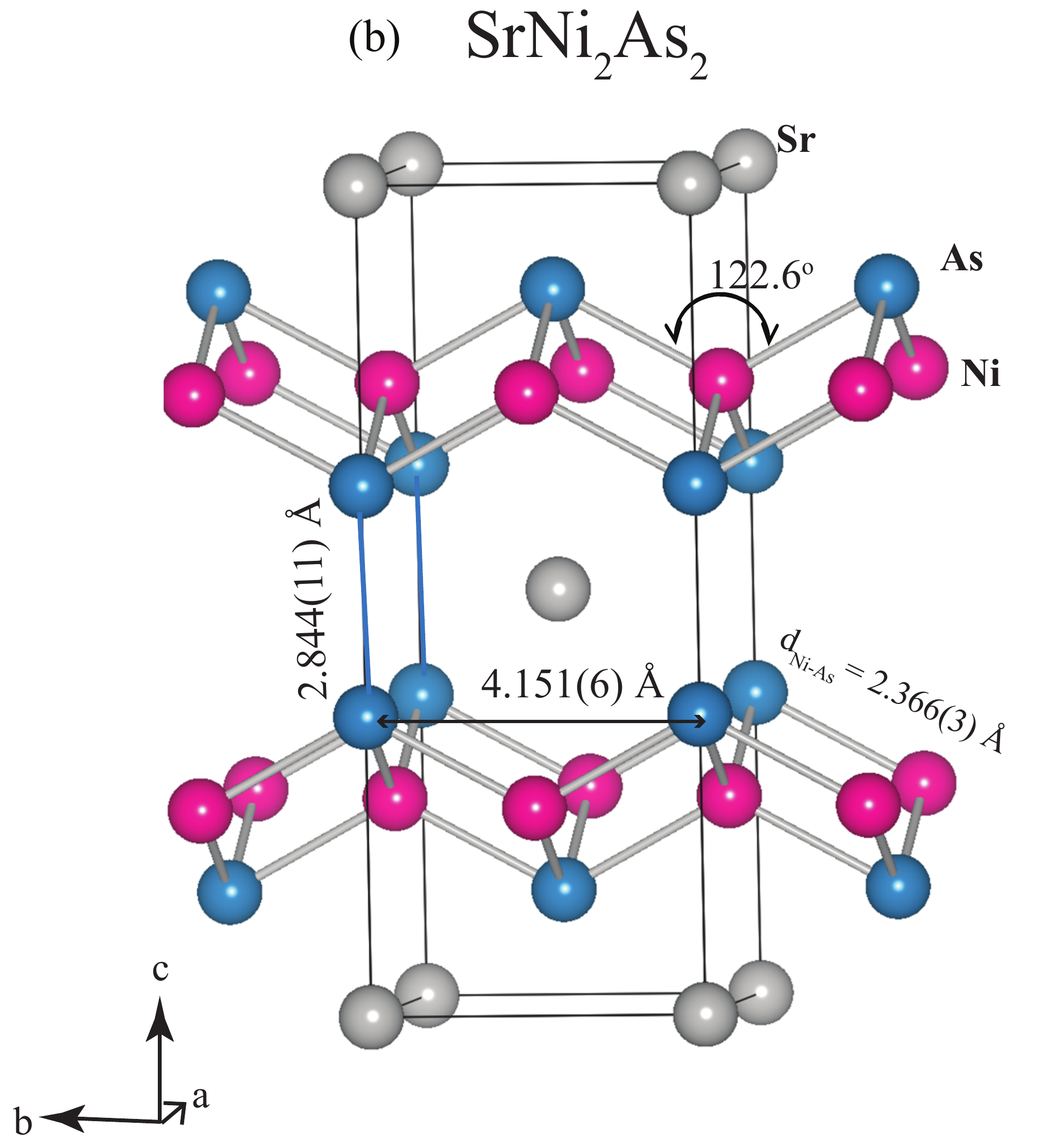}
\caption{Unit cell of the crystal structures of \sca\ and \sna\ together with selected structural parameters.}
\label{unitcell}
\end{figure} 

\begin{figure}
\includegraphics[width=2.5in]{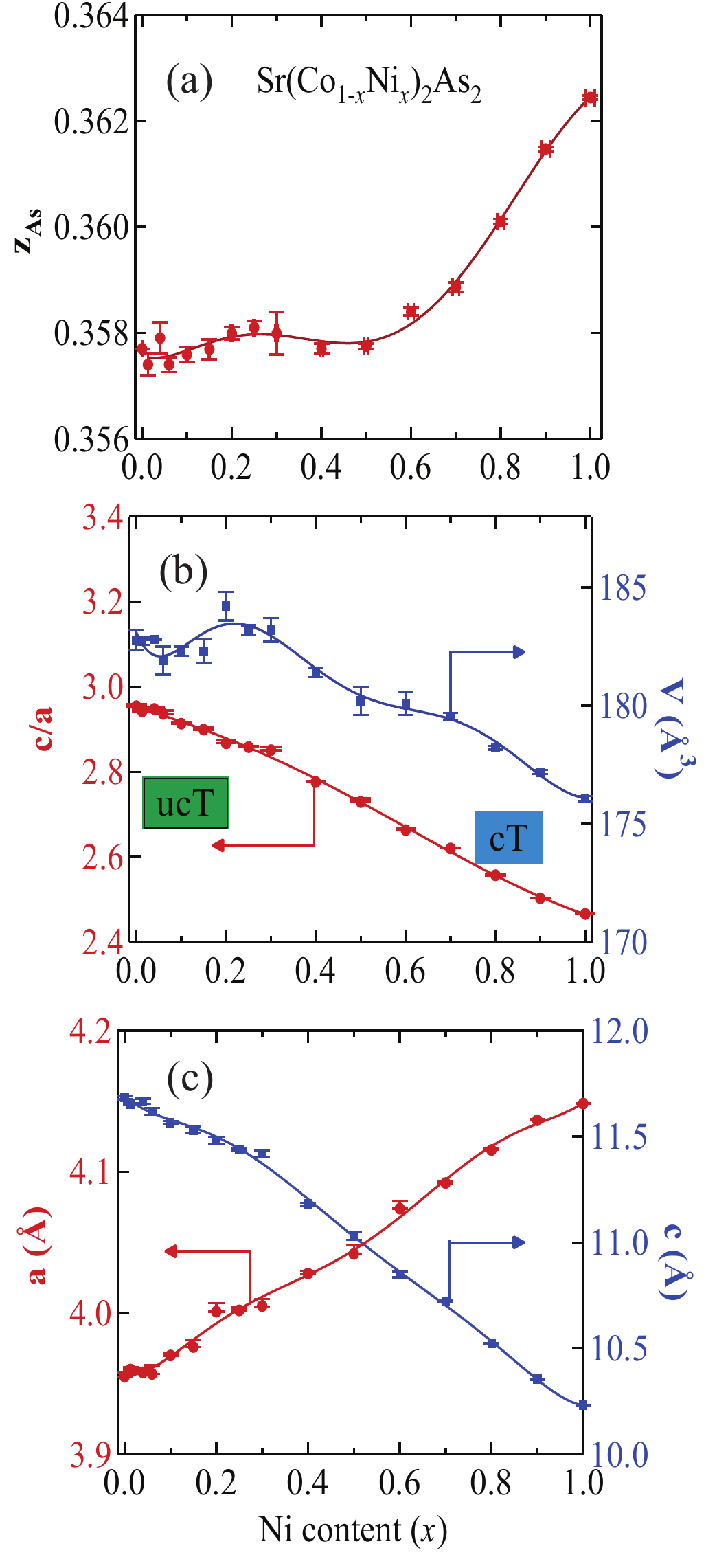}
\caption{Crystallographic parameters for \scna\ crystals versus composition~$x$, including (a)~$z\rm_{As}$, the As \mbox{$c$-axis} position parameter (b)~the $c/a$ ratio and unit-cell volume $V_{\rm cell}$, and (c)~the $a$ and~$c$ lattice parameters.  The solid curves are guides to the eye.  A smooth crossover occurs from the uncollapsed-tetragonal (ucT) to collapsed-tetragonal (cT) structures with increasing~$x$.}
\label{latticeparameter}
\end{figure}

Both \sca\ and \sna\ adopt the tetragonal \tcs\,-type crystal structure and the prospective view of its unit cell with selected lattice parameters are shown in Fig.~\ref{unitcell}. The bonding in the crystal structure has an important impact on the structural and physical properties. The \sca\ structure contains a weak As--As interlayer bonding  due to the long As--As  bond length along the $c$-axis [$d_{\rm As-As}=$ 3.3323(3)~\AA] which results in a two-dimensional (2D) Co-As layered structure separated by Sr atoms as shown in Fig.~\ref{unitcell}(a). Thus \sca\ has an uncollapsed-tetragonal structure (ucT).  On the other hand, \sna\ has a collapsed-tetragonal crystal structure (cT) as it shows a three-dimensional (3D) Ni-As network via the formation of strong interlayer As-As bonds with $d_{\rm As-As}=$ 2.844(11)~\AA\ along the $c$~axis and with a $c/a$ ratio of 2.466, as listed in Table~\ref{CrystalData}.  When the \mbox{As-As} interlayer bonding distance approaches the elemental convalent As-As bond distance or when the $c/a$ ratio is less than 2.67, taking $\rm CaFe_2As_2$ as a reference compound, the system is considered to be in the collapsed-tetragonal phase \cite{Wu2008, Goldman2009, Anand2012}. 

Crystallographic data for \scna\ with \mbox{$x=0$ to~1} obtained from the single-crystal XRD measurements at room temperature are given in Table~\ref{CrystalData}. The tetragonal lattice parameters $a$, $c$, the $c$-axis As positional parameter $z_{\rm As}$, the ratio $c/a$, and the unit cell volume $V_{\rm cell}=a^2c$ are plotted versus~$x$ in Fig.~\ref{latticeparameter}. The  $c$ lattice parameter strongly but nonlinearly decreases, whereas the $a$ lattice parameter strongly increases approximately linearly with~$x$.  These behaviors give rise to the strongly nonlinear variation of $V_{\rm cell}$ with~$x$.  Furthermore, whereas the $c/a$ ratio smoothly and almost linearly decreases with~$x$, the $z_{\rm As}$ varies strongly nonlinearly, suggesting important changes to the electronic properties

The smooth decrease in $c/a$ with $x$ indicates the occurrence of  a continuous structural crossover from the ucT structure at $x=0$ to the cT structure at $x=1$. The values of $c/a$ for $x=0.5$ (2.729) and $x=0.6$ (2.663) are in the vicinity of the crossover ($\approx 2.67$) between the ucT and cT structures \cite{Anand2012}. Therefore, we infer that a structural crossover occurs in \scna\ at $x\approx 0.5$.  It is of interest to study how these structural changes due to electron doping of \sca\ correlate with the ground state properties of the system, as investigated in the following sections.

\section{\label{Sec:magresult} Magnetization and magnetic susceptibility Data}

\begin{figure}
\includegraphics[width=3.45in]{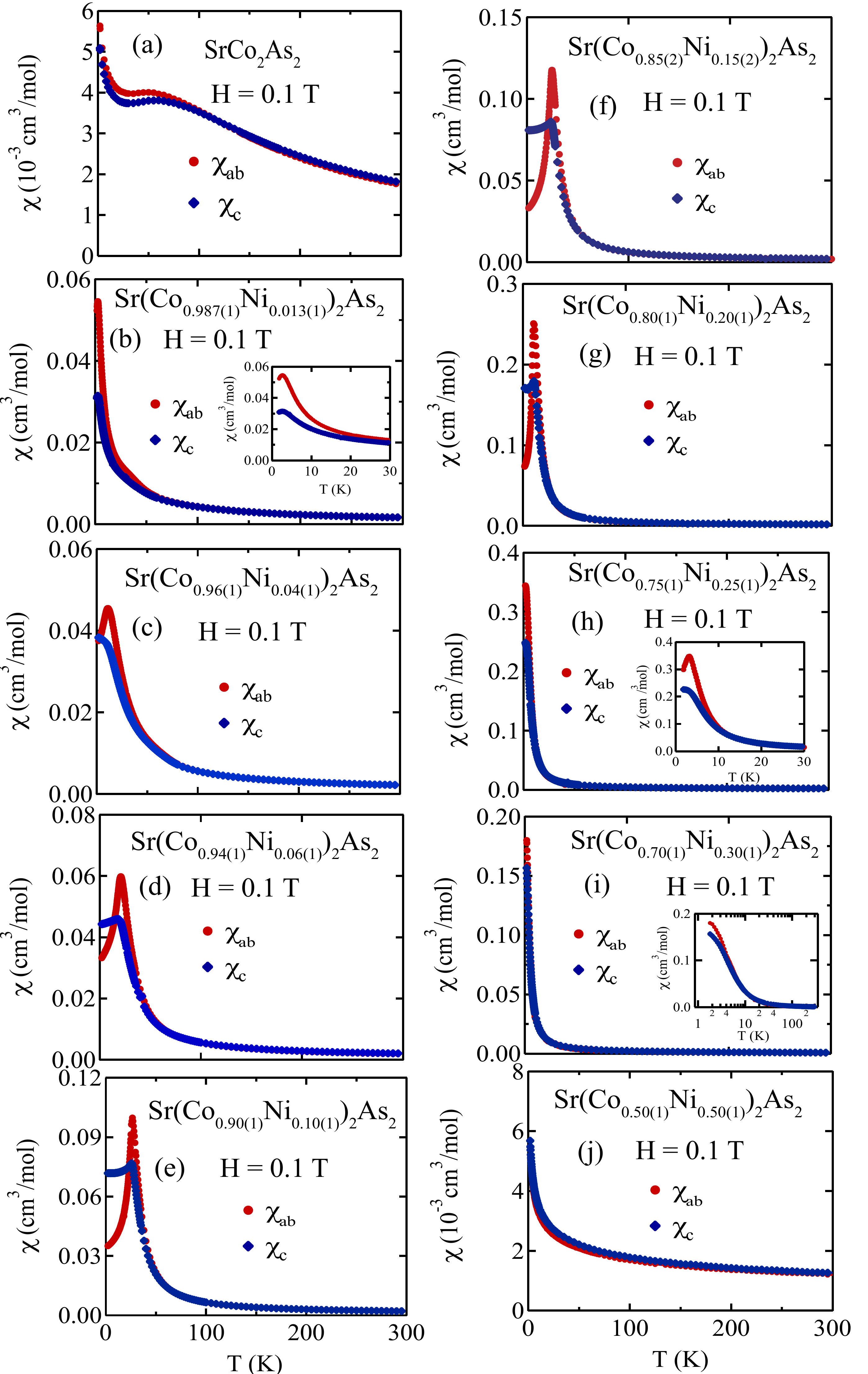}
\caption{Zero-field-cooled (ZFC) magnetic susceptibility $\chi \equiv M/H$ for \scna\ with $x =$ 0, 0.013, 0.04, 0.06, 0.1, 0.15, 0.20, 0.25, 0.30, and 0.50 single crystals as a function of temperature~$T$ from 1.8 to 300 K measured in magnetic fields~$H = 0.1$~T applied in the $ab$~plane ($\chi_{ab}$) and along the $c$~axis ($\chi_c$). } 
\label{Fig.chi}
\end{figure}

\begin{figure}
\includegraphics[width=3in]{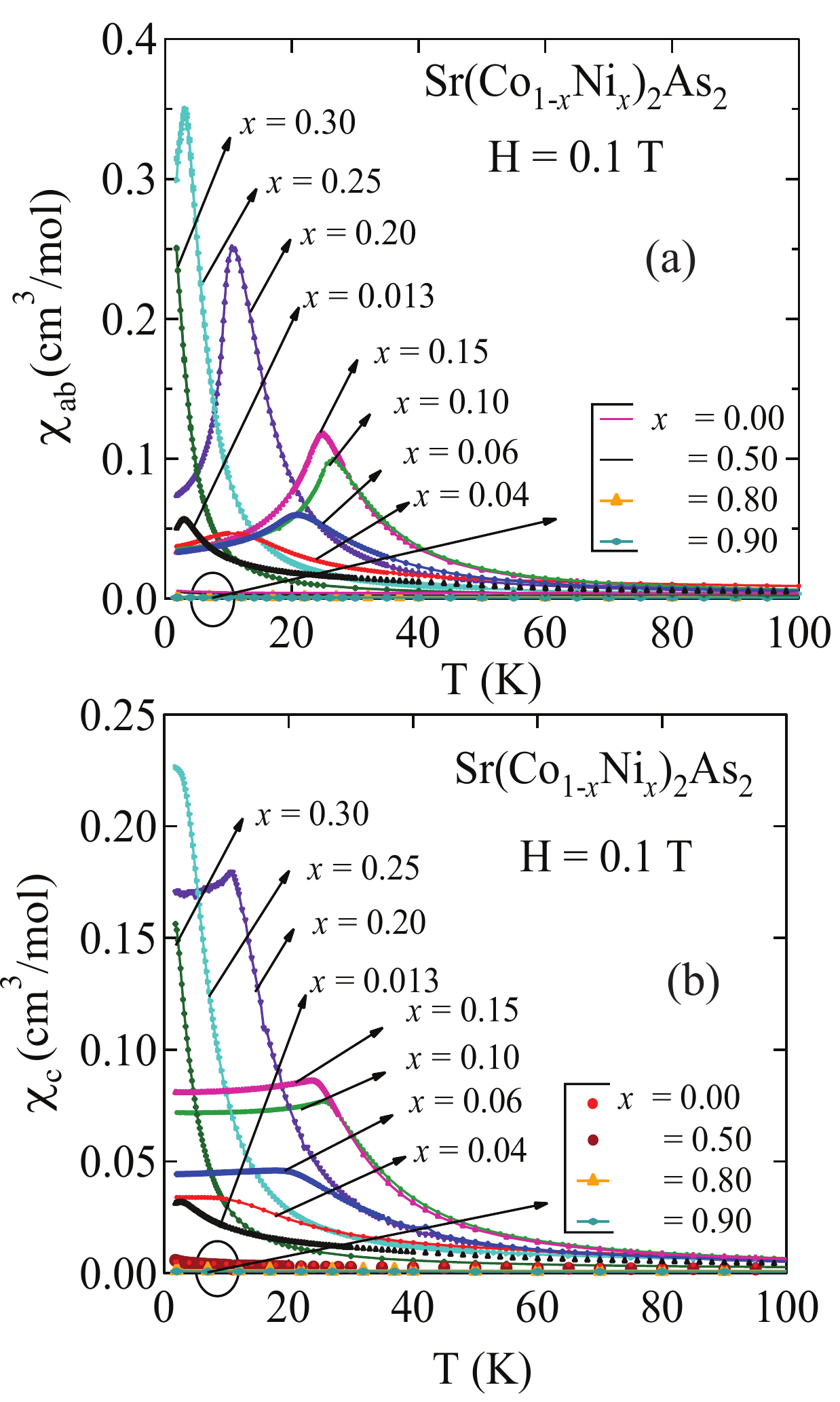}
\caption{Comparison of the temperature~$T$ dependences of the magnetic susceptibility $\chi$ with the applied magnetic field (a)~$H \parallel ab$ and (b)~$H \parallel c$ for \scna\ crystals with $x=0$ to 0.90.}
\label{Fig:chi2}
\end{figure}

The temperature dependence of the magnetic susceptibility $\chi(T) \equiv M(T)/H$ for \scna\ crystals with $x=0$, 0.013, 0.04, 0.06, 0.10, 0.15, 0.20, 0.25, 0.30, and 0.50 measured in an applied magnetic field $H = 0.1$~T for both $H~\parallel~c$ and $H~\parallel~ab$ are shown in Fig.~\ref{Fig.chi}. The self-flux-grown \sca\ crystal in Fig.~\ref{Fig.chi}(a) has a PM ground state with no observable magnetic ordering transitions down to 0.05~K [Y. Furukawa, private communication], supplementing previous data that established this observation for Sn-flux-grown crystals down to 1.8~K~\cite{Pandey2013}. However, our self-flux-grown crystal shows a significantly larger \mbox{low-$T$} upturn than Sn-flux-grown crystals do, suggesting a higher level of magnetic defects in our crystal.  This is consistent with expectation, since our flux was separated from the crystals by centrifuging at 1180~$^\circ$C, whereas the Sn-flux-grown crystals were centrifuged at the much lower temperature of $\sim700~^\circ$C\@.  Probably related to this difference is that the broad maximum in $\chi(T)$ we see at $\sim 60$~K is significantly lower than the temperature of $\approx 115$~K of the broad maximum observed~\cite{Pandey2013} for Sn-flux-grown crystals.

A small peak appears in the $\chi(T)$ data in the inset of Fig.~\ref{Fig.chi}(b) even for the low doping level $x=0.013$, ascribed to an AFM transition at $T\rm_N\approx 3$~K\@. This small peak evolves into larger peaks as the Ni doping increases. The anisotropic $\chi(T)$ data  below $T\rm_N$ for $0.04 \leq x \leq 0.20$ in Figs.~\ref{Fig.chi}(d)--\ref{Fig.chi}(g), where $\chi_c$ is nearly independent of $T$ and $\chi_{ab}$ strongly decreases for $T\rightarrow0$, indicating that the AFM ordered moment lies in the $ab$ plane.  According to molecular-field theory (MFT), the variable nonzero limits of $\chi_{ab}(T\rightarrow 0)/\chi_c(T_{\rm N})$ suggest that the AFM ordering is either an intrinsic coplanar noncollinear $ab$-plane ordering or intrinsic collinear $ab$-plane ordering with multiple AFM domains \cite{{Johnston2015},{Johnston2012}}.  Comparisons of the $ab$-plane and $c$-axis susceptibilities for the crystals are shown in Figs.~\ref{Fig:chi2}(a) and~\ref{Fig:chi2}(b), respectively.

The AFM transition temperatures $T\rm_N$ of \scna\ with $x=0.013$ to 0.25 were estimated from the temperatures of the maxima of $d(\chi T)/dT$ \cite{Fisher1962} and are listed in Table~\ref{Tab.chidata2}. With increasing~$x$, it is seen that $T\rm_N$ initially increases, reaches a maximum of 26.5~K for $x=0.10$, decreases to 3.1~K for $x=0.25$, and finally disappears below our low-$T$ limit of 1.8~K for $x\geq0.30$. The small peak in $\chi(T)$ at $T_{\rm N}\approx 3$~K for $x=0.25$ in the inset of Fig.~\ref{Fig.chi}(h) is similar to that for $x=0.013$ in the inset of Fig.~\ref{Fig.chi}(b). Thus the ground state of \scna\ changes from PM to AFM with only a small amount of Ni substitution ($x=0.013$), which shows that the magnetism of \sca\ is extremely sensitive to electron doping.

It is also apparent from Figs.~\ref{Fig.chi} and~\ref{Fig:chi2} that the $\chi_{ab}(T)$ data  near $T\rm_N$ for the crystals with $x=0.04$ to~0.20 exhibit a strong increase in magnitude compared to crystals with compositions outside this range.  This strong increase may due to the occurrence of strong anisotropic FM spin fluctuations beyond MFT as previously found in the \csca\ system~\cite{Sangeetha2017a} although single-ion uniaxial anisotropy could also contribute to it.  When the ground state of the the system changes from AFM to PM for $x\geq0.50$, the ordinate scale returns approximately to that shown for $x=0$ in Fig.~\ref{Fig.chi}(a).

The crystal-structure studies of \scna\ in Sec.~\ref{Sec:crystalstudy} above showed that the composition $x\sim0.30$ at which AFM order disappears with increasing~$x$ approximately coincides with the composition of the crossover between the ucT and cT structures. Moreover, previous band-structure calculations for \sca\ indicated that the system is magnetically metastable due to the presence of a sharp peak in the DOS near $E_{\rm F}$~\cite{Pandey2013}. We later show in Sec.~\ref{Sec:ElecStruct:Magnetism} that electron doping further increases the magnitude of the DOS peak due to the presence of a flat band at an energy just above $E_{\rm F}(x=0)$. 

\begin{figure}
\includegraphics[width=3.4in]{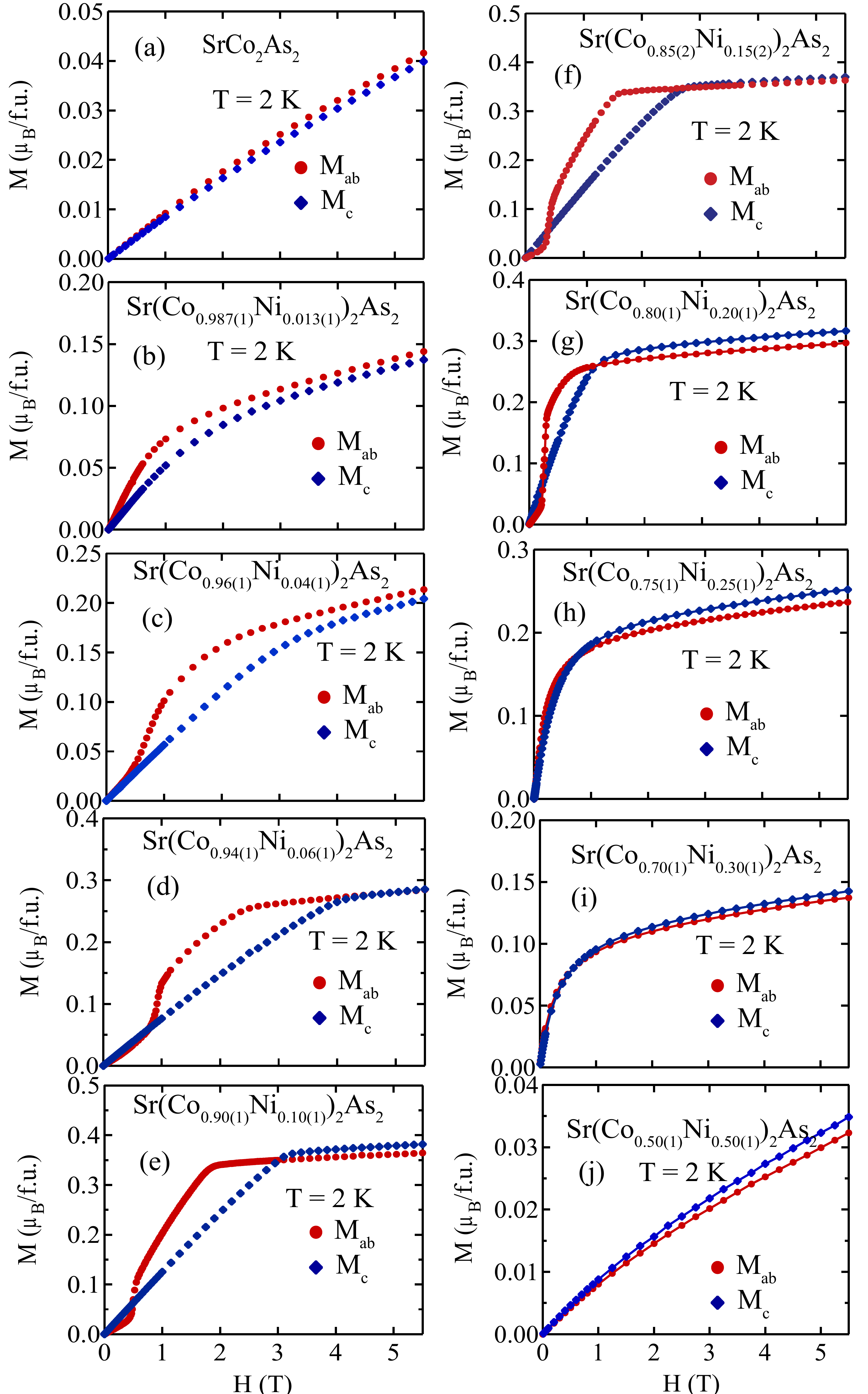}
\caption{ Magnetization $M$ of \scna\ ($x =$ 0, 0.013, 0.04, 0.06, 0.1, 0.15, 0.20, 0.25, 0.30, and 0.50) single crystals  at $T = 2$~K as a function of field $H$ in the range of 0 to 5.5 T applied in the $ab$~plane ($M_{ab}$) and along the $c$~axis ($M_c$).}
\label{Fig.MH}
\end{figure}

\begin{figure}
\includegraphics[width=3in]{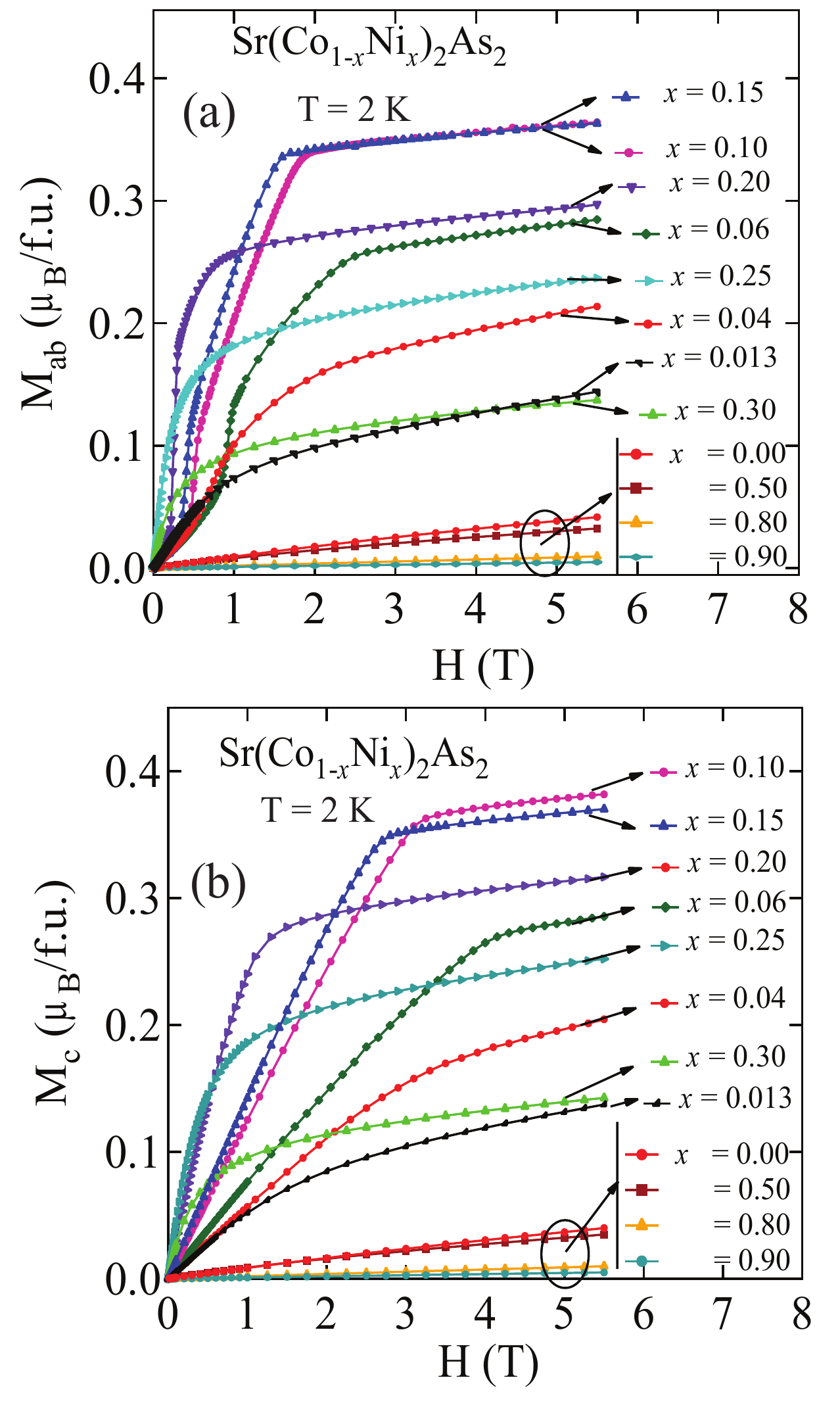}
\caption{Comparisons of the applied magnetic-field dependences of the magnetization with fields \mbox{(a)~$H \parallel ab$} and \mbox{(b)~$H \parallel c$} for \scna\ crystals with $x=0$ to~0.90. }
\label{Fig:MH2}
\end{figure}

In order to clarify the nature of the magnetic behavior observed in Figs.~\ref{Fig.chi} and~\ref{Fig:chi2} and the variation of the magnetic ground state with Ni doping, magnetization versus applied magnetic field $M(H)$ isotherms were measured at $T=2$~K for \scna\ crystals with compositions $x=0$ to~0.5 with both $H~||~ab$ and $H~||~c$ as shown in Fig.~\ref{Fig.MH}. $M$ is proportional to $H$ for $x=0$, consistent with results from the previous report~\cite{Pandey2013}. From the $\chi(T)$ data in Fig.~\ref{Fig.chi}, the system orders antiferromagnetically already at $x=0.013$ which is supported by the strongly nonlinear $M(H)$ data in Fig.~\ref{Fig.MH}(b). Furthermore, the crystals with $x=0.04$, 0.06, 0.10, 0.15, and 0.20 show clear evidence for spin-flop transitions, indicative of $ab$-plane AFM ordering in agreement with our discussion of the anisotropic $\chi(T)$ above.

Comparisons of the magnetization data for the different crystals are shown for $ab$-plane and $c$-axis fields in Fig.~\ref{Fig:MH2}.  One sees that the maximum magnitude of $M_{ab}$ up to $H=5.5$~T increases upon Ni doping  and reaches a maximum at $x=0.15$, similar to our $\chi(T)$ data in Fig.~\ref{Fig:chi2}, then decreases with further Ni doping. For $x\geq0.25$, the system is paramagnetic at all tempertures. 

\begin{figure}
\includegraphics[width=2.75in]{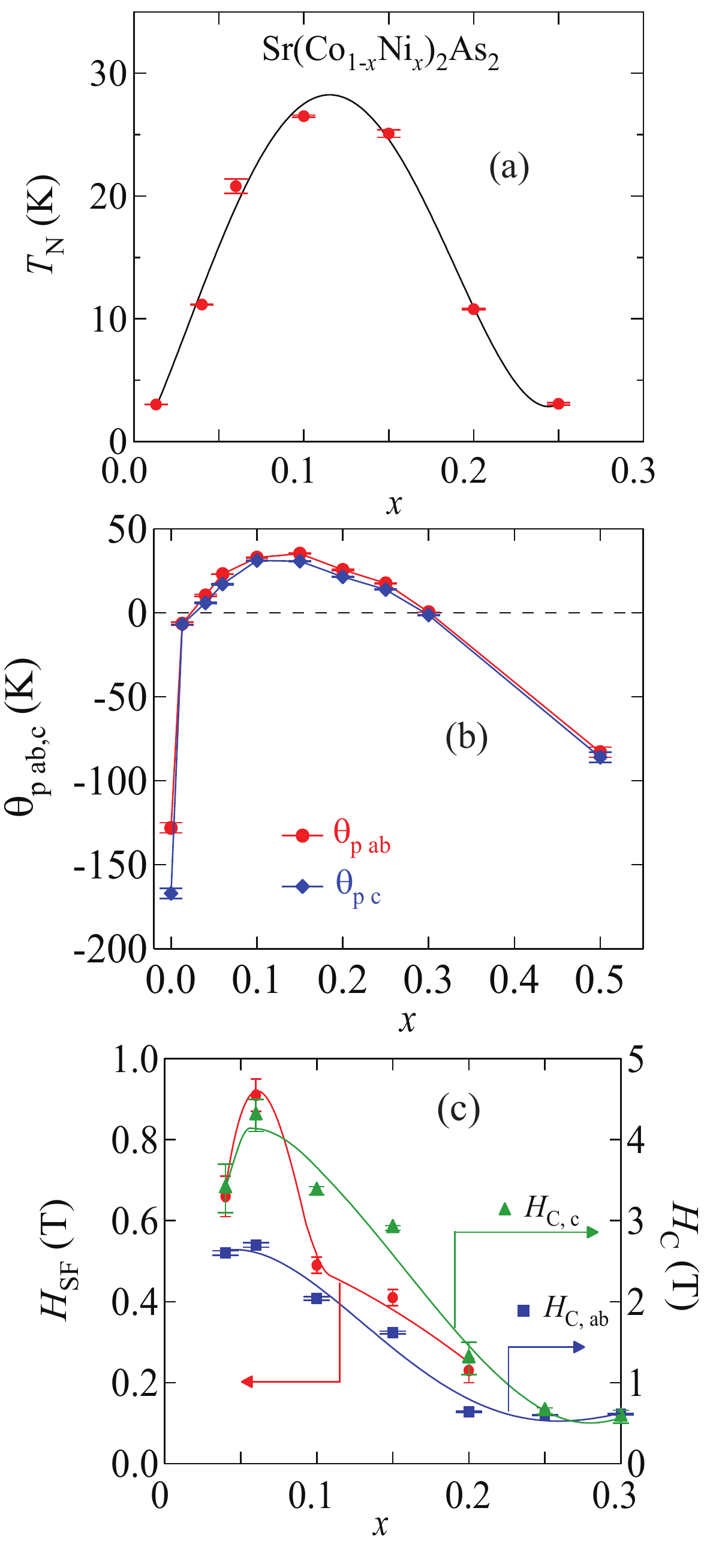}
\caption{Composition $x$ dependences of the (a)~AFM transition temperature $T_{\rm N}$, (b)~Weiss temperatures $\theta_{{\rm p},ab}$ and $\theta_{{\rm p},c}$, and~(c)~spin-flop field $H_{\rm SF}$ and critical field $H_{\rm c}$ for \scna\ crystals.  The lines are guides to the eye.}
\label{Fig:MPars1}
\end{figure}

\begin{figure}
\includegraphics[width=2.75in]{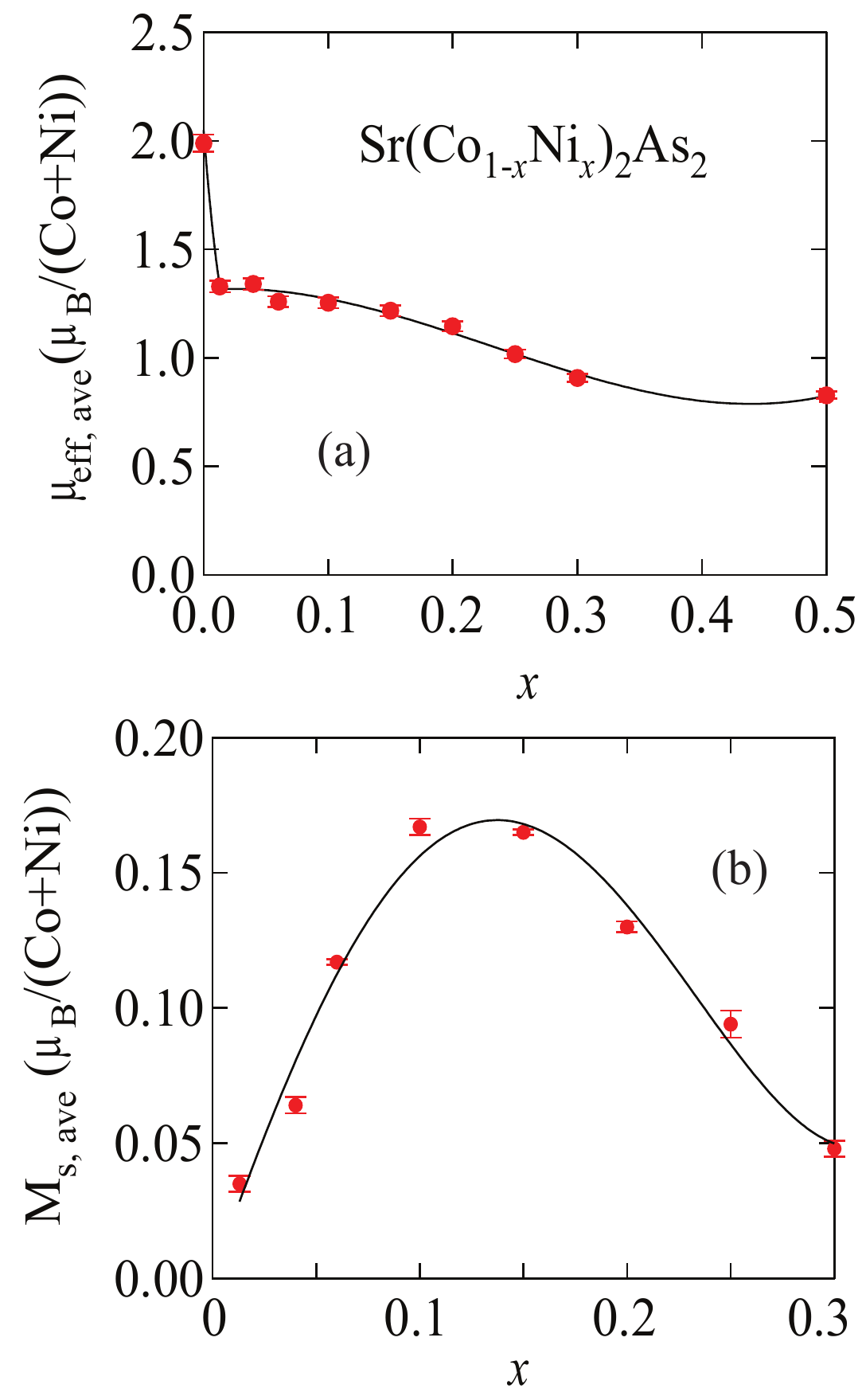}
\caption{Composition $x$ dependences of the angle-averaged (a)~effective moment per transition metal atom $\mu_{\rm eff,ave}$ and (b)~saturation moment per transition metal atom $M_{\rm sat,ave}$ for \scna\ crystals. The lines are guides to the eye.}
\label{Fig:MPars2}
\end{figure}

The parameters obtained from the $M(H)$ data are the spin-flop field $H_{\rm SF}$, the critical field $H_{\rm c}$ at which $M_{ab}$ becomes approximately equal to $M_c$ with increasing~$H$, the saturation magnetic moment $M_{\rm sat}$ obtained from a linear exprapolation of $M(H)$ at $T=2$~K at high fields to $H=0$, and the angle-averaged saturation moment $M_{\rm sat,ave}$ as listed in Table~\ref{Tab.MH} and plotted in Figs.~\ref{Fig:MPars1} and~ ~\ref{Fig:MPars2}. Besides, the spin-flop field $H_{\rm SF}$ first increases from 0.66~T ($x=0.04$) to 0.91~T ($x=0.06$), then decreases to 0.23~T ($x=0.20$) and gradually disappears for $x\geq0.20$. 

In the following sections the $\chi(T)$ and $M(H)$ isotherm data presented above together with additional $M(H)$ isotherm data up to $H=14$~T are analyzed first from a local-moment perspective and then from an itinerant magnetism point of view.

\section{\label{Sec:LocMomMag} Local-Moment Magnetism Model}

\subsection{\label{Sec:CW} Curie-Weiss Law for the Paramagnetic State}

Here we analyze the magnetic susceptibility $\chi_\alpha(T)$ for \scna\ at high temperatures (in the PM state) within the local-moment Heisenberg model in terms of the modified Curie-Weiss law
\begin{equation}
\chi_{\alpha}(T) =\chi_0 + \frac{C_{\alpha}}{T-\theta_{\rm p\alpha}} \quad (\alpha = ab,\ c),
\label{Eq:ChiFit}
\end{equation}
where $\chi_0$ is a $T$-independent contribution, $C_\alpha$ is the Curie constant with the field in the $\alpha$ principal-axis direction where $\alpha$ is included to take into account possible magnetic anisotropies, and $\theta_{\rm p\alpha}$ is the Weiss temperature arising from interactions between the spins.  For local Heisenberg spins with spin angular momentum quantum number~$S$ in units of $\hbar$, the Curie constant is given in Gaussian cgs units by
\bse
\begin{equation}
C\left[{\rm{\frac{cm^3\,K}{mol~spins}}}\right] = 0.5002\,S(S+1).
\label{Eq:Ccalc}
\end{equation}
Thus for the minimum quantum spin $S=1/2$ one obtains
\begin{equation}
C(S=1/2) = 0.375\,{\rm cm^3\,K/(mol~spins)}.
\label{Eq:CS12}
\end{equation}
\ese

\begin{figure}
\includegraphics[width=3.4in]{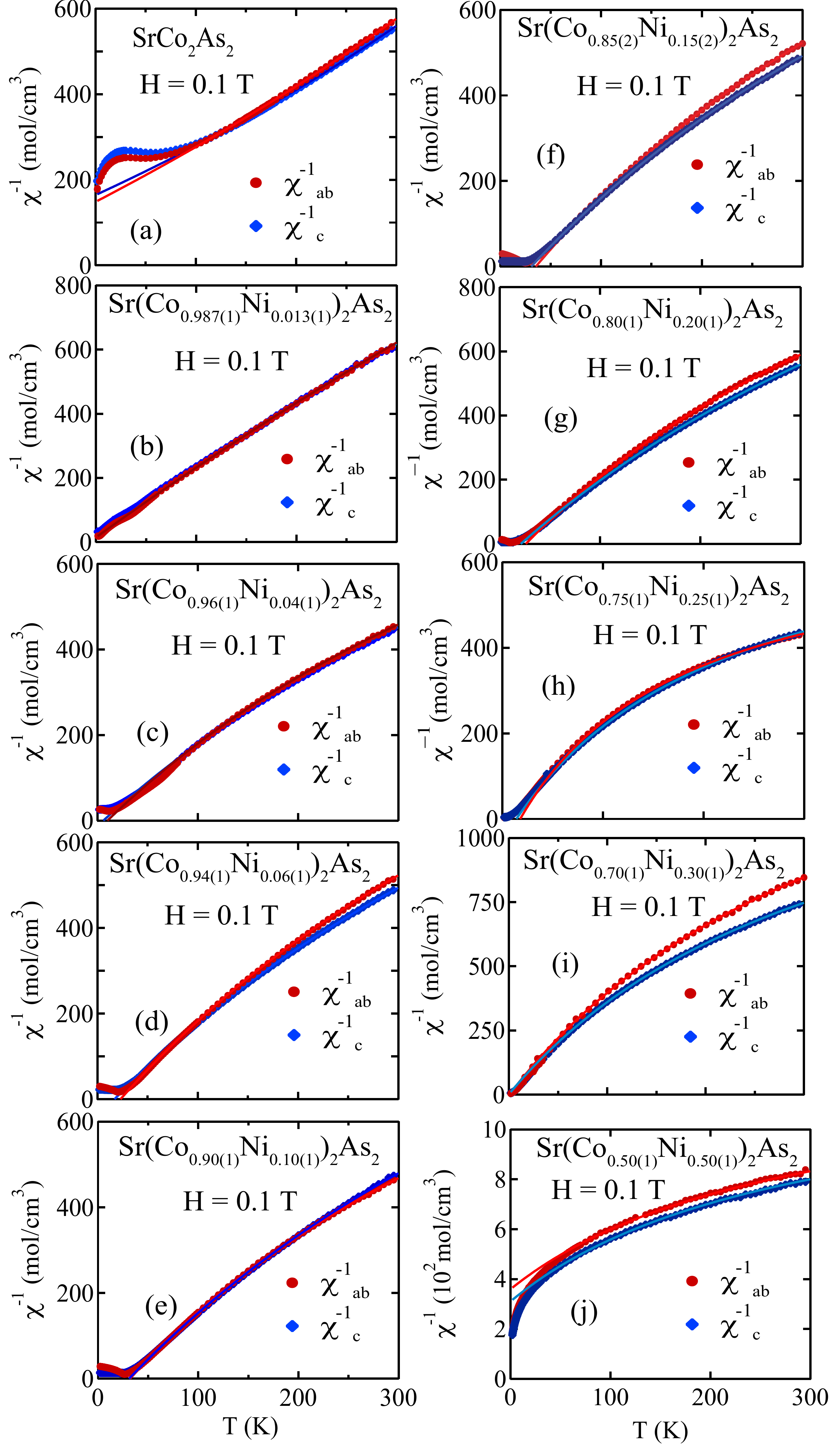}
\caption{Zero-field-cooled (ZFC) magnetic susceptibility $\chi \equiv M/H$ for \scna\  with $x =$ 0, 0.013, 0.04, 0.06, 0.1, 0.15, 0.20, 0.25, 0.30, and 0.50 single crystals as a function of temperature~$T$ from 1.8 to 300 K measured in magnetic fields~$H = 0.1$~T applied in the $ab$~plane ($\chi_{ab}$) and along the $c$~axis ($\chi_c$).}
\label{Fig.Invchi}
\end{figure}

Figure \ref{Fig.Invchi} shows the inverse susceptibility $\chi^{-1}(T)$ at $H~=$~0.1~T applied along the $c$ axis $(\chi^{-1}_c$) and in the $ab$ plane ($\chi^{-1}_{ab}$) as a function of $T$ from 1.8 to 300~K\@. As one can see from the figures, the $\chi^{-1}(T)$ plots are slightly curved, which can be accounted for by a nonzero $\chi_0$ in Eq.~(\ref{Eq:ChiFit}).  The high-temperature (100 K $<T<$ 300 K) $\chi(T)$ data for \scna\ with $x=0$ to 0.50 for both field directions fitted by Eq.~(\ref{Eq:ChiFit}) are shown as the solid lines in Fig.~\ref{Fig.Invchi}. The fitting parameters $\chi_0$, $C_{\alpha}$, and $\theta_{\rm p \alpha}$ are listed in Tables~\ref{Tab.chidata1} and \ref{Tab.chidata2} together with the effective Bohr magneton number per transition-metal atom $p_{\rm eff\alpha}$ calculated from $C_{\alpha}$ using Eq.~(\ref{Eq:peffalpha}) below. The parameters obtained for \sca\ are consistent with the previous single crystal results \cite{Pandey2013}. The temperature independent susceptibility $\chi_0$ is negative for \sca\ but becomes positive upon Ni doping.  For \scna , a Co(Ni) atom in the Co$^{+1}$(Ni$^{+2}$) state is expected to have $S = 1$. As a result, the Curie constant $C=0.5002~ g^2 S(S+1)/4~ {\rm cm^3 K/mol} = 1~ {\rm cm^3 K/mol}~(g=2)$. On the other hand, a low spin state $S=\frac{1}{2}$  for Co$^{+2}$ (Ni$^{+1}$), $C = 0.375~{\rm cm^3\,K/mol}$ which is close to the observed values for $0.013\leq x \leq 0.20$. 

The Curie constant $C_\alpha$ is conventionally given per mole of formula units by
\bse
\label{Eqs:Calpha}
\begin{equation}
C_{\alpha} = \frac{N_{\rm A} g_\alpha^2 S(S+1)\mu^2_{\rm B}}{3k_{\rm B}},
\label{Eq.Cvalue1}
\end{equation}
where $N_{\rm A}$ is Avogadro's number, $g$ is the spectroscopic splitting factor ($g$ factor), $S$ is the spin quantum number of the local moment, $\mu_{\rm B}$ is the Bohr magneton, and $k_{\rm B}$ is Boltzmann's constant. A convenient parameterization of the Curie constant is in terms of the effective moment per formula unit $\mu_{\rm eff}^{\rm f.u.}$, given by
\begin{equation}
C_\alpha = \frac{N_{\rm A}{\mu_{\rm eff}^{\rm f.u.}}^2}{3k_{\rm B}} = \frac{N_{\rm A}{p_{\rm eff}^{\rm f.u.}}^2\mu_{\rm B}^2}{3k_{\rm B}},
\label{Eq:peff}
\end{equation}
\ese
where $p_{\rm eff}^{\rm f.u.}$ is the effective Bohr magneton number per formula unit.  If there are $n$ magnetic atoms per f.u., then the effective Bohr magneton number per magnetic atom is given by Eqs.~(\ref{Eq.Cvalue1}) and~(\ref{Eq:peff}) as
\begin{equation}
p_{\rm eff\alpha} = \sqrt{\frac{3k_{\rm B}C_\alpha}{nN_{\rm A}\mu_{\rm B}^2}} \approx \sqrt{\frac{8C_\alpha}{n}},
\label{Eq:peffalpha}
\end{equation}
where $C_\alpha$ is in cgs units of ${\rm cm^3\,K/mol~f.u.}$ and the cgs values of the fundamental constants were used.  In this paper $n=2$ for two transition metal atoms per f.u.\ of \scna.  Using Eq.~(\ref{Eq:CS12}), the value of $p_{\rm eff}$ for a local spin-1/2 with $g=2$ is
\begin{equation}
p_{\rm eff}(S=1/2) = 1.73.
\label{Eq:peffS12}
\end{equation}

The values of $p_{\rm eff\alpha}$ obtained from the respective Curie constants are listed in Tables~\ref{Tab.chidata1} and \ref{Tab.chidata2}.  With the exception of pure ${\rm SrCo_2As_2}$, the values of $p_{\rm eff\alpha}$ for our \scna\ crystals are all significantly smaller than the value in Eq.~(\ref{Eq:peffS12}), suggesting itinerant magnetism instead of local-moment magnetism for these compositions.

For later discussion, from Eqs.~(\ref{Eqs:Calpha}) one has
\begin{equation}
p_{\rm eff\alpha} = g_\alpha\sqrt{S(S+1)}.
\end{equation}
Taking a spherical average eliminates the effects of most sources of magnetic anisotropy and one obtains
\begin{equation}
p_{\rm eff} = g\sqrt{S(S+1)} = \sqrt{gS(gS+g)}.
\end{equation}
The saturation Bohr magneton number $p_{\rm sat}$ (low-$T$ ordered moment divided by $\mu_{\rm B}$)  is
\begin{equation}
p_{\rm sat}=gS,
\end{equation}
yielding
\bse
\begin{equation}
p_{\rm eff} =\sqrt{p_{\rm sat}(p_{\rm sat}+g)},
\end{equation}
from which one obtains 
\begin{equation}
p_{\rm sat} = \frac{g}{2}\left(\sqrt{1+\frac{4}{g^2}p_{\rm eff}^2}-1\right).
\end{equation}
If one sets $g=2$ for itinerant carriers and also for Heisenberg spins, one obtains the relation
\begin{equation}
p_{\rm sat} = \sqrt{1+p_{\rm eff}^2}-1\qquad (g=2).
\end{equation}
In the weak itinerant FM analyses discussed below in Sec.~\ref{Sec:itinerantMag}, one defines the quantity
\begin{equation}
p_{\rm c} = \sqrt{1+p_{\rm eff}^2}-1\qquad (g=2),
\label{Eq:pcDef}
\end{equation}
\ese
where $p_{\rm eff}$ is obtained by fitting the PM susceptibility by the modified Curie-Weiss law~(\ref{Eq:ChiFit}).  Thus  $p_{\rm c}$ is equal to the measured saturation moment $p_{\rm sat}$ for a local-moment system but not generally for an itinerant-electron FM\@.  Hence the ratio $p_{\rm c}/p_{\rm sat}$ is an indicator of where a particular FM lies in the local-moment/itinerant-moment continuum.

From Table~\ref{Tab.chidata1}, \sca\ has a large negative Weiss temperature, suggesting that AFM interations are dominant in the system which is consistent with inelastic neutron scattering measurements where the presence of stripe-type AFM fluctuations was found \cite{Jayasekara2013}. However, the Weiss temperature changes rapidly from $\theta_{\rm p, ave}= -141$~K ($x=0$) to $\theta_{\rm p,ave}=-6.6(2)$~K with a small amount of Ni doping $x=0.013$, but is still negative. Remarkably, as $x$ increases into the range $0.04\leq x \leq0.25$, the Weiss temperature becomes positive (FM-like) and the spherical average increases from  $\theta_{\rm p, ave}=8.9$~K ($x=0.04$) to 33.7~K ($x =0.15$) and then decreases to 16.3~K for $x =0.25$. The positive Weiss temperatures indicate the presence of predominantly FM exchange interactions, in particular for $x=0.10$ at which $T_{\rm N}$ reaches its maximum value.  For $x\geq0.30$, where the system is not magnetically ordered, $\theta_{\rm p,ave}$ returns to negative values.  Paradoxically, \scna\ thus exhibits the maximum AFM ordering temperatures at compositions at which FM exchange interactions dominate over AFM interactions.  The anisotropies in the Curie constants $C_\alpha$ and in the Weiss temperatures~$\theta_{\rm p\alpha}$ in Table~\ref{Tab.chidata1} may arise from $g$~anisotropy, single-ion uniaxial anisotropy, anisotropic magnetic dipole interactions, and/or from exchange anisotropy.

\subsection{Molecular-Field-Theory Analysis of Magnetic Susceptibility in the Ordered State in Terms of Helical AFM Ordering}

In this section we use the so-called unified molecular field theory (MFT) for local-moment AFMs developed by one of the authors~\cite{Johnston2012, Johnston2015}. This theory applies to systems of identical crystallographically-equivalent Heisenberg spins interacting by Heisenberg exchange and  does not use the concept of magnetic sublattices. Instead, the zero-field magnetic properties are calculated solely from the Heisenberg exchange interactions between an arbitrary spin and its neighbors. This formulation will be used to calculate the anisotropic $\chi(T)$ at temperatures $T\leq T_{\rm N}$ arising from planar helical AFM ordering with the ordered moments aligned in the $ab$ plane where the helix axis is the $c$~axis.

\begin{figure}
\includegraphics [width=2in]{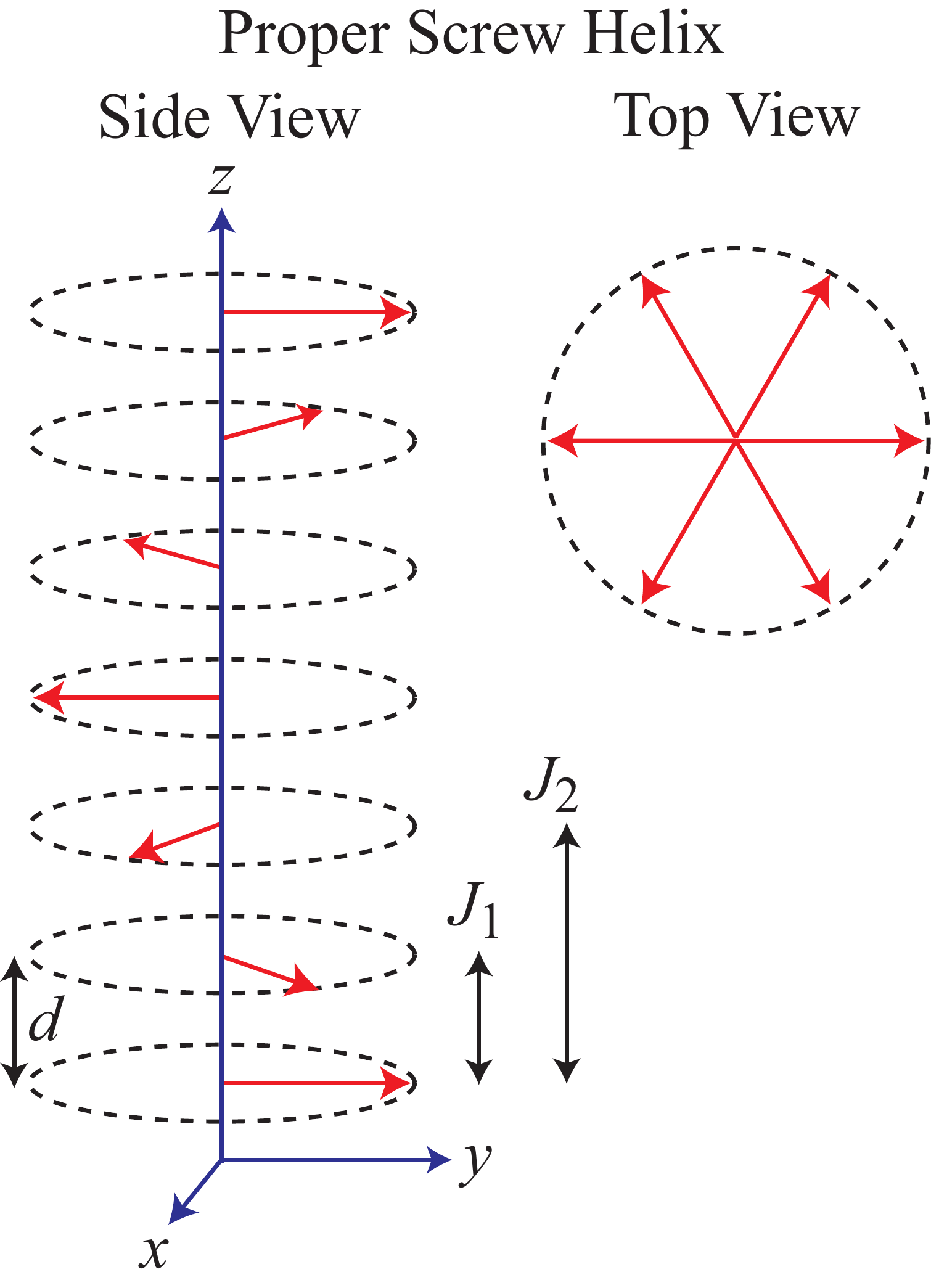}
\caption {Generic helical AFM structure \cite{Johnston2012}.  Each arrow represents a layer of moments perpendicular to the $z$~axis that are ferromagnetically aligned within the $xy$ plane and with interlayer separation $d$.  The wave vector {\bf k} of the helix is directed along the $z$~axis.  The magnetic moment turn angle between adjacent magnetic layers is $kd$.  The nearest-layer and next-nearest-layer exchange interactions $J_{1}$ and $J_{2}$, respectively, within the $J_0$-$J_{1}$-$J_{2}$ Heisenberg MFT model are indicated.  The top view is a hodograph of the magnetic moments. }
\label{Fig:J0_Jz1_Jz2_model_helix}
\end{figure}

A picture of a $z$-axis helix is shown in Fig.~\ref{Fig:J0_Jz1_Jz2_model_helix}, where here the $xy$~plane corresponds to the $ab$~plane and the $z$~axis to the $c$~axis.  Within the classical $J_0$-$J_1$-$J_2$ local-moment mean-field Heisenberg model described in the figure caption, the turn angle $kd$ between adjacent FM-aligned planes is given by~\cite{Johnston2012, Johnston2015}
\begin{equation}
kd = {\rm arccos}\left(-\frac{J_1}{4J_2}\right),
\label{Eq:kdJ1J2}
\end{equation}
where $J_2$ must be positive (AFM) for a helical structure, whereas $J_1$ can be either negative (FM) or positive.  Within MFT, the $c$-axis susceptibility $\chi_{Jc}$ of the helix is independent of temperature for $T\leq T_{\rm N}$, whereas the ratio of the in-plane zero-field susceptibility $\chi_{Jab}$ at $T=0$ to $\chi_J(T_{\rm N})$ is~\cite{Johnston2012, Johnston2015}
\begin{equation}
\frac{\chi_{Jab}(T=0)}{\chi_J(T_{\rm N})}=\frac{1}{2\left[1+2~{\rm cos}(kd)+2~{\rm cos}^2(kd)\right]},
\label{Eq:kd}
\end{equation}
and the subscript~$J$ indicates that anisotropy and \mbox{$T$-independent} terms in $\chi(T\geq T_{\rm N})$ are not present in the data or have been corrected for.  In our case the anisotropy in the PM state is so small it can be ignored here.

\begin{figure}
\includegraphics[width=3.45in]{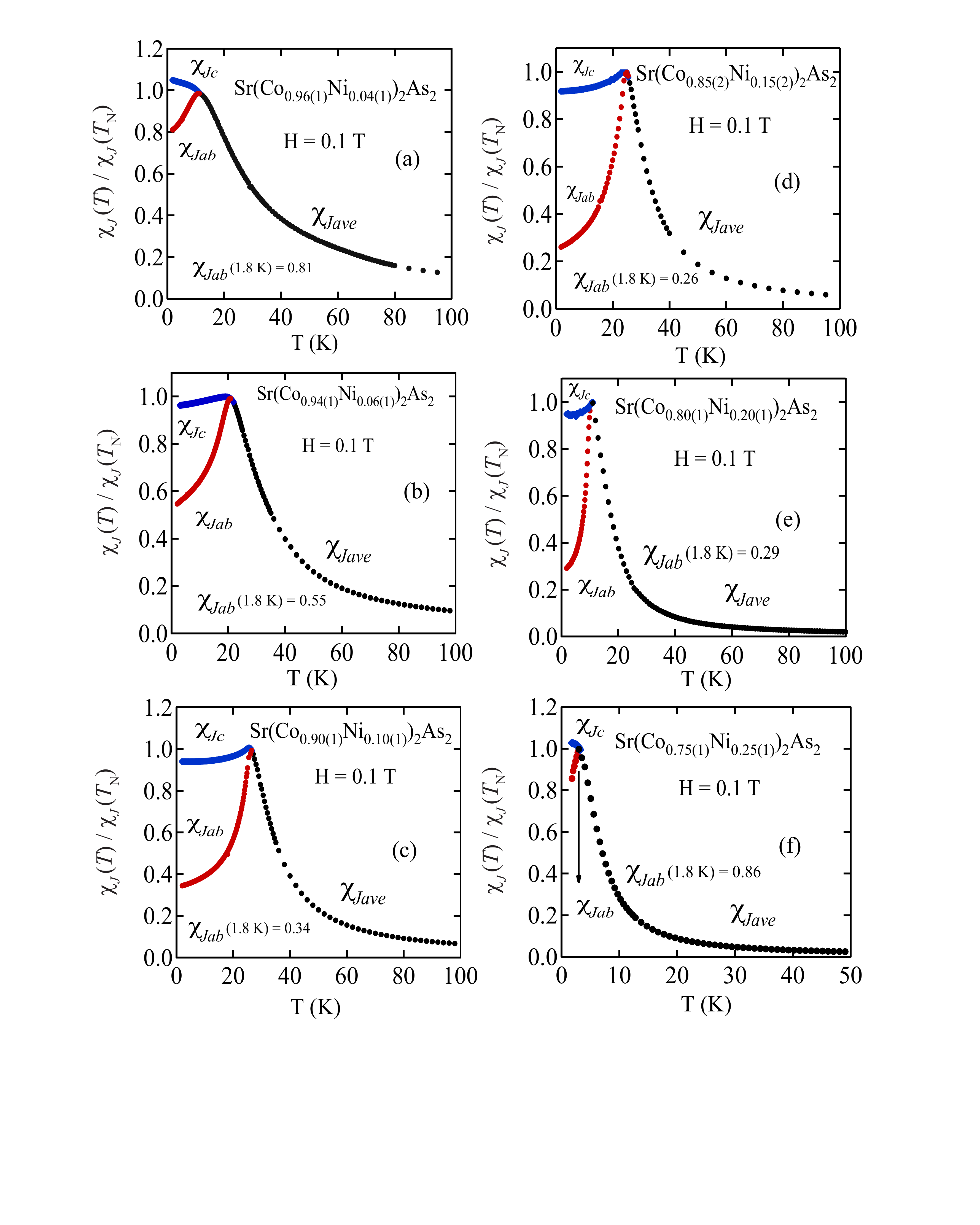}
\caption{Normalized susceptibility $\chi_J(T)/\chi_J(T_{\rm N})$ versus $T$ of \scna\ for $x =0.04$, 0.06, 0.1, 0.15, 0.20, and 0.25 at applied field $H=0.1$ T in the $ab$ plane $\chi_{Jab}$ and along the $c$ axis $\chi_{Jc}$.}
\label{Fig.chiTNN}
\end{figure}

In order to compare our $\chi(T\leq T_{\rm N})$ data in Fig.~\ref{Fig.chi} with Eq.~(\ref{Eq:kd}), we first correct for the $T$-independent susceptibility $\chi_0$ and the anisotropy at $T> T_{\rm N}$.  $\chi_0$ is taken into account at all temperatures according to
\bse
\begin{equation}
\chi_{J\alpha}(T)=\chi_{\alpha}(T) - \chi_{0\alpha}\qquad(\alpha=ab~{\rm or}~c),
\end{equation}
where $\chi_{\alpha}(T)$ is the measured susceptibility and the $\chi_{0\alpha}$ values are given in Table~\ref{Tab.chidata1}. The remainig anisotropy in the PM state is likely due to magnetic dipole interactions and/or single-ion quantum uniaxial anisotropies, for which the magnetic susceptibility tenser is traceless in the PM state. Then one obtains the Heisenberg susceptibility $\chi_J$ in the PM state according to
\begin{equation}
\chi_J(T\geq T_{\rm N}) = \frac{1}{3}[2\chi_{ab}(T)+\chi_c(T)].
\end{equation}
\ese

Figure \ref{Fig.chiTNN} shows the normalized $\chi_{Jab}(T)/\chi_J(T_{\rm N})$ for \scna\ with $x=0.04$, 0.06, 0.10, 0.15, 0.20, and 0.25 after correcting for $\chi_0$ and the anisotropy in the PM state as discussed above.  One can now estimate the helix turn angle $kd$ from $\chi_J(T=1.8~{\rm K})/\chi_J(T_{\rm N})$ using Eq.~(\ref{Eq:kd}) and the results are listed in Table~\ref{Tab.chidata2}, where $kd$ is seen to strongly depend on the Ni content~$x$. The magnetic structure is evidently incommensurate because the values of $kd$ are irrational.

The above discussion of helical ordering is necessarily only of qualitative significance, since the MFT was formulated for local-moment systems instead of itinerant-electron antiferromagnets.  In any case, magnetic neutron diffraction studies would be very interesting to test the above suggestions about the magnetic structure.

\section{\label{Sec:itinerantMag}Itinerant Ferromagnetism Model}

The $\chi^{-1}(T)$ data in Fig.~\ref{Fig.Invchi} for our \scna\ crystals follow Curie-Weiss-like behavior in the PM state with $T>T_{\rm N}$ with $p_{\rm eff}$ values as given in Table~\ref{Tab.chidata1} which are smaller than expected for the minimum local spin $S=1/2$ in Eq.~(\ref{Eq:peffS12}), suggestive of itinerant moments. Furthermore, the small angle-averaged saturation moments $\mu_{\rm sat,ave} \sim 0.1~\mu_{\rm B}$ per transition metal atom in Table~\ref{Tab.MH} also suggest itinerant magnetism.  Thus the magnetism in \scna\ is likely itinerant rather than arising from local moments.  In addition, the Weiss temperatures for the crystals with the larger $T_{\rm N}$ values are positive, indicating dominant FM interactions.  Therefore in this section we analyze the high-field $M(H)$ data and the $\chi(T)$ data in the PM regime for \scna\ using Takahashi's spin-fluctuation theory of weak (small ordered moment) itinerant ferromagnets~\cite{Takahashi2013}. 

\subsection{$M^2$ vs $H/M$ Isotherms,  $M^4$ vs $H/M$ Isotherms, and the Curie Temperature for Weak Itinerant Ferromagnetism in \scna}

\begin{figure}
\includegraphics[width=3.45in]{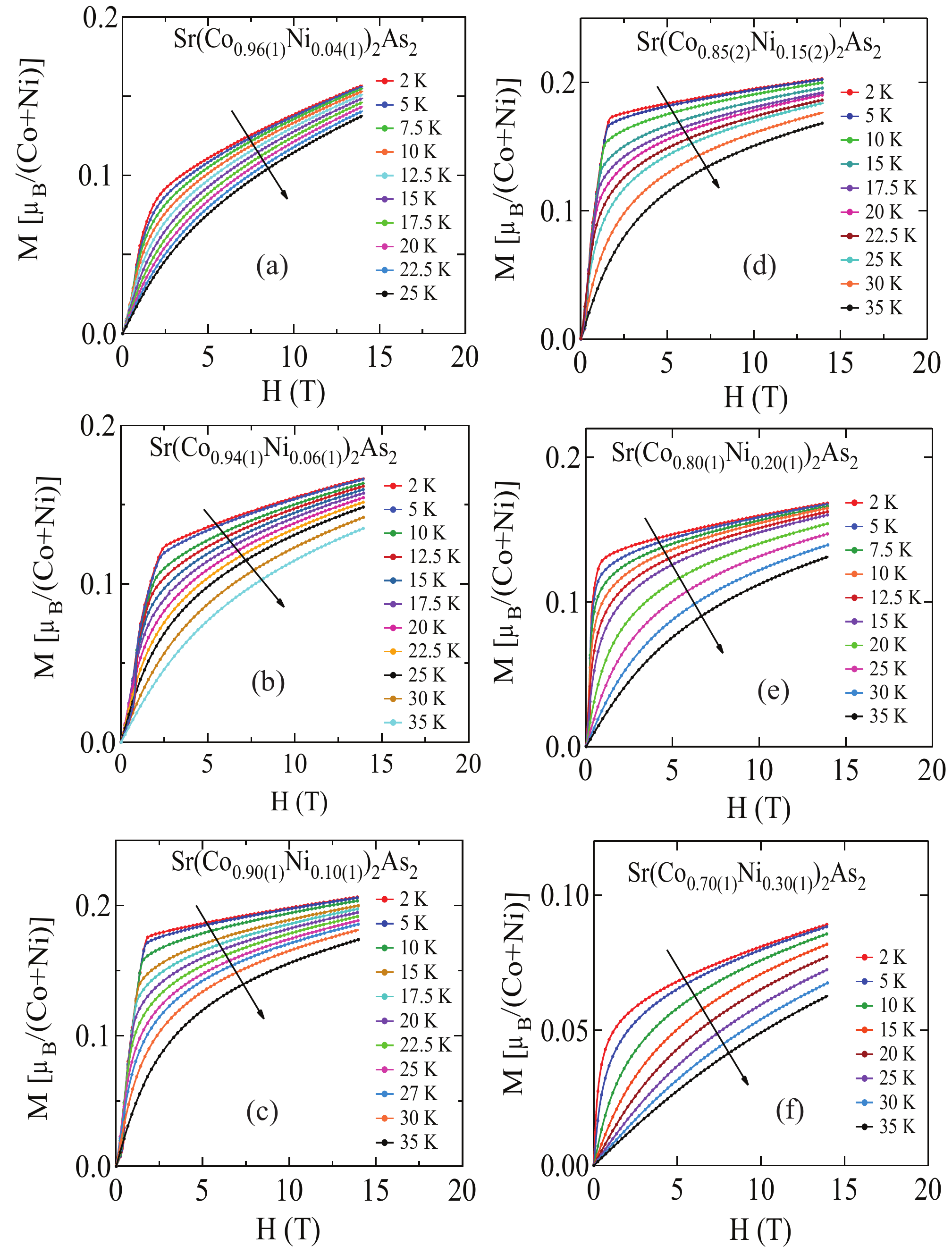}
\caption{Isothermal magnetization $M$ vs $H$ of \scna\ crystals with $x=0.04$, 0.06, 1, 0.15, 0.20, and 0.3 as a function of magnetic field $H$ (0 to 14~T) applied in the $ab$ plane at the indicated temperatures.}
\label{Fig:MH-14T}
\end{figure} 

\begin{figure*}
\includegraphics[width=3.5in]{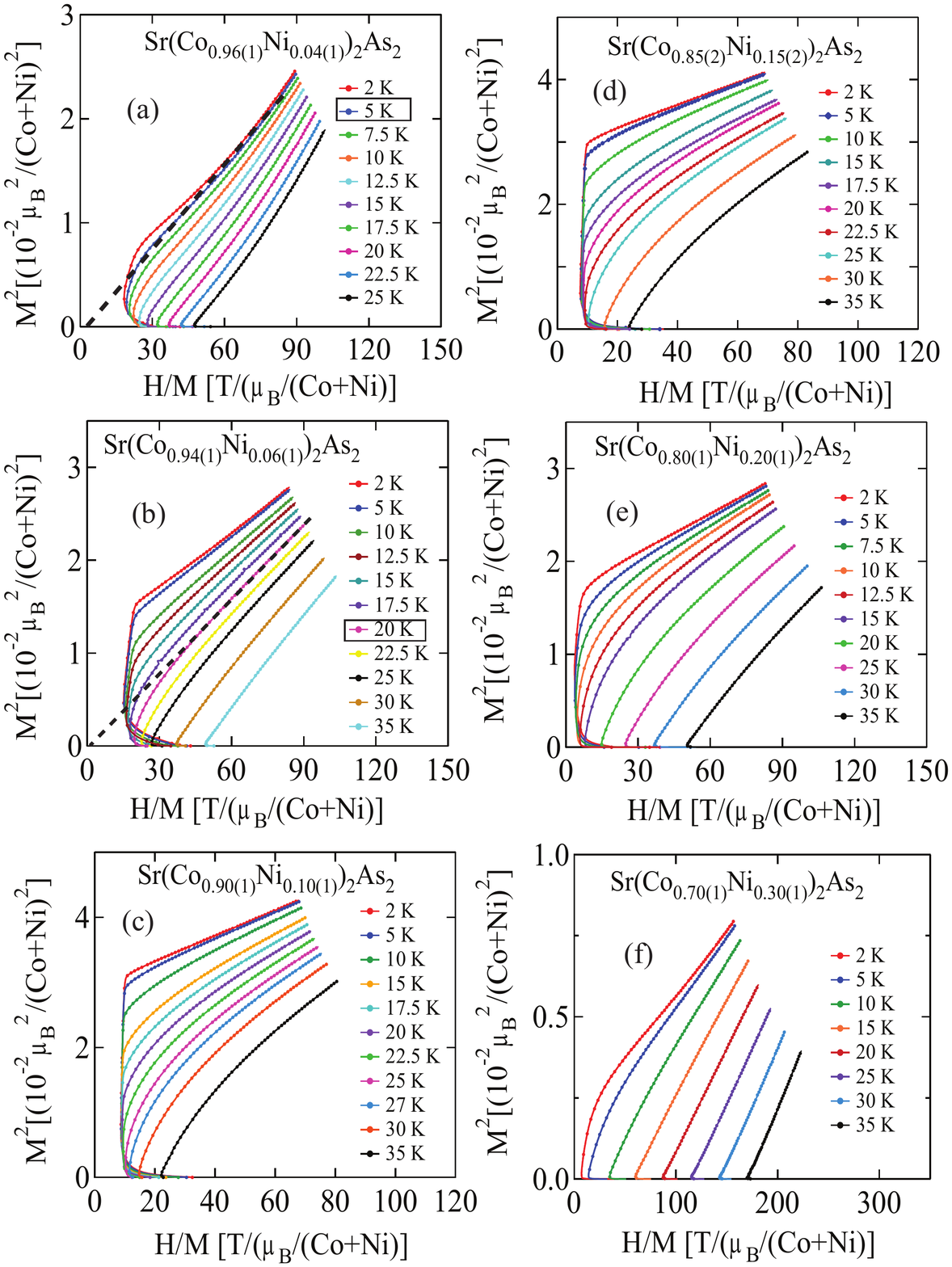}\includegraphics[width=3.5in]{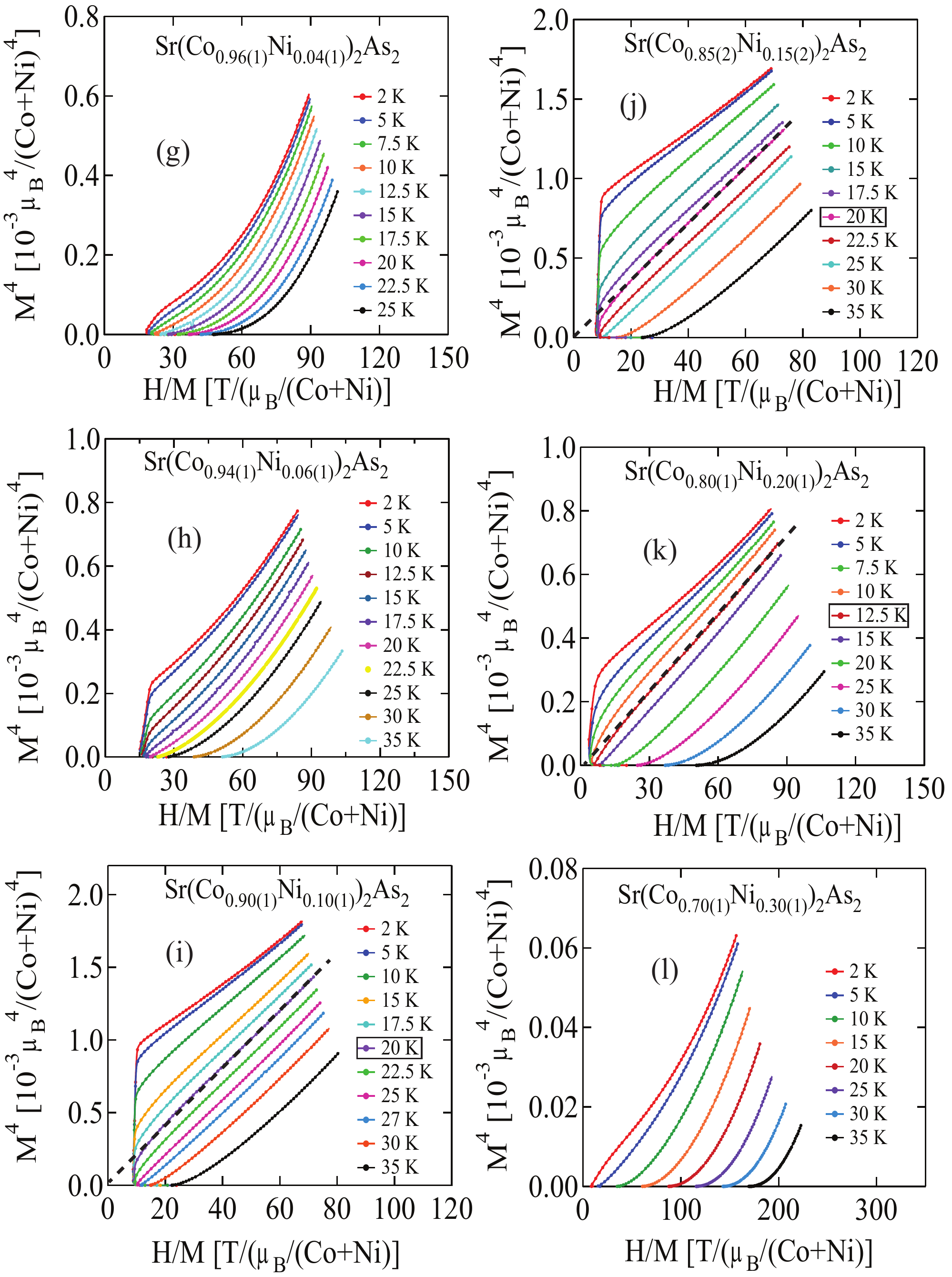}
\caption{(a)--(f): Isothermal magnetization $M^2$ vs $H/M$ plots (Arrott plots) of \scna\ crystals with compositions $x=0.04$, 0.06, 1, 0.15, 0.20, and 0.3 as a function of magnetic field $H$ applied in the $ab$ plane at the indicated temperatures.  Only the higher-field data are relevant due to the presence of spin-flop transitions at lower fields.  The temperatures of the isotherms for which dashed straight line extrapolations to low field pass through the origin in panels (a) and~(b)  with $x =0.04$ and~0.06, respectively, are inferred to be the Curie temperatures $T_{\rm C}$ for the respective samples.  (g)--(l): Isothermal magnetization $M^4$ vs $H/M$ for the same \scna\ crystals as a function of magnetic field $H$ applied in the $ab$ plane at the indicated temperatures.}
\label{Fig:M2-HM}
\end{figure*}

\begin{figure*}
\includegraphics[width=6in]{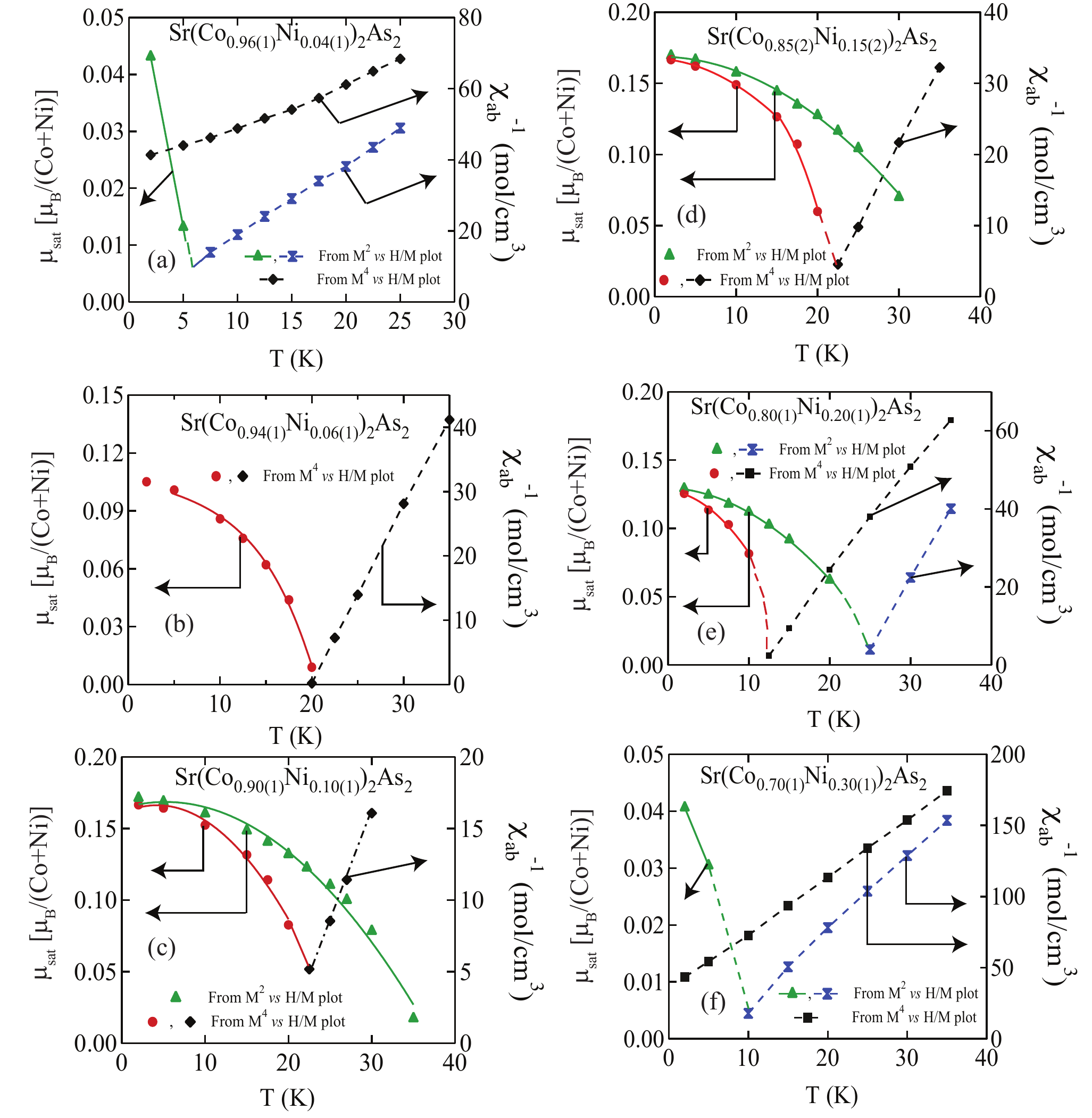}
\caption{ Saturation moment $\mu_{\rm sat}$ and inverse molar susceptibility $\chi_{ab}^{-1}$ extracted from the $M^2$ vs $H/M$ and $M^4$ vs $H/M$ measurements in Fig.~\ref{Fig:M2-HM} plotted with left and right ordinate scales, respectively.}
\label{Fig:AP}
\end{figure*}

Figure~\ref{Fig:MH-14T} displays $M(H)$ isotherms with $H$ up to 14~T for six \scna\ crystals at the indicated temperatures. It is seen that the saturation moment obtained from an exrapolation of the high-field data to $H=0$ is very small, of order $0.1~\mu_{\rm B}$/(Co+Ni), and that the $M(H)$ curves at $T = 2$~K continue to increase with increasing magnetic field and do not saturate up to 14~T, as expected for an itinerant magnetic system. We find that our data can be well understood in terms of Takahashi's theory of spin-fluctuation-driven weak itinerant ferromagnetism~\cite{Takahashi2013}, as follows.

In Takahashi's theory, it is assumed that the total local spin fluctuation amplitude, i.e., the sum of zero point and thermally-induced spin fluctuation amplitudes, is globally conserved for weak ferromagnets.  The Landau expansion for free energy $F(M,T)$ of a ferromagnet in terms of the magnetization per unit volume~$M$ at temperature~$T$ is \cite{Takahashi2013}
\bea
F(M,T)&=&F(0,T)+\frac{1}{2}a_1(T)M^2+\frac{1}{4}a_2(T)M^4+....\nonumber \\*
a_1&=&\frac{1}{(g\mu_B)^2\chi(0)}, \\* 
a_2&=&\frac{F_1}{(g\mu_B)^4N_0^3} \nonumber
\label{Eq:TSF1}
\eea
where the $a_i$ are expansion coefficients related to the electronic density of states and its derivatives near $E_{\rm F}$. The magnetic field $H$-dependent $M$ to ${\cal O}(M^3)$ is obtained as
\begin{equation}
H=\partial F/\partial M=a_1(T)M+a_2(T)M^3.
\label{Eq:TSF2}
\end{equation}
The $M$ in the ground state is expressed as~\cite{Takahashi2013}
\bea
H&=&\frac{F_1}{(g\mu_{\rm B})^4N_0^3}(M^2-M_0^2)M \nonumber, \\*
F_1&=&\frac{2k_{\rm B}T_{\rm A}^2}{15cT_0},
\label{Eq:TSF3}
\eea
where $F_1$ (the mode-mode coupling term) is the coefficient of $M^4$ in the Landau free-energy expansion, $c=1/2$, and $N_0$ is the number of magnetic atoms [here the number of Co/Ni transition metal atoms in \scna]. One sees that in the ground state, the magnetic isotherm is influenced by the presence of zero-point spin fluctuations because $F_1$ is expressed in terms of the spectral spin fluctuation parameters $T_0$ and $T_{\rm A}$ in temperature units (K). The parameter $T_{\rm A}$ is a measure of the width of the distribution of the dynamical susceptibility fluctuations in wave vector space, whereas $T_0$ represents the energy width of the spin fluctuation spectrum in energy (frequency) space corresponding to the stiffness of spin density amplitude. It can be calculated experimentally from the inverse slope of Arrott plots ($M^2$ versus $H/M$)~\cite{Arrott2010} at low temperatures as follows.

Utilizing the values of $k_{\rm B}$ and $\mu_{\rm B}$ in cgs units with $g=2$, Eq.~(\ref{Eq:TSF3}) gives
\begin{equation}
H({\rm Oe})=\left(248.2\frac{\rm Oe}{\rm K}\right) \left(\frac{T_{\rm A}^2}{T_0}\right)\frac{(M^2-M_0^2)M}{(N_0\mu_{\rm B})^3}.
\end{equation}
Now expressing $M$ and $M_0$ in units of $\mu_{\rm B}$ per magnetic atom gives
\begin{equation}
H({\rm Oe})=\left(248.2\frac{\rm Oe}{\rm K}\right) \left(\frac{T_{\rm A}^2}{T_0}\right) (M^2-M_0^2)M,
\end{equation}
which yields the Arrott-like expression
\bse
\label{Eqs:Arrott}
\begin{equation}
M^2-M_0^2=\frac{1}{\left(248.2\frac{\rm Oe}{\rm K}\right) \left(\frac{T_{\rm A}^2}{T_0}\right)} \frac{H}{M},
\end{equation}
where $M$ and $M_0$ are now dimensionless because their units have been divided out and $H$ is in cgs units of Oe. Thus a plot of $M^2$ vs $H/M$ is predicted to be linear with $y$-intercept $M_0$ and slope
\begin{equation}
\frac{M^2-M_0^2}{H/M}=\frac{1}{\left(248.2\frac{\rm Oe}{\rm K}\right) \left(\frac{T_{\rm A}^2}{T_0}\right)}.
\label{Eq:M2}
\end{equation}

The ordered moment $M_0$ in the ground state, again in dimenisonless units of $\mu_{\rm B}$ per magnetic atom, is given by~\cite{Takahashi2013}
\begin{equation}
M_0\approx 2 \sqrt{C_{4/3}\frac{5T_0}{T_{\rm A}}}\left(\frac{T_{\rm C}}{T_{\rm A}}\right)^{2/3},
\label{Eq:M0}
\end{equation}
\ese
where $C_{4/3}$ is a constant ($C_{4/3}$ = 1.006\,089...). Equations~(\ref{Eq:M2}) and (\ref{Eq:M0}) allow both spectral parameters $T\rm_A$ and $T\rm_0$ to be estimated from the ground-state $M(H)$ isotherm, provided a value for the inferred Curie temperature $T\rm_C$ is known. Here we expect Eqs.~(\ref{Eqs:Arrott}) to be valid for a \scna\ crystal if our minimum \mbox{measurement} temperature of 2~K for the high-field $M(H)$ isotherms is much less than the measured $T_{\rm N}$ or the inferred value of the Curie temperature $T_{\rm C}$, whichever is smaller.

Thus the low-temperature ground-state isotherm of $M^2$ versus $H/M$ is predicted to be linear at high fields.  In contrast, the ``critical isotherm'' at $T_{\rm C}$ has the different form $M^4$ versus $H/M$ which shows a proportional behavior with zero $y$~intercept~\cite{Takahashi2013}. This form arises because the critical magnetic isotherm is predominantly influenced by  temperature-induced spin fluctuations instead of zero-point spin fluctuations.

The critical isotherm at $T\rm_C$ is described by \cite{Takahashi2013} 
\bea
H&=&\frac{T_{\rm_A}^3k\rm_B/\mu\rm_B}{2[3\pi(2+\sqrt{5})T\rm_C]^2} \left(\frac{M}{N_0\mu\rm_B}\right)^5,\\*
&=& \left(4.671\frac{\rm Oe}{\rm K}\right)\left(\frac{T\rm_ A^3}{T\rm_C^2}\right) \left(\frac{M}{N_0\mu\rm_B}\right)^5.
\eea
Expressing $M$ in units of $\mu\rm_B$ per magnetic atom makes $M$ dimensionless and gives the proportional relation in Gaussian cgs units as
\begin{equation}
M^4=\frac{1}{{4.671\frac{\rm Oe}{\rm K}}\left(\frac{T\rm_A^3}{T\rm _C^2}\right)} \frac{H}{M}
\end{equation}
with slope
\begin{equation}
\frac{M^4}{H/M}=\frac{1}{{4.671\frac{\rm Oe}{\rm K}}\left(\frac{T_{\rm A}^3}{T_{\rm C}^2}\right)}. 
\label{Eq:M4}
\end{equation}
This slope is expressed in terms of $T\rm_A$, which allows  $T\rm_A$ to be determined since  $T\rm_C$ can be determined from the temperature at which an extrapolation of the high-field linear behavior passes through the origin on an $M^4$ versus $H/M$ plot as discussed above. 

We analyze our $ab$-plane magnetization data at fields up to $H=14$~T for \scna\ crystals with compositions $x=0.04$, 0.06, 0.10, 0.15, 0.20, and 0.30 in Fig.~\ref{Fig:M2-HM} in terms of $M^2$ vs $H/M$ (Arrott) and $M^4$ versus $H/M$ plots, respectively. The data at the higher magnetic fields was used for the analyses since spin-flop transitions occur at lower fields (see Table~\ref{Tab.MH}). The  temperatures at which the extrapolations of the high-field data to $H=0$ pass through the origin are indicated by solid dashed lines in Figs.~\ref{Fig:M2-HM}(i), \ref{Fig:M2-HM}(j), and~\ref{Fig:M2-HM}(k) for $x=0.10$, 0.15, and~0.20, respectively. The corresponding values of the saturation moment $\mu_{\rm sat}$ and the inverse susceptibility $\chi_{ab}^{-1}$ obtained from the extrapolated intercepts of the linear fits to the $y$ and $x$ axes, repectively, are shown in Fig.~\ref{Fig:AP}. Here, the temperature at which the extrapolated $\chi_{ab}^{-1}(T)$ passes through zero determines the Curie temperature $T\rm_C$.

As one can see in Fig.~\ref{Fig:M2-HM}, Arrott plots are only relevant for $x=0.04$, 0.06, and 0.30 since they show reasonably linear behaviors at high fields. At the same time, the Arrott plots for $x=0.10$, 0.15, and 0.20 show strong curvature around $T\rm_C$. In this case, one needs to analyze the isothermal $M(H)$ data in terms of $M^4$ vs $H/M$ plots, shown in Fig.~\ref{Fig:M2-HM}~(g)--(l), where these plots for $x=0.10$, 0.15, and 0.20 show linear behaviors at high fields. Conversely, the $M^4$ vs $H/M$ plots for $x=0.04$, 0.06, and 0.30 show an upward curvature for temperatures around $T\rm_C$\@. The values of $T\rm_C$ for which the extrapolated data pass through the origin are thus determined from the Arrott plots for $x=0.04$, 0.06 and from the $M^4$ versus $H/M$ plots for $x=0.10$, 0.15, and 0.20 and are listed in Table~\ref{Tab.SFT}. These differences indicate that temperature-induced spin fluctuations are not as important for the compositions $x=0.04$ and~0.06 as for $x=0.10$, 0.15, and 0.20.  In this context, it is interesting that the critical isotherm at $T=30$~K for the itinerant ferromagnet SrCo$_2$(Ge$_{1-x}$P$_x)_2$ with $x=0.6$ was linear on an Arrott plot~\cite{Jia2011}, which is analogous to our crystals with $x = 0.04$, 0.06, and~0.30.

The spin fluctuation parameters  $T\rm_A$ and $T\rm_0$ are calculated based on Arrott plots using Eqs.~(\ref{Eq:M2}) and~(\ref{Eq:M0}), whereas Eqs.~(\ref{Eq:M0}) and (\ref{Eq:M4}) are used to calculate $T\rm_A$ and $T\rm_0$ from the $M^4$ vs $H/M$ isotherms at $T_{\rm C}$ and all the resulting derived parameters are listed at the top and bottom of Table~\ref{Tab.SFT}, respectively. The parameter $F_1$ is calculated using Eq.~(\ref{Eq:TSF3}).

\subsection{Rhodes-Wohlfarth Plots}

For weak itinerant ferromagnets, the spontaneous (ordered) magnetic moment is much smaller than expected for local-moment magnets, {\it i.e.}, significantly smaller than minimum value of 1~$\mu_{\rm B}$ per magnetic atom for $S=1/2$ and $g=2$. The effective moment can also be smaller than the minimum value of $1.73\,\mu_{\rm B}$ per magnetic atom for $S=1/2$ in Eq.~(\ref{Eq:peffS12}). To systematize these behaviors, Rhodes and Wohlfarth proposed to plot the ratio $p{\rm_c}/p_{\rm sat}$ versus the Curie temperature $T_{\rm C}$~\cite{Rhodes1963}, now called Rhodes-Wohlfarth plot, where $p_{\rm c}$ is defined above in Eq.~(\ref{Eq:pcDef}) in terms of the measured effective moment $\mu_{\rm eff}$ obtained at $T>T_{\rm C}$ from the Curie constant in the Curie-Weiss law describing the magnetic susceptibility in the PM phase.

\begin{figure}
\includegraphics[width=2.5in]{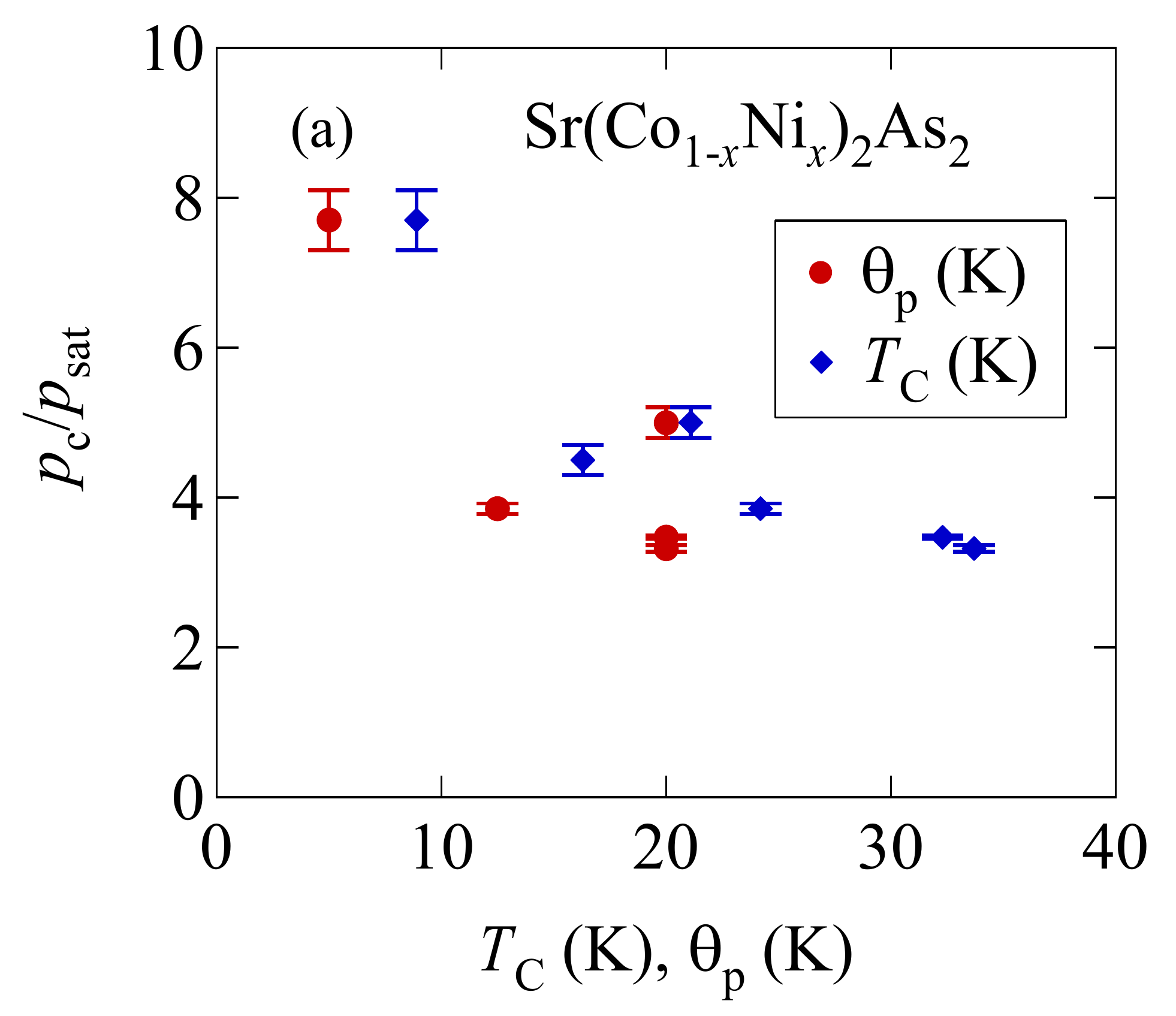}
\includegraphics[width=2.5in]{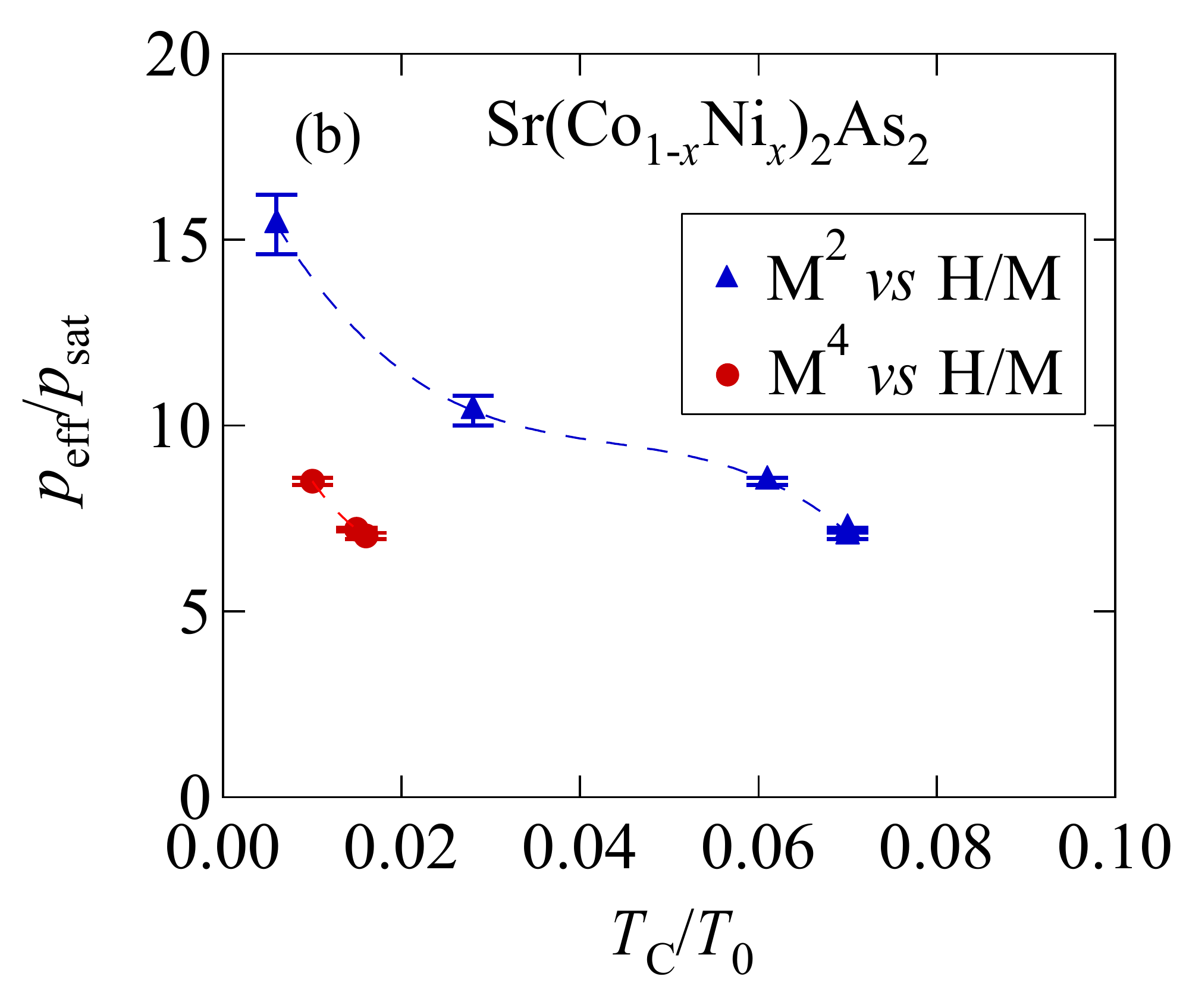}
\caption{(a) Rhodes-Wohlfarth plot $p{\rm_c}/p_{\rm sat}$ vs Curie temperature $T\rm_C$ and average Weiss temperature $\theta_{\rm p,ave}$.  (b) Deguchi-Takahashi plot of $p{\rm_{eff}}/p_{\rm sat}$ vs $T_{\rm C}/T_0$ for \scna\ crystals with $x$~=~0.04, 0.06, 0.10, 0.15, 0.20, and 0.25.}
\label{Fig_RW}
\end{figure}

As previously noted in Sec.~\ref{Sec:CW}, the ratio \mbox{$p{\rm_c}/p_{\rm sat} = 1$} for local-moment magnets, whereas for itinerant-moment systems, one obtains  $p{\rm_c}/p_{\rm sat}>1$.  Listed in Table~\ref{Tab.RW} are the Rhodes-Wohlfarth ratios $p{\rm_c}/p_{\rm sat}$ calculated for our \scna\ crystals with compositions $x=0.04$, 0.06, 0.10, 0.15, 0.20, 0.25, and~0.30 from the data in Tables~\ref{Tab.chidata2} and~\ref{Tab.SFT}.  Figure~\ref{Fig_RW}(a) shows a plot of $p{\rm_c}/p_{\rm sat}$ versus both $T\rm_C$ and the angle-averaged Weiss temperature $\theta_{\rm p,ave}$ for \scna.  One sees that $p{\rm_c}/p_{\rm sat}>1$ for \scna\ as expected. For $0.10\leq x \leq 0.25$  where \scna\ orders magnetically, the value of $p{\rm_c}/p_{\rm sat}$ is close to that for the famous weak itinerant ferromagnet ZrZn$_2$ \cite{Wohlfarth1968}. For $x=0.30$, the ratio increases to 6.1.

Another generalized plot of $p{\rm_{eff}}/p_{\rm sat}$ vs $T{\rm_C}/T_0$, known as the Deguchi-Takahashi plot, is also based on the spin-fluctuation theory. The ratio $T{\rm_C}/T_0$  characterizes the degree of itinerancy of the magnetic electrons in the spin fluctuation theory. The value of $T{\rm_C}/T_0$ becomes small for weak itinerant ferromagnets with a large temperature-induced spin fluctuation amplitude.  A value $T{\rm_C}/T_0\sim 1$ is the localized-moment limit. The obtained values of $T{\rm_C}/T_0$  estimated from the $M^2$ vs $H/M$ and $M^4$ vs $H/M$ isotherms are listed in Table~\ref{Tab.RW} and a plot of $p{\rm_{eff}}/p_{\rm sat}$  versus $T{\rm_C}/T_0$ is shown in Fig.~\ref{Fig_RW}(b). These data indicate that the magnetically-ordered samples of \scna\ are close to the itinerant-magnetism limit of the itinerant magnetism-local moment magnetism continuum.


\subsection{Magnetic Susceptibility of \scna\ above $T\rm_C$ in Terms of Takahahshi's Spin-Fluctuation Theory for Weak Itinerant Ferromagnets}

The parameter $y$ in Takahashi's theory of the thermal-fluctuation-induced magnetic susceptibility~$\chi$ above $T_{\rm C}$ is given by~\cite{Takahashi2013}
\bse
\label{Eqs:yCalc}
\begin{equation}
y = \frac{N_{\rm A}g^2\mu_{\rm B}^2}{2k_{\rm B}T_{\rm A}\chi},
\label{Eq:yDef}
\end{equation}
where $N_{\rm A}$ is Avogadro's number, $g = 2$ and the Gaussian cgs system of units is used.  From Eq.~(\ref{Eq:yDef}) one obtains
\begin{equation}
\chi^{-1} = \left(\frac{2k_{\rm B}T_{\rm A}}{N_{\rm A}g^2\mu_{\rm B}^2}\right)y.
\label{Eq:chim12}
\end{equation}

The temperature-induced amplitude $A(y,t)$ of the spin fluctuations is given by~\cite{Takahashi2013}
\begin{equation}
A(y,t) = \int_0^1 \left\{\ln[u(x)]-\frac{1}{2u(x)}-\psi[u(x)]\right\}x^3dx,
\end{equation}
where
\begin{equation}
u(x) \equiv x(x^2+y)/t,\qquad t=T/T_0,
\end{equation}
and $\psi(z)$ is the digamma function which is the logarithmic derivative of the gamma function $\Gamma(z)$.  The parameter  $y$ in Eq.~(\ref{Eq:chim12}) is numerically calculated from~\cite{Takahashi2013}
\begin{equation}
y(t) = \frac{1}{c}[A(y,t) - A(0,t_{\rm C})],
\label{Eq:Findy(t)}
\end{equation}
\ese
where $t_{\rm C}\equiv T_{\rm C}/T_0$ and $c=1/2$.

\begin{figure}
\includegraphics[width=2.75in]{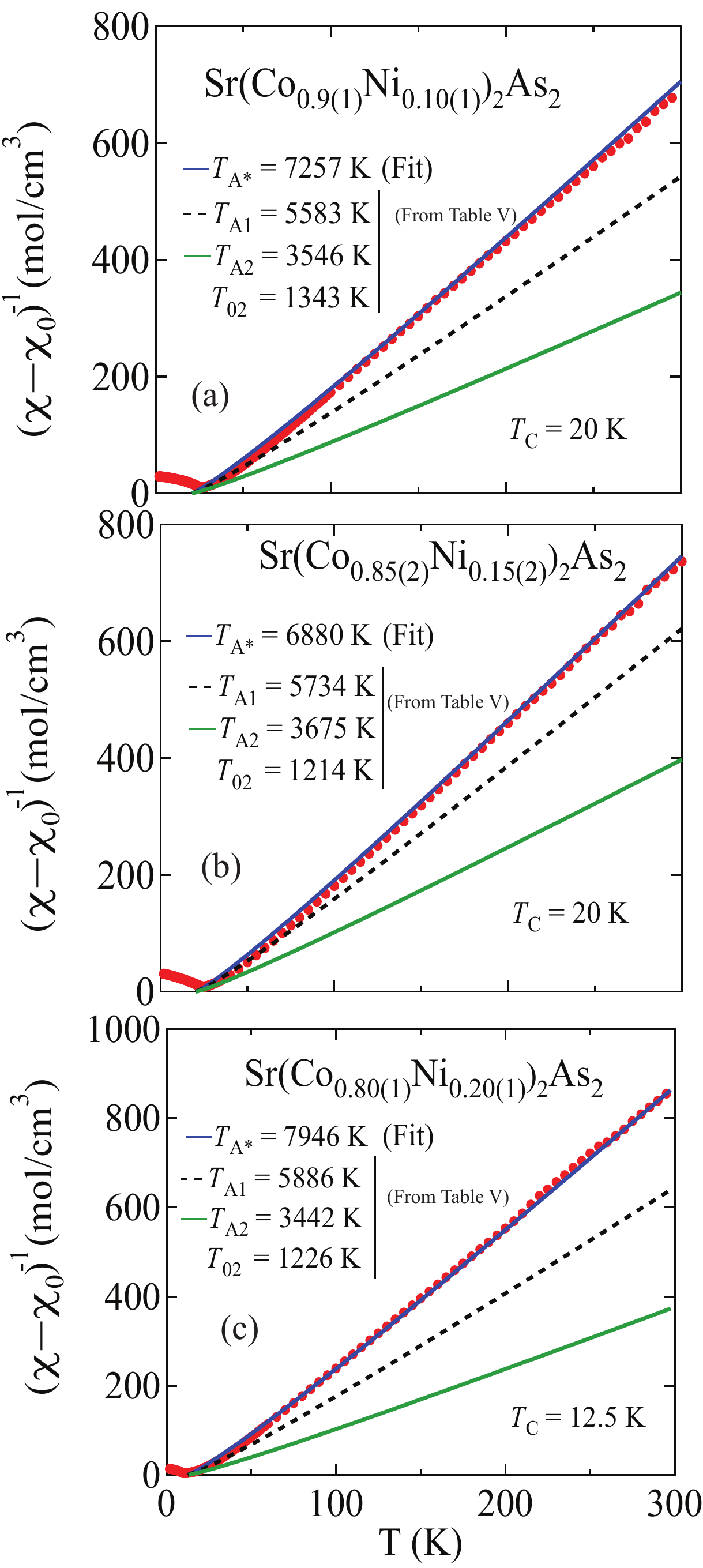}
\caption{Temperature dependence of the inverse susceptibility $(\chi-\chi_0)^{-1}$ for \scna\ with $x=0.10$, 0.15, and 0.20. The red circles represent experimental data.  The black dashed lines and the green lines are data calculated using the $T_{\rm A}$ and $T_0$ values listed in each figure that were taken from Table~\ref{Tab.SFT}, whereas the solid blue lines are fits of the data by Eq.~(\ref{Eqs:yCalc}) using $T\rm_{A}$, labeled as $T\rm_{A*}$, as an adjustable parameter. }
\label{Fig:Invchi_itfit}
\end{figure}

The inverse susceptibility data for $x=0.15$, 0.20, and 0.25 as a function of $T$ were fitted by Eqs.~(\ref{Eqs:yCalc}) as shown in Fig.~\ref{Fig:Invchi_itfit}. Here the $T$-dependent inverse susceptibility $[\chi(T)-\chi_0]^{-1}$ was fitted where $\chi(T)$ is the measured susceptibility and $\chi_0$ is the temperature-independent contribution with values given in Table~\ref{Tab.chidata1}. The black dashed lines and the solid green lines are the results obtained using the spin fluctuation parameters $T_{\rm A1}$, $T_{\rm A2}$ and $T_{02}$ given in Table~\ref{Tab.SFT}, whereas the solid blue lines are fits of the data by Eqs.~(\ref{Eqs:yCalc}) using $T\rm_{A}$, labeled as $T_{\rm{A^*}}$, as an adjustable parameter.  The values of $T\rm_C$ utilized in the fits are given in Table~\ref{Tab.SFT}.  The calculated inverse susceptibility is seen to follow a Curie-Weiss-type law rather well, and is found to be depend to the value of $T_{\rm A}$.

\subsection{\label{Sec:ElecStruct:Magnetism} Electronic Structure Calculations: Magnetism}

\begin{figure*}
\includegraphics [width=7in]{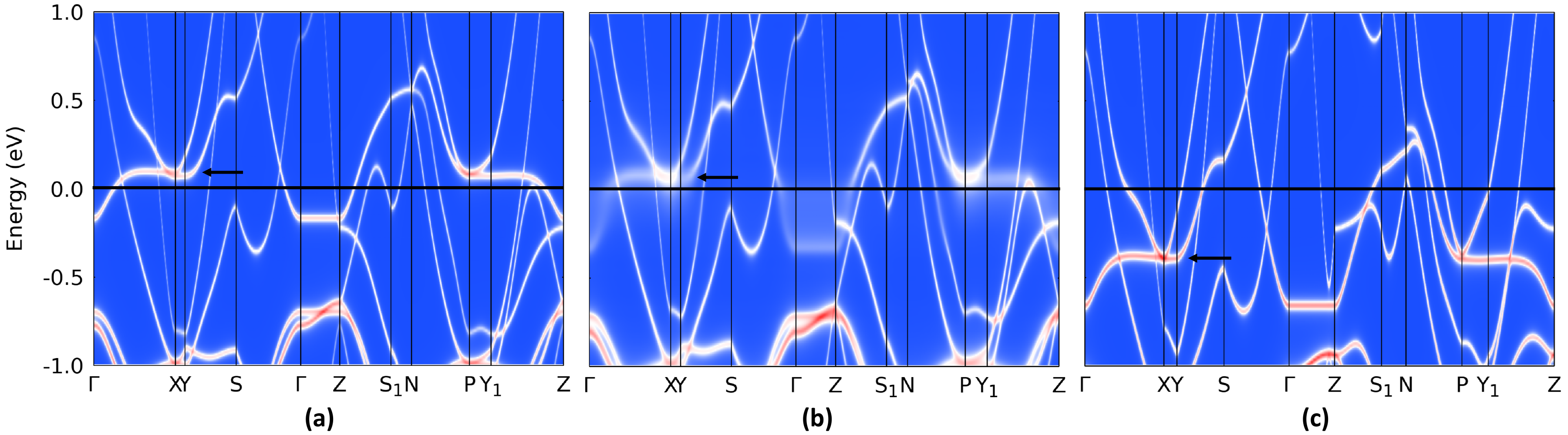}
\caption {KKR-CPA electronic dispersions (spectral functions) calculated for nonmagnetic \scna,  at $x$ values of (a)~0, (b)~0.15 and (c)~1. Color-coded intensity scale is high (red), medium (white), and low (blue), with flat bands (black arrows) indicated along $\Gamma-X-Y$ and $P-Y_1-Z$ directions. }
\label{Fig:SCNA_Band_Struct}
\end{figure*}

\begin{figure}
\includegraphics [width=3.3in]{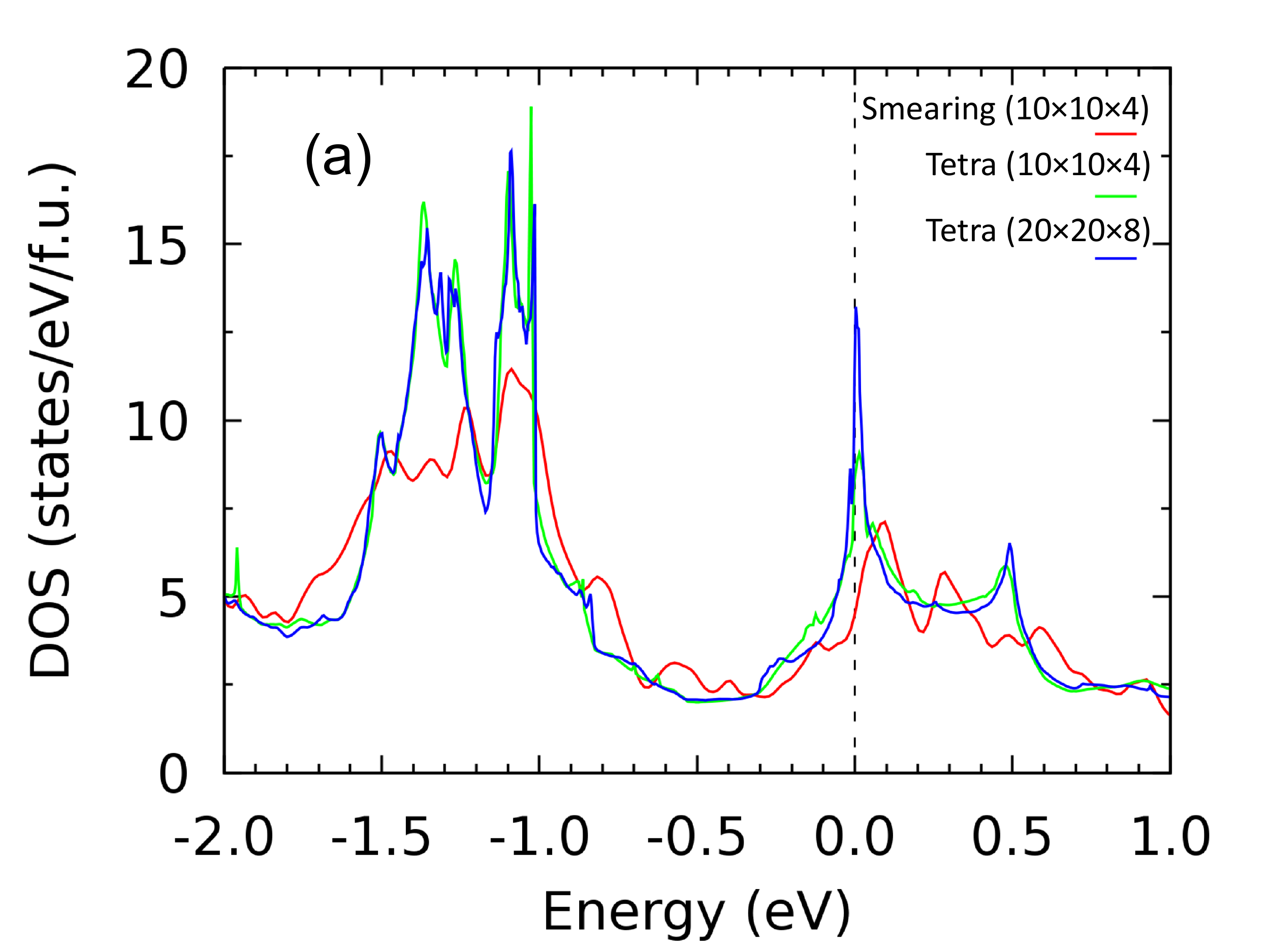}
\includegraphics [width=3.3in]{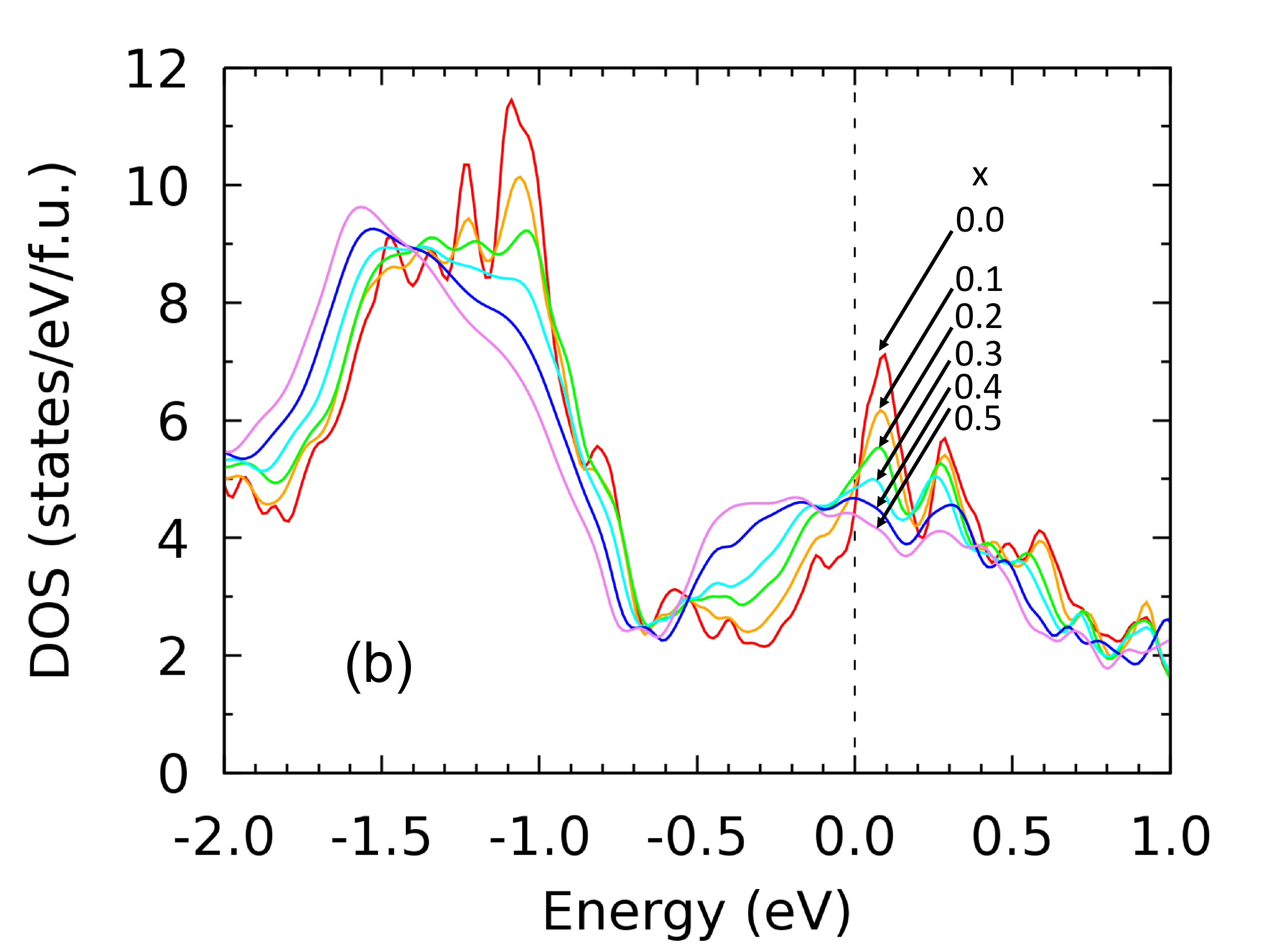}
\caption {(a) Electronic density of states (DOS) including both spin directions versus energy of nonmagnetic \sca\ calculated in KKR with a Gaussian smearing of 0.04~eV on a {\bf k}-point mesh of ($10\times10\times4$) in comparison to the DOS calculated with the tetrahedron method on {\bf k}-point meshes of ($10\times10\times4$) and ($20\times20\times8$), respectively. (b) KKR-CPA DOS including both spin directions versus energy calculated for nonmagnetic \scna\  with compositions from $x=0.0$ to~0.5. The Fermi energy $E_{\rm F}$ is at energy zero.  The DOS($E_{\rm F}$) first increases then decreases with increasing~$x$.}
\label{Fig:DOSSCNA}
\end{figure}

To explain the observed magnetic ordering in \scna\ from an itinerant magnetism picture, we calculated the spectral functions (electronic dispersion, or ÒbandÓ structure) of the alloyed compounds via the KKR-CPA method~\cite{1, 2, 3}, which handles chemical and magnetic disorder~\cite{Johnson1986, Johnson1990, 4, 5, 6}. For pure \sca\ [$x=0$, Fig.~\ref{Fig:SCNA_Band_Struct}(a)], segments of flat bands along the $\Gamma-X-Y$ and $P-Y_1-Z$ directions (or $M-\Gamma$ and $A-Z$ directions in the conventional cell) are just 100~meV above the Fermi energy $E_{\rm F}$\@. These flat bands are derived mostly from Co $d_{xy}$ orbitals.

Substituting Ni on the Co sites corresponds to electron doping. For $x=0.15$, the flat bands along $\Gamma-X$ and $P-Y_1-Z$ directions [Fig.~\ref{Fig:SCNA_Band_Struct}(b)] already touch $E_{\rm F}$, which can drive Fermi-surface nesting and Stoner-type instabilities. However, it is not a simple rigid-band shift. In addition to shifting to lower energy, these flat bands have much more broadening due to chemical disorder than other bands, which reflects the change in electron scattering length for states around $E_{\rm F}$ and alters transport properties, for example. An increase in the density of states (DOS) at $E_{\rm F}$ drives a magnetic instability according to the Stoner criterion [9, 10] that favors long-range order.

As our experimental data indicate, there is non-Fermi-liquid behavior in the alloying range of 20--30\% Ni, which coincides with the region of large dispersion broadening near $E_{\rm F}$ from chemical disorder found in KKR-CPA Bloch spectral functions, see Fig.~(\ref{Fig:SCNA_Band_Struct}), where, for $\sim 20$--30\% Ni, $k$ is a ``less-good'' quantum number for those Co $d_{xy}$ states, with much shorter electron-scattering length.  Then, above 30\% Ni, the dispersion becomes dramatically less broadened as those states move below $E_{\rm F}$, when $k$ again becomes a ``good'' quantum number, as expected for a Fermi liquid. This effect is reflected in the DOS, see below, and therefore the heat capacity (next section).

For example, for pure \sna\ [$x=1$, Fig.~\ref{Fig:SCNA_Band_Struct}(c)], the flat bands of Co $d_{xy}$ character associated with magnetic instability are now far below $E_{\rm F}$\@. Thus, in the Ni-rich region, the paramagnetic state (randomly-oriented finite local moments) prevails, while in the Co-rich region, the flat bands just above $E_{\rm F}$ can drive long-range magnetic order from electron doping upon substitution of a small amount of Ni for Co. Yet, the total energy difference between nonmagnetic (NM), ferromagnetic (FM) and A-type antiferromagnetic (AFMA) states is very small \mbox{($\sim0.1$~meV/f.u.)} with a magnetic moment of 0.3~$\mu_{\rm B}$ at the Co sites in the magnetically-ordered cases.

From the electronic dispersion in Fig.~\ref{Fig:SCNA_Band_Struct}(a), the total density of states (DOS) of nonmagnetic \sca\ calculated from KKR is plotted versus energy in Fig.~\ref{Fig:DOSSCNA}(a). To validate our DOS results, the DOS calculated from a plane-wave basis set with the projector-augmented-wave (PAW) method in VASP~\cite{Kresse1996, Kresse1996b} on two {\bf k}-point meshes are plotted for comparison. The DOS curves over a large energy range agree including the peak just above the Fermi energy ($E_{\rm F}$) due to the flat band of Co~$d_{xy}$ character. The DOS($E_{\rm F}$) value of 11~states/(eV\,f.u.) including both spin directions calculated on the dense ($20\times20\times8$) {\bf k}-point mesh agrees with the previous study~\cite{Pandey2013}. With the additional enhancements from electron-phonon coupling and many-body renormalization of the band structure, this value can be compared directly with the result from specific heat measurements. In KKR-CPA, in order to control computational cost, we used the ($10\times10\times4$) {\bf k}-point mesh with a Gaussian smearing of 0.04~eV, which already gives good total energy convergence, but the DOS($E_{\rm F}$) value becomes smaller from this numerical device. [Increasing the {\bf k}-mesh size improves the results for DOS($E_{\rm F}$), as shown in Fig.~\ref{Fig:DOSSCNA}(a).] However, the KKR-CPA result for DOS($E_{\rm F}$) does reflect correct changes due to chemical disorder upon alloying with Ni and shows an interesting behavior. As seen in Fig.~\ref{Fig:DOSSCNA}(b) which shows the DOS versus energy zoomed in around $E_{\rm F}$ for the compositions from $x=0.0$ to~0.5, the main peak due to the flat band of Co~$d_{xy}$ character is increasingly broadened as $x$ increases due to chemical disorder. At the same time, the DOS below $E_{\rm F}$ (at $E_{\rm F}-0.2$~eV) increases. For DOS($E_{\rm F}$), the change is not monotonic; it first increases with alloying and reaches a maximum at $x=0.2$, while the main DOS peak is still well above the Fermi energy, then it decreases upon further alloying. This behavior agrees with the trend in specific-heat measurements (see below) and shows that the chemical-disorder effects on the electronic dispersion do not correspond to a rigid-band-shifting picture.

\section{\label{Sec:heatcapacity} Heat Capacity}

\subsection{Experimental Data and Fits}

\begin{figure}
\includegraphics[width=3.4in]{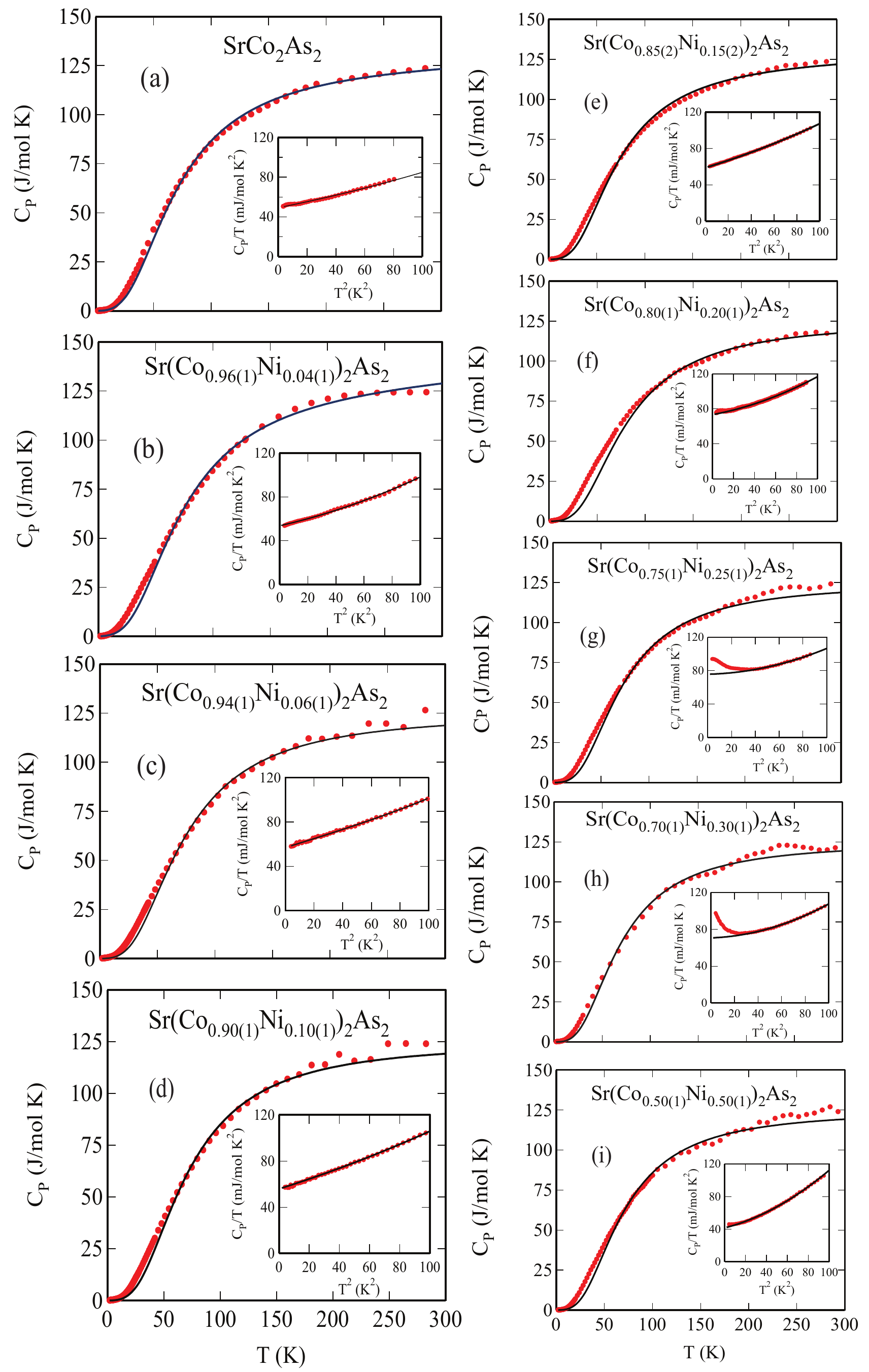}
\caption{Heat capacity $C_{\rm p}$ for \scna\ with $x=0$, 0.04, 0.06, 0.10, 0.15, 0.20, 0.25, 0.30, and 0.50 single crystals versus temperature $T$\@. The solid curves in the main panels are fits of the data by Eqs.~(\ref{Eq:Debye_Fit}). Insets: $C_{\rm p}/T$ vs $T^2$ below 10~K. The solid lines are fits by Eq.~(\ref{Eq.CpFit}) where the $T$ ranges of the fits are 1.8~K to 10~K for all $x$~values except for $x=0.20$, 0.25, and 0.30 for which the fit range was 6~K to 10~K, where extrapolations of the fits to 1.8~K are shown.  One sees upturns in the low-temperature data for $x = 0.20$, 0.25, and 0.30 associated with proximity to a magnetic quantum critical point at $x\approx0.30$.}
\label{Fig:Heat_Capacity}
\end{figure}

Figure~\ref{Fig:Heat_Capacity} shows the heat capacity $C_{\rm p}$ versus $T$ for the \scna\ crystals with compositions $x=0$, 0.04, 0.06, 0.10, 0.15, 0.20, 0.25, 0.30, and 0.50 over the entire $T$ range of the measurements from 1.8 to 300~K in zero applied magnetic field. The $C_{\rm p}(T)$ values at 300~K of \scna\ are $\approx$ 125~J/mol~K which is  close to the classical Dulong-Petit value $C_{\rm V} = 3nR = 124.7 $~J/mol~K, where $n = 5$ is the number of atoms per f.u.\ and  $R$ is the molar gas constant.

The low-temperature $C_{\rm p}/T$ vs $T^2$ data are shown in the insets of Fig.~\ref{Fig:Heat_Capacity}.  The data were fitted over the temperature ranges given in the figure caption by
\begin{equation}
\frac{C_{\rm p}(T)}{T} = \gamma + \beta T^2 +\delta T^4,
\label{Eq.CpFit}
\end{equation}
which is the sum of the electronic ($\gamma$) and low-$T$ lattice ($\beta T^2+\delta T^4$) heat capacity contributions. We fitted the data for $x=0.20$, 0.25 and $x=0.30$ only to illustrate the presence of low-$T$ upturns in the data and are not used henceforth.  The  fitted values of the Sommerfeld coefficient $\gamma$ and the lattice-vibration parameters $\beta$ and $\delta$ for $x=0$ to 0.15 and for $x=0.50$ are listed in Table~\ref{Tab.HC}. The Debye temperature $\Theta_{\rm D}$ is obtained from the value of $\beta$ according to 
\bea
\Theta_{\rm D} = \left(\frac{12\pi^4Rn}{5\beta}\right),
\label{Eq:thetaD}
\eea
where $n$ is the number of atoms per formula unit ($n=5$ here). The values of $\Theta_{\rm D}$ obtained from the $\beta$ values are listed in Table~\ref{Tab.HC}.

The $C{\rm_p}(T)$ data over the temperature range 100 to 280~K were fitted for $x=0$ to 0.15 and for $x=0.50$ by~\cite{Goetsch2012}
\bea
C_{\rm p}(T) &=& \gamma T+ nC_{\rm V\,Debye}(T),\label{Eq:Debye_Fit} \\*
C_{\rm V\,Debye}(T) &=& 9R \left(\frac{T}{\Theta_{\rm D}}\right)^3\int_{0}^{\Theta_{\rm D}/T}\frac{x^4e^x}{(e^x-1)^2} dx,\nonumber
\eea
where $n=5$ is the number of atoms per formula unit and $C_{\rm V\,Debye}(T)$ is the Debye lattice heat capacity at constant volume per mole of atoms. The $\gamma$ values in Table~\ref{Tab.HC} obtained from the low-$T$ fits were used.  The fits are shown as the black solid curves in Fig.~\ref{Fig:Heat_Capacity} and the fitted values of the Debye temperature $\Theta_{\rm D}$ are listed in Table~\ref{Tab.HC}.  One sees that the values of $\Theta_{\rm D}$ obtained from the fits by Eqs.~(\ref{Eq.CpFit}) and~(\ref{Eq:Debye_Fit}) are reasonably close to each other for $x=0$ to 0.15 and for $x=0.50$. 

The density of degenerate conduction carrier states at the Fermi energy $E\rm_F$ for both spin directions $D_\gamma$($E\rm_F$) is obtained from $\gamma$ according to
\bse
\bea
{\cal D}_\gamma(E\rm_F)=\frac{3\gamma}{\pi^2k_B^2}
\label{Eq:DOS1}
\eea
\rm{which gives}
\bea
{\cal D}_\gamma(E\rm_F)\left[\frac{states}{eV~f.u.}\right] = \frac{1}{2.359}~\gamma\left[\frac{mJ}{mol~ K^2}\right],
\label{Eq:DOS} 
\eea
\ese
where ``mol'' refers to a mole of formula units (f.u.).  The ${\cal D}_\gamma$($E\rm_F$) values calculated for our \scna\ crystals from their $\gamma$ values using Eq.~(\ref{Eq:DOS}) are listed in Table~\ref{Tab.HC}.

\subsection{Non-Fermi-Liquid Behaviors in the Heat Capacity at Low Temperatures}

\begin{figure}
\includegraphics[width=3.45in]{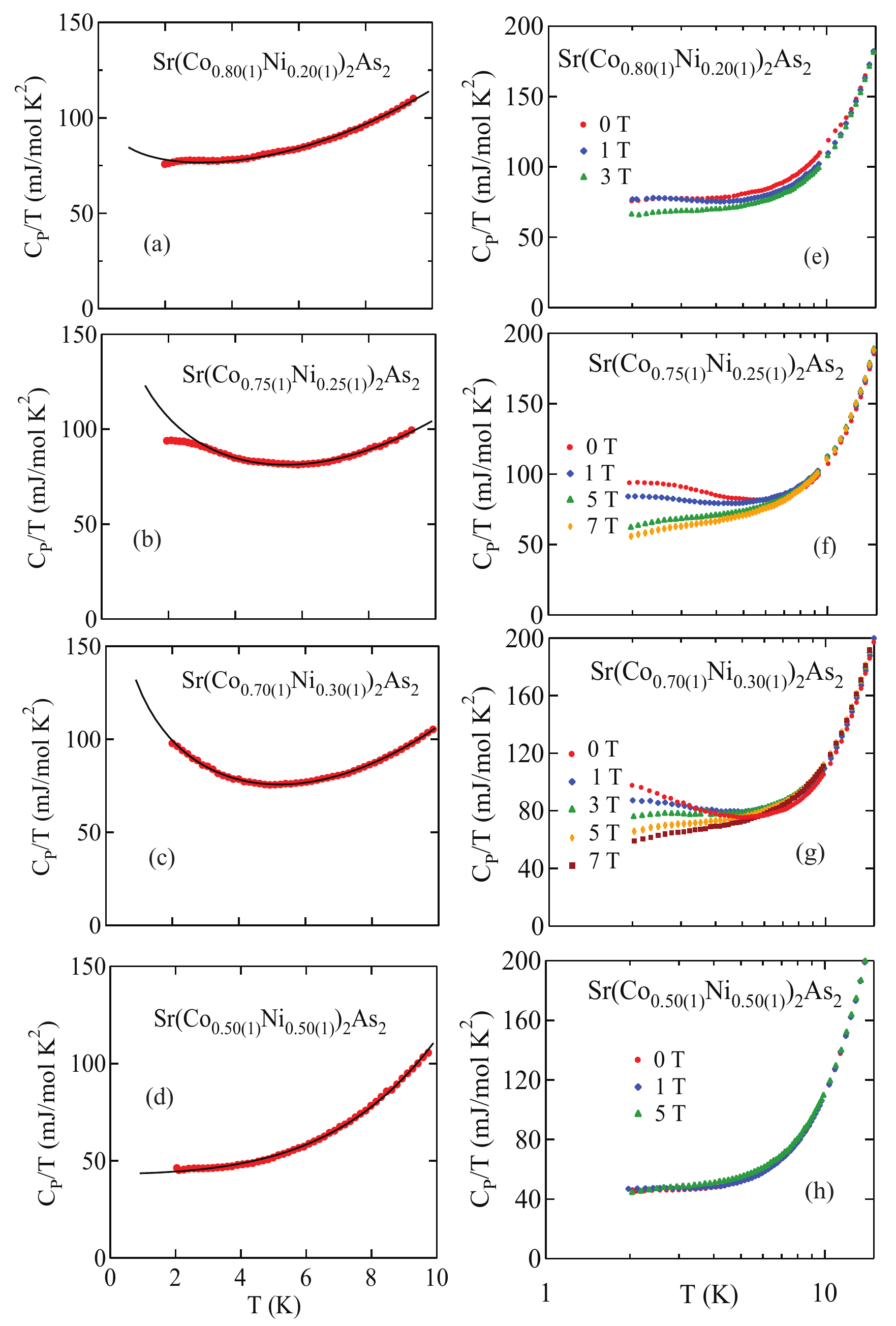}
\caption{(a)--(d) Heat capacity $C_{\rm p}/T$ vs $T$ data at zero field from 1.8 to 10~K for \scna\ with $x=0.20$, 0.25, 0.30, and 0.50, respectively. The solid black curves in (a)--(c) are fits by Eq.~(\ref{Eq:Cp_SF_Fit2}). The plot and fit in~(d) are reproduced from the inset of Fig.~\ref{Fig:Heat_Capacity}(f).  (e)--(h) $C_{\rm p}/T$ vs $T$ at various fields on logarithmic temperature axes.}
\label{Fig:HC_SF}
\end{figure}

As seen in the insets of Fig.~\ref{Fig:Heat_Capacity}, the $C_{\rm p}(T)/T$ vs $T^2$ plots at low~$T$ show linear (Fermi-liquid) behaviors for $ x=0.0$, 0.04, 0.06, 0.10, 0.15, and 0.50. However, for $x=0.20$, 0.25, and 0.30, $C_{\rm p}(T)/T$ vs $T^2$ departs from linearity below 10~K\@.  In particular, the data for the latter two compositions exhibit clear upturns with decreasing~$T$\@.  The upturns in $C_{\rm p}(T)/T$ at low~$T$ for $x=0.25$ and 0.30 cannot result from high-$T$ tails of Schottky anomalies arising from local-moment paramagnetic impurities because in that case the size of the upturns would {\it increase} with field, rather than decrease as demonstrated in Figs.~\ref{Fig:HC_SF}(a)--(c).  Similarly, in a \mbox{metallic} spin glass such as Cu:Mn containing 2790 ppm of Mn and with a spin-glass transition temperature of 3.9~K, a 7~T field strongly {\it increases} the magnitude of the high-$T$ tail of the broad peak in the heat capacity~\cite{Brodale1983}.

Because FM interactions dominate AFM interactions in \scna\ for compositions of interest here as shown in previous sections, we fitted the $C_{\rm p}(T)/T$ versus~$T$ data for $x=0.20$, 0.25, and 0.30 from our minimum temperature of 2~K to 10~K by a model for FM quantum-critical behavior.  For a 3D ferromagnet in the classical temperature regime at the QCP composition~\cite{Millis1993}, or for a 2D ferromagnet treated using scaling theory~\cite{Wu2014}, a contribution $\ln T$ to $C/T$ is found, yielding the fit function
\begin{equation}
\frac{C_{\rm p}(T)}{T} = \gamma + \beta T^2  + \delta T^4 + \kappa \ln(T/T_{\rm SF}),
\label{Eq:Cp_SF_Fit2}
\end{equation}
where $\gamma$ and \{$\beta$, $\delta$\} again represent the coefficients of the degenerate electronic and lattice heat-capacity contributions, respectively, $\kappa$ represents the strength of the FM spin fluctuations, and $T_{\rm SF}$ is a characteristic spin-fluctuation temperature.  The $C/T\sim \ln T$ contribution has been observed for quantum-critical fluctuations in the layered ferromagnet ${\rm YFe_2Al_{10}}$~\cite{Wu2014} and in the 3D ferromagnet \underline{Pd}Ni~\cite{Nicklas1999}.  We find that Eq.~(\ref{Eq:Cp_SF_Fit2}) fits our low-temperature $C_{\rm p}/T$ versus $T$ data for $x = 0.20$, 0.25, and 0.30 at low temperatures reasonably well, as shown in Figs.~\ref{Fig:HC_SF}(a), \ref{Fig:HC_SF}(b), and \ref{Fig:HC_SF}(c), respectively, where the composition $x=0.30$ is close to the quantum-critical value.  The fitting parameters  $\gamma$,  $\beta$,  $\delta$, $\kappa$, $T_{\rm SF}$, and the calculated values of $\Theta_{\rm D}$ and ${\cal D}_\gamma (E\rm_F)$ obtained from $\beta$ and $\gamma$ utilizing Eqs.~(\ref{Eq:thetaD}) and~(\ref{Eq:DOS}), respectively, are listed in Table~\ref{Tab.HC_SF}.  

\begin{figure}
\includegraphics[width=2.75in]{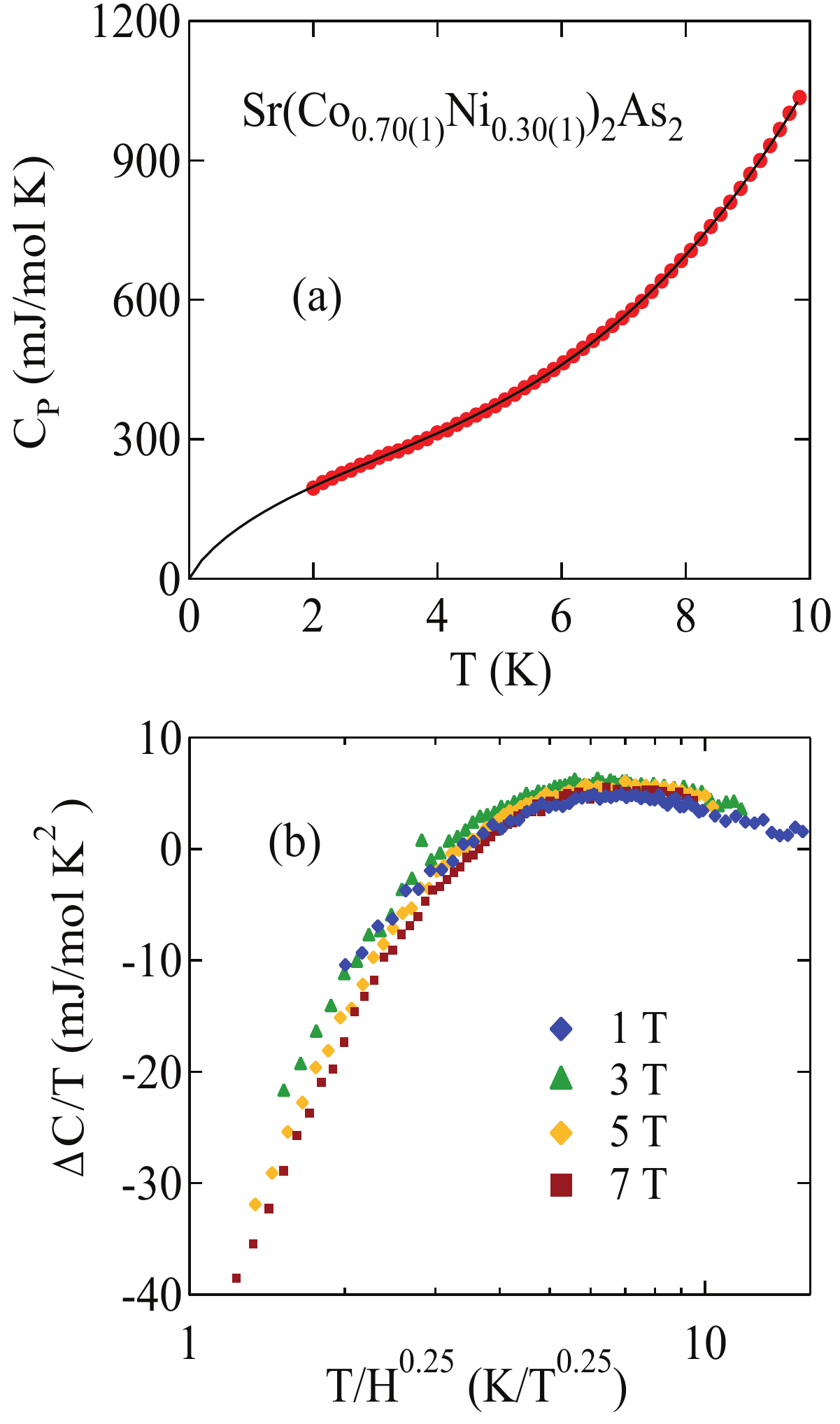}
\caption{(a)~Heat capacity $C_{\rm p}$ vs temperature~$T$ in zero magnetic field from 1.8 to 10~K for \scna\ with the approximate quantum-critical composition $x=0.30$. The solid black curve is a fit of the data using  Eq.~(\ref{Eq:Cp_SF_Fit2}) and the parameters in Table~\ref{Tab.HC_SF}, with an extrapolation of the fit to zero temperature. (b)~Scaling plot of the $\Delta C/T$ data in Fig.~\ref{Fig:HC_SF}(g) versus $\log_{10}(T/H^{0.5})$.}
\label{Fig:Cp_QCP_Fit_Ni0p30Per}
\end{figure}

If the quantum-critical behavior for $x=0.30$ is indeed due to FM fluctuations, one would expect them to be suppressed by applied magnetic fields~\cite{Jia2009, Wu2014}.  Such was found to be the case, as shown in Fig.~\ref{Fig:HC_SF}(g).  After subtracting the $H=0$ data from the $H>0$ data in Fig.~\ref{Fig:HC_SF}(g), one obtains the quantity
\begin{equation}
\frac{\Delta C}{T} \equiv \frac{C_{\rm p}(H)}{T} - \frac{C_{\rm p}(H=0)}{T},
\label{Eq:DeltaC/T}
\end{equation}
which is plotted versus log$_{10}(T/H^{0.25}$) in Fig.~\ref{Fig:Cp_QCP_Fit_Ni0p30Per}(b), where the $H$ exponent was adjusted to obtain the best scaling.  A reasonably good scaling is seen.  A very good scaling of $\Delta C/T$ versus log$_{10}(T/B^{0.59}$) was previously found for the quasi-2D ferromagnet ${\rm YFe_2Al_{10}}$ in Ref.~\cite{Wu2014}, where the compound is found to be intrinsically quantum-critical.

\subsection{\label{Sec:ElecStruct} Comparison of Calculated Electronic-Structure DOS with Experimental Heat Capacity DOS}

\begin{figure}
\includegraphics [width=3.in]{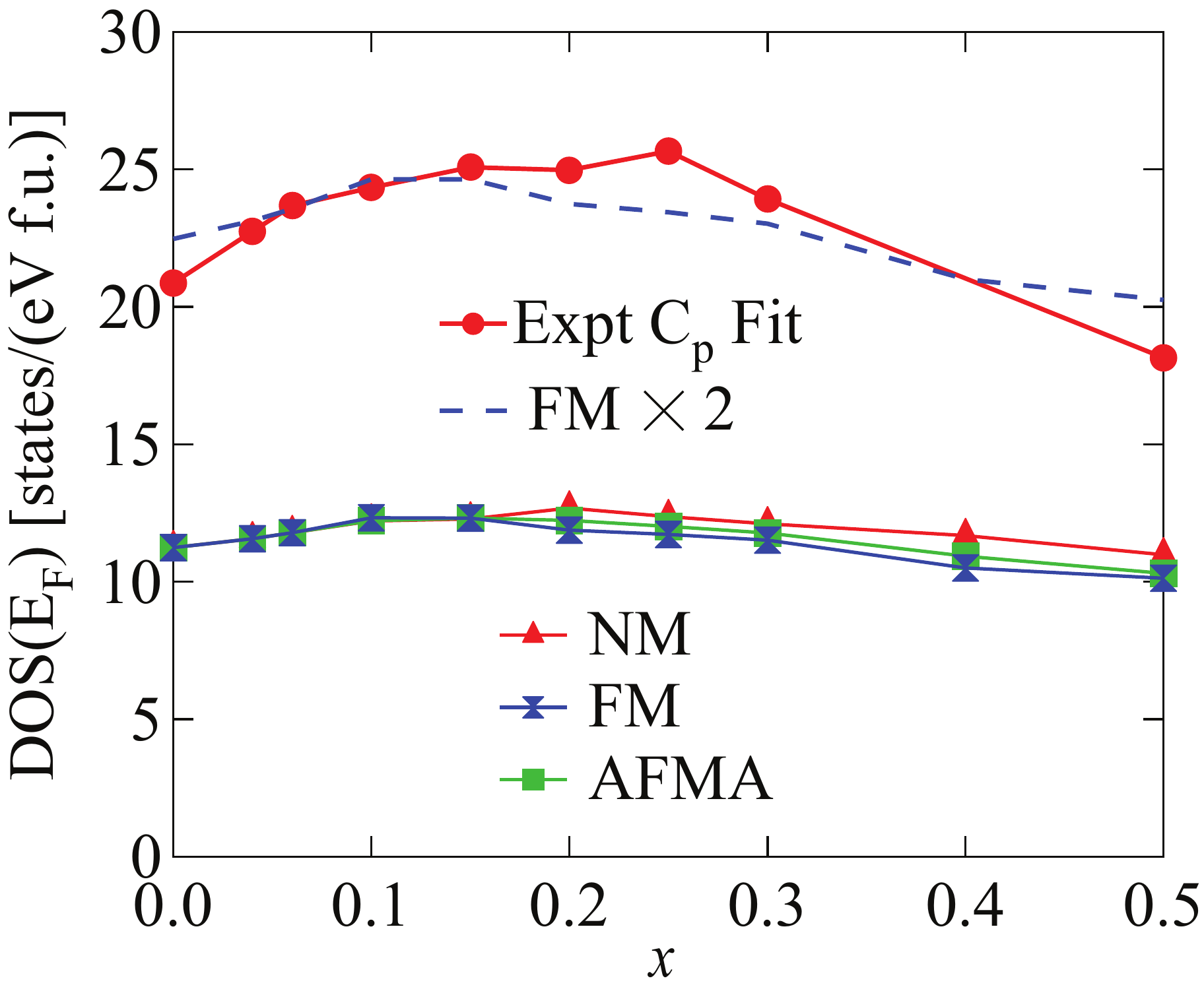}
\caption {KKR-CPA electronic density of states at the Fermi energy [DOS$(E_{\rm F})]$ for both spin directions calculated for nonmagnetic (NM), ferromagnetic (FM) and A-type antiferromagnetic (AFMA) \scna\ with compositions $x=0.0$ to~0.5.  The experimental densities of states DOS$(E_{\rm F})$ (Expt C$_{\rm p}$ Fit) in Tables~\ref{Tab.HC} and~\ref{Tab.HC_SF} obtained from the electronic heat capacity coefficients~$\gamma$ using Eq.~(\ref{Eq:DOS}) are plotted for comparison.  The blue dashed line is the theoretical FM DOS$(E_{\rm F})$ versus~$x$ in the lower part of the figure multiplied by two.}
\label{Fig:DOSEFSCNA}
\end{figure}

Figure~\ref{Fig:DOSEFSCNA} shows the KKR-CPA density of states (DOS) at $E_{\rm F}$ versus electron doping~$x$ for the nonmagnetic (NM), ferromagnetic~(FM), and A-type antiferromagnetic (AFMA) states, with similar results.  The DOS is seen to first increase up to $x\approx 0.2$ and then slowly decrease as a function of Ni substitution~$x$ on the Co site (electron-doping), while the main DOS peak is still above $E_{\rm F}$ [green line in Fig.~\ref{Fig:DOSSCNA}(b)]. This change in the DOS$(E_{\rm F})$ results from the significant broadening of the flat bands around~$E_{\rm F}$. The variation in DOS$(E_{\rm F})$ with~$x$ is compared in  Fig.~\ref{Fig:DOSEFSCNA} with the results obtained from the experimental DOS$(E_{\rm F})$ values in Tables~\ref{Tab.HC} and~\ref{Tab.HC_SF} obtained by fitting the heat capacity data.  One sees that the $x$~dependence of the observed and calculated DOS$(E_{\rm F})$ behaviors are very similar, but the observed values obtained from the heat capacity are shifted upwards.  This difference can be accounted for by the presence of an electron-phonon coupling constant~$\lambda_{\rm ep}$ and a many-body renormalization of the band structure via electron-electron interactions represented by an electron-electron coupling constant~$\lambda_{\rm ee}$ according to~\cite{Anand2014}
\begin{equation}
{\rm DOS(E_{\rm F})^{obs} = DOS(E_{\rm F})^{calc}(1 + \lambda_{ep})(1 + \lambda_{ee})},
\end{equation}
where typically $0<\lambda_{\rm ep}\lesssim 1$.  We infer from Fig.~\ref{Fig:DOSEFSCNA} that the net enhancement of the band-structure density of states is about a factor of two, as shown by the dashed blue line in Fig.~\ref{Fig:DOSEFSCNA}.  This enhancement is about the same as in \bca~\cite{Pandey2013}.  Hence, the behavior is explained well by the disorder-induced effects that lead to competing magnetic states (NM, FM, and AFMA), see Sec.~\ref{Sec:ElecStruct:Magnetism}.

\section{\label{Sec:Res} Electrical Resistivity}

\begin{figure}
\includegraphics[width=3.3in]{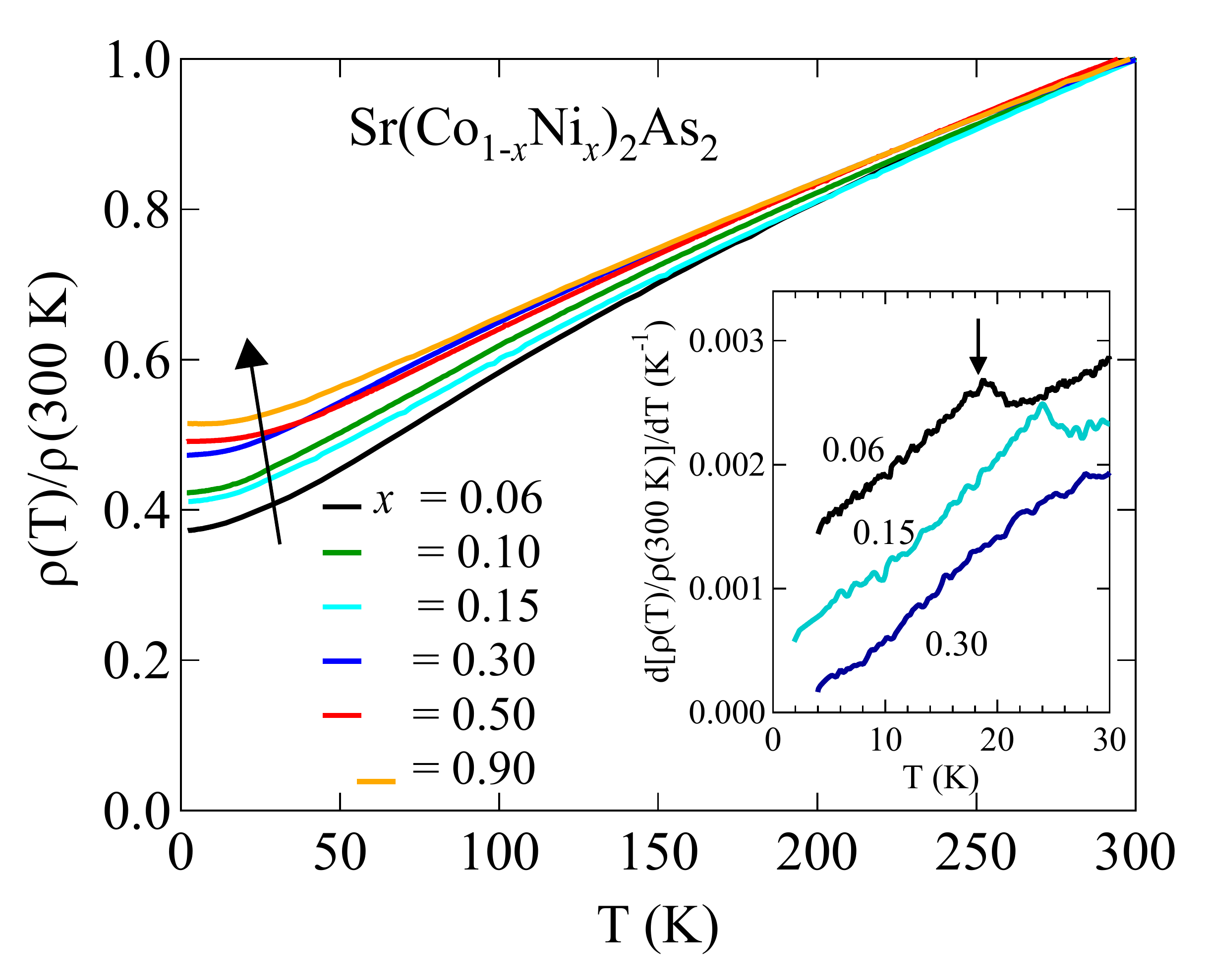}
\caption{Normalized electrical resistivity $\rho(T)/\rho(300~{\rm K})$ for \scna\ crystals with $x=0.06,$ 0.10, 0.15, 0.30, 0.50, and 0.90. The inset shows the temperature derivative of the normalized resistivity in the vicinity of $T_{\rm N}$\@. The derivative curves for $x=0.06$ and 0.15 reveal sharp features at magnetic transitions (indicated with an arrow for $x=0.06$) and smooth evolution without sharp features in $x=0.30$.}
\label{Fig:Res2}
\end{figure}

\begin{figure}
\includegraphics[width=2.5in]{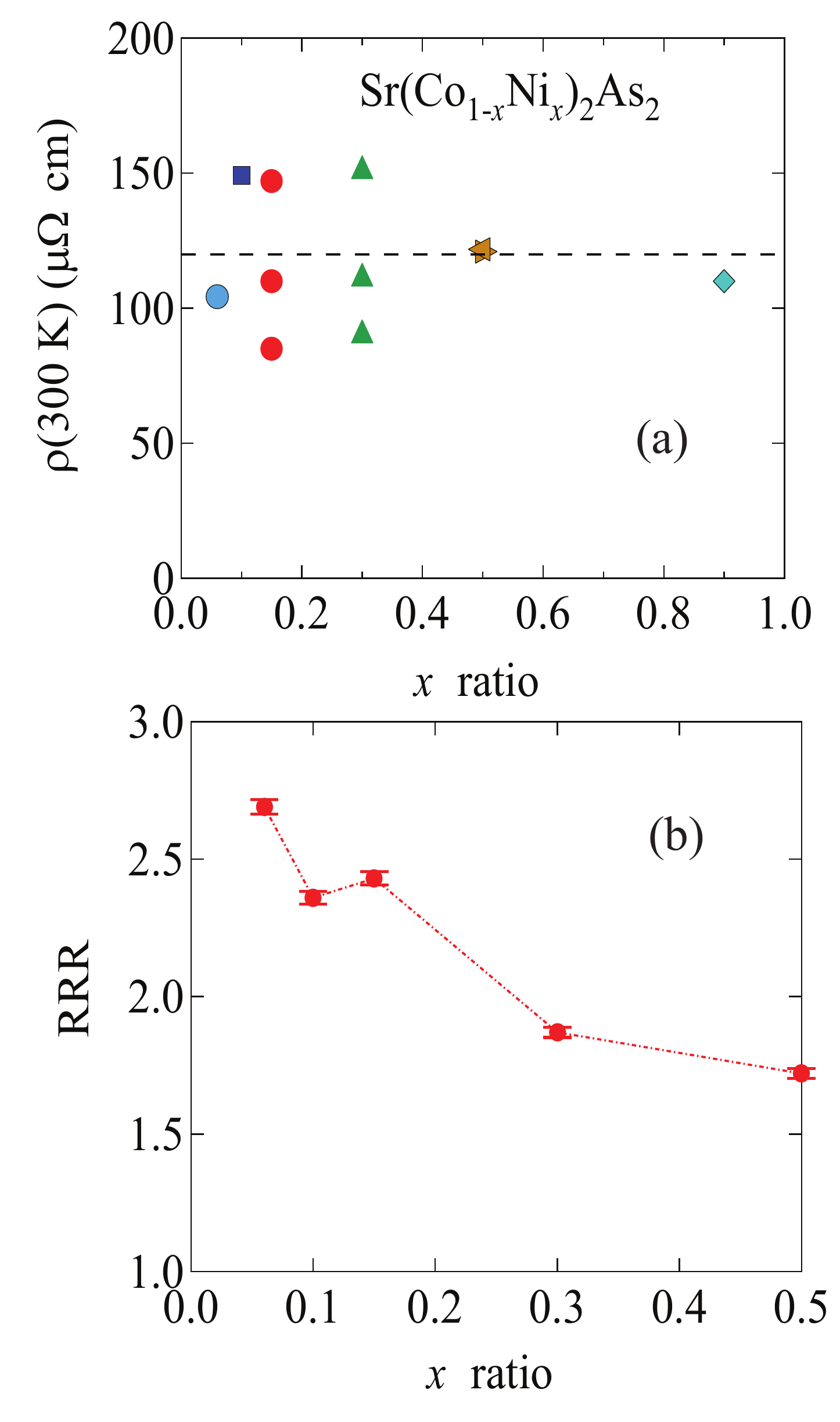}
\caption{(a)~Electrical resistivity $\rho$ at $T=300$~K and (b)~residual resistivity ratio ${\rm RRR} \equiv\rho(2~{\rm K})/\rho(300$~K) for \scna\ crystals with $x=0.06,$ 0.10, 0.15, 0.30, and 0.50.}
\label{Fig:Res1}
\end{figure}

\begin{figure}
\includegraphics[width=2.75in]{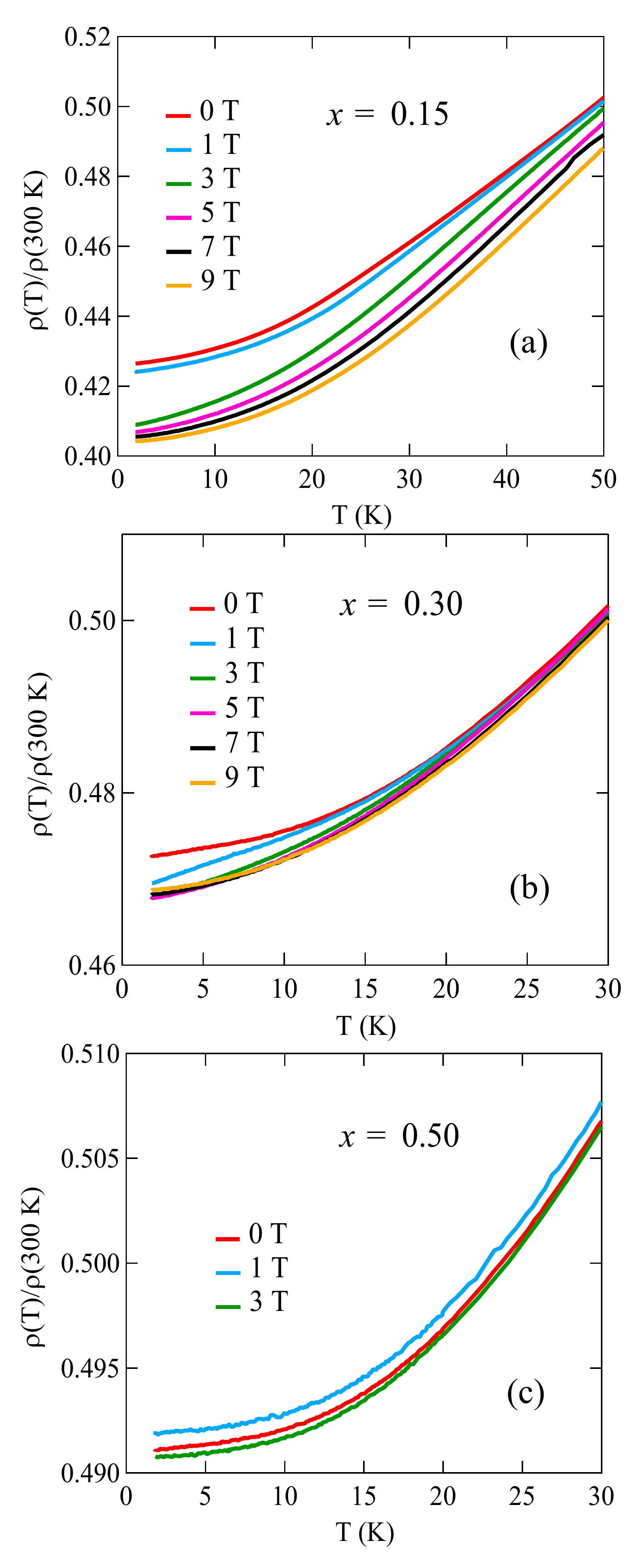}
\caption{In-plane normalized resistivity $\rho(T)/\rho(300$~K) for \scna\ crystals with compositions $x=0.15, 0.30$, and 0.50 measured in the indicated magnetic fields $H \parallel c$ as a function of temperature~$T$.}
\label{Fig:Res3}
\end{figure}

\begin{figure}
\includegraphics[width=3.45in]{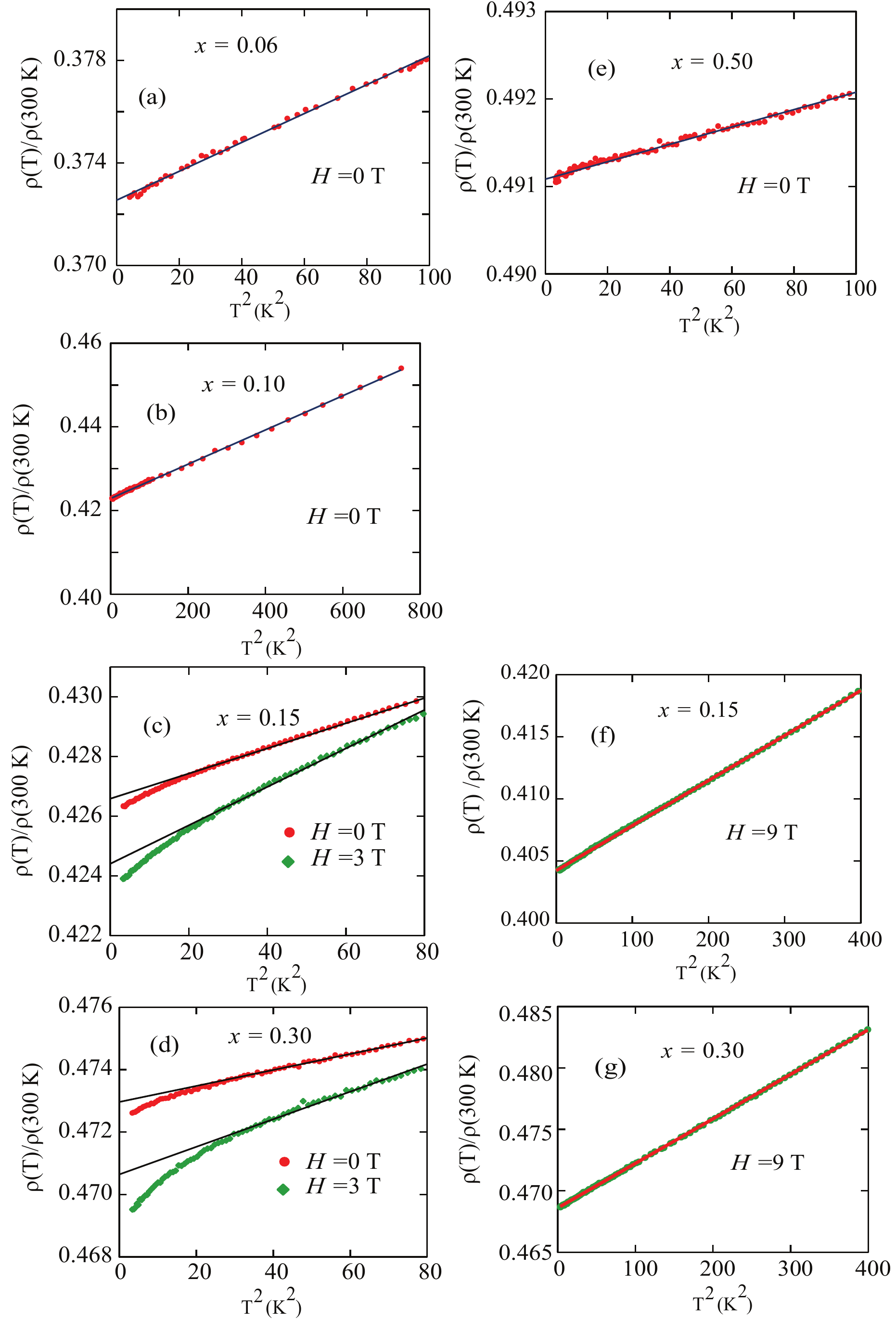}
\caption{(a)--(e)~In-plane normalized resistivity $\rho(T)/\rho(300$~K) vs $T^2$ for \scna\ crystals with compositions $x=0.06,$ 0.10, 0.15, 0.30, and 0.50 in magnetic fields $H_c = 0$ and/or 3~T, as indicated. (f,~g)~$\rho(T)/\rho(300$~K) vs $T^2$ for $x=0.15$ and~0.30 in $H_c=9$~T\@.}
\label{Fig:Res4}
\end{figure}

\begin{figure}
\includegraphics[width=2.25in]{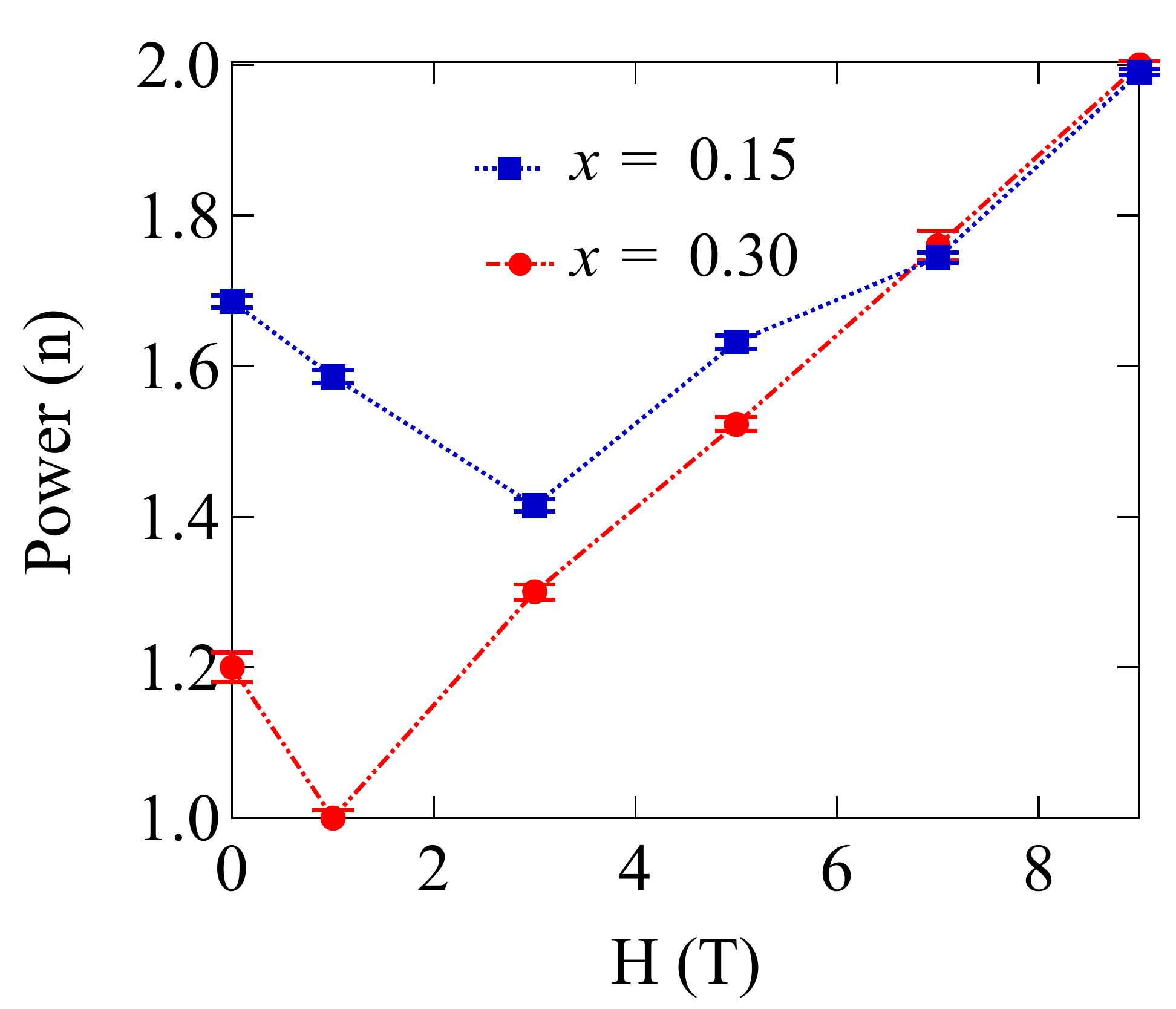}
\caption{ The power~$n$ in the expression $\rho(T)=\rho_0+AT^n$ for \scna\ crystals with $x=0.15$ and~0.30 at low temperatures~$1.8~{\rm K}\lesssim T \lesssim 9$~K versus applied magnetic field~$H_c$. These fits are not related to the linear fits shown in Fig.~\ref{Fig:Res4}.}
\label{Fig:Res5}
\end{figure}

The in-plane electrical resistivity of the \scna\ crystals was studied for compositions with $x=0.06$, 0.10, 0.15, 0.30, 0.50, and 0.90. All compositions show metallic temperature-dependent resistivity in zero applied field, as shown in Fig.~\ref{Fig:Res2}. The resistivity value at room temperature was determined to be  $\approx 120\pm30~\mu\Omega$\,cm as seen in Fig.~\ref{Fig:Res1}(a), and did not show any systematic trend with~$x$. The large error bars reflect the fact that the measurements were not performed on a statistically significant array of samples of each composition and are known to suffer from potential cracks \cite{anisotropy} affecting the correct determination of the sample geometric factor. Nonetheless, this behavior is notably different from the evolution of the resistivity at room temperature in electron-doped Ba(Fe$_{1-x}$Co$_x$)$_2$As$_2$, showing a factor of three decrease in a similar doping range~\cite{pseudogap}. The behavior in \scna\ is rather similar to the doping-independent value of $\rho(300~{\rm K})$ found in hole-doped Ba$_{1-x}$K$_x$Fe$_2$As$_2$~\cite{YLiucrystals}.  To exclude the effect of a systematic error in the comparison of quantitative resistivity measurements, we analyze all resistivity data using normalized $\rho(T)/\rho(300{\rm K})$ values.

The expected anomalies in the temperature-dependent resistivity at the magnetic transitions \cite{FisherLanger} are very weak in \scna\ and can be noticed only in the derivative plots, $d\rho(T) /dT$, shown in the inset of Fig.~\ref{Fig:Res2}. Note that the derivative exhibits sharp peaks at around 20~K at the respective N\'eel temperatures for $x=0.06$ and $x=0.15$.  The derivative for the $x=0.30$ crystal with no magnetic ordering shows no such maximum.  The resistivity in Fig.~\ref{Fig:Res2} at low temperatures is determined mostly by substitutional disorder, with the residual resistivity ratio $\rho(300~{\rm K})/\rho(1.8~{\rm K})$ in Fig.~\ref{Fig:Res1}(b) decreasing from 20 in pure SrCo$_2$As$_2$ \cite{Pandey2013} to 2.7 for $x=0.06$ and 1.8 for $x=0.50$.

The resistivity notably decreases on application of a magnetic field parallel to the $c$-axis as shown in Fig.~\ref{Fig:Res3}. This negative magnetoresistance clearly shows a contribution of spin-disorder scattering in the samples. A more detailed inspection of the low-temperature part of the temperature-dependent resistivity is made using $\rho$ versus $T^2$ plots as shown in  Fig.~\ref{Fig:Res4}. In such plots the temperature-dependent resistivity expected for a Fermi liquid, $\rho= \rho(0) +AT^n$ with $n=2$, would be a straight line. This is indeed observed in zero magnetic field for the crystals with $x=0.06$, 0.10, and 0.50 in panels (a), (b), and (e), respectively, consistent with the heat capacity data in the respective insets of Fig~\ref{Fig:Heat_Capacity}.

For crystals with $x=0.15$ and 0.30, downward deviations from the linear dependence are found in Fig.~\ref{Fig:Res4} at the lowest temperatures, suggesting that the exponent~$n$ of the power-law function becomes smaller than 2, thus revealing non-Fermi-liquid behavior \cite{Schofield1999, Stewart2001}. The exponent $n$ found from fits to the resistivity data between 1.8~K and 9~K (not shown) exhibits a notable dependence on magnetic field (see Fig.~\ref{Fig:Res5}), as is observed in systems with magnetic-field-tuned quantum-critical points such as  YbRh$_2$Si$_2$~\cite{YRS}, CeCoIn$_5$~\cite{PaglioneQCP}, YbAgGe~\cite{YbAgGe}, and YbPtBi~\cite{YbPtBi}. The minimum in the $n(H)$ dependence is observed at 3~T for the crystal with $x=0.15$ and at 1~T for $x=0.30$. In the latter case the exponent at the $n(H)$ minimum is close to $n=$1, as observed in cuprates, organics and heavy fermion systems at their quantum critical points \cite{Louisreview}. The fact that the characteristic field at which the exponent $n$ is minimum decreases with increasing $x$ suggests that quantum-critical behavior may be observed in zero field for a composition near $x=0.30$, as we observe in the heat capacity. 

\section{\label{Sec:NFLdiscuss} Discussion of Non-Fermi-Liquid Behaviors in S\lowercase{r}(C\lowercase{o}$_{1-x}$N\lowercase{i}$_x)_2$A\lowercase{s}$_2$}

Non-Fermi-liquid (NFL) behaviors in $d$- and $f$-band metals have been intensively studied since the 1990's~\cite{Schofield1999, Stewart2001}, where the early studies emphasized theoretical and experimental studies of compounds containing lanthanides and/or uranium.  Studies of $d$-band compounds other than those presented here have previously been reported, such as studies of ruthanates~\cite{Laad2001}, high-$T_{\rm c}$ cuprates, and iron pnictides~\cite{Ishida2010}, where materials showing NFL characteristics are sometimes dubbed ``bad metals'' or ``strange metals''.  NFL behavior has been documented in P-doped 122-type iron arsenide crystals ${\rm BaFe_2}$(As$_{1-x}$P$_x)_2$~\cite{Analytis2014}.  For optimally hole-doped \bkfa\ crystals with a superconducting onset temperature $T_{\rm c}=39$~K, high-resolution laser-based angle-resolved photoemission spectroscopy (ARPES) measurements indicated that the normal state of this material has a well-defined Fermi surface but no quasiparticles~\cite{Huang2019}.  Furthermore, the superconducting condensate is found to be coherent but the temperature dependence of the superconducting gap dramatically differs from the Bardeen-Cooper-Schrieffer (BCS) prediction~\cite{Huang2019}.  Such behaviors originate from criticality~\cite{Coleman2005}, where quantum-mechanical fluctuations drive phase transitions at $T=0$ upon varying a control parameter such as pressure, magnetic field, or a material's composition, but quantum criticality can also have important observable finite-temperature manifestations~\cite{Stewart2001, Coleman2005}.

A special case of this phenomenon is the \emph{local} marginal Fermi liquid as originally applied to the high-$T_{\rm c}$ cuprates, in which the energy scale for the low-energy particle-hole excitations is set by the temperature~\cite{Varma1989, Abrahams1996}.  Some of the predicted properties are in agreement with measurements on these cuprates, such as the $T$-linear normal-state electrical resistivity over a wide range of temperature above~$T_{\rm c}$, in contrast to the $T^2$ behavior expected for a Fermi liquid, and the optical conductivity.  For our \scna\ crystals, an example of $\rho\sim T^n$ behavior with $n=1$ was found for $x=0.30$ in $H=1$~T (see Fig.~\ref{Fig:Res5}); for other fields with $0\leq H \leq 7$~T we obtained $1.2 \leq n\leq 1.8$ for both $x=0.15$ and 0.30, but with both crystals showing Fermi-liquid behavior with $n=2$ at the higher field $H=9$~T\@.

Itinerant weak ferromagnetism was discovered in a polycrystalline sample of ZrZn$_2$ in 1957~\cite{Matthias1957}.  This was surprising because both Zr and Zn are superconducting and are nonmagnetic elements.  The Curie temperature and the low-$T$ saturation moment of their polycrystalline sample were estimated as $T_{\rm C} = 35$~K and $\mu_{\rm sat} = 0.13~\mu_{\rm B}$/f.u., respectively~\cite{Matthias1957}. Subsequent work on high-quality single crystals of ZrZn$_2$~\cite{Pfleiderer2001, Uhlarz2004, Smith2008, Zou2012, Sutherland2012} yielded $T_{\rm C} = 28.5$~K and $\mu_{\rm sat} = 0.17~\mu_{\rm B}$/f.u.~\cite{Uhlarz2004}.  Here $\rho\propto T^{5/3}$, attributed to the presence of a \emph{nonlocal} marginal Fermi liquid state in ZrZn$_2$~\cite{Smith2008}.  Our $\rho(T)$ data in Fig.~\ref{Fig:Res5} for a \scna\ crystal with $x=0.15$ has a temperature exponent $n\sim5/3$ for $H=0$, 1, 5, and 7~T, and for $x=0.30$ at $H=7$~T\@.  We note that the Sommerfeld coefficient of ZrZn$_2$ in $H=0$ is large, $\gamma = 45~{\rm mJ/mol~K^2}$~\cite{Pfleiderer2001}, of the same order as the values for \scna\ in Tables~\ref{Tab.HC} and~\ref{Tab.HC_SF}.  However, the $C_{\rm p}(T)$ data for ZrZn$_2$~\cite{Zou2012} do not show the upturns we see for \scna\ crystals with $x\geq0.25$ and~30 in the insets of Fig.~\ref{Fig:Heat_Capacity} and in Fig.~\ref{Fig:HC_SF} because ZrZn$_2$ is not close to a quantum critical point. On the other hand, $C_{\rm p}(T)$ at low~$T$ for ZrZn$_2$ is suppressed with increasing $H$~\cite{Pfleiderer2001} as we also observe in Fig.~\ref{Fig:HC_SF}, which likely arises from the suppression of FM fluctuations with field in both systems.  The low-$T$ resistivity of ZrZn$_2$ increases with increasing~$H$~\cite{Zou2012}, whereas we observe a decrease in the normalized resistivity for $x=0.15$ and~0.30 in Figs.~\ref{Fig:Res3} and~\ref{Fig:Res4}.

Thus there appears to be a field- and composition-dependent crossover in the nature of the non-Fermi liquid when it occurs in the \scna\ system.  


\section{\label{Sec:Sum} Summary}

The physical properties of Ni-doped \scna\ single crystals were investigated for compositiions $x= 0$ to~0.9. Single-crystal x-ray diffraction studies at room temperature showed a continuous crossover from the ucT (uncollapsed tetragonal) to cT (collapsed tetragonal) structure with increasing~$x$. The temperature~$T$ dependences of the magnetic susceptibilities $\chi_{ab}$ and $\chi\rm_c$ demonstrated that the paramagnetic (PM) ground state of \sca\ transforms to an antiferromagnetic (AFM) state with only a trace amount of Ni doping ($x = 0.013$). This appears to be  consistent with our {\it ab~initio} electronic-structure calculations for $x=0$ that exhibit a flat-band peak in the density of states just above the Fermi energy that contibutes to magnetic ordering at small doping levels.  The AFM order persists up to $x=0.25$, above which the PM state is observed for $x=0.30$ and $x=0.50$.

Based on  molecular field theory (MFT) for local magnetic moments, the $\chi(T)$ data at temperatures $T$ less than the N\'eel temperature $T\rm_N$ indicate that the ordered moments are aligned within the $ab$ plane. The nonzero limits of $\chi_{ab}(T < T\rm_N)$ suggest a planar helical AFM structure. We calculated the turn angle $kd$ between adjacent $ab$-plane ordered moments along the helix $c$-axis within MFT from the ratio $\chi_{ab}(T=0)/\chi(T_{\rm N})$ for the crystals showing AFM ordering and found that the helix turn angle varies with Ni concentration.  Neutron diffraction or magnetic x-ray scattering measurements of the magnetic structures are needed to clarify the AFM structure and test our model for the structure.

The $\chi^{-1}(T)$ data at $T>T\rm_N$ follow Curie-Weiss-like behavior. The Weiss temperature and $T\rm_N$ follow the same trend, initially increasing and then decreasing with increasing Ni doping. The positive Weiss temperatures in the range $x=0.04-0.25$ indicate that ferromagnetic exchange interactions are dominant in these AFM crystals. In terms of our helical AFM structure hypothesis, the positive Weiss temperatures reflect the FM intralayer interactions needed to line up the spins within a layer of the helix, and that these are significantly stronger than the interlayer interactions which must include AFM interactions to drive the AFM order.

The magnetization versus field $M(H)$ data below $T\rm_N$ with {\bf H} in the $ab$~plane exhibit a spin-flop-like transition followed by a second-order transition to the PM state for $x=0.04$ to~0.20, whereas $c$-axis measurements reveal only the expected second-order AFM to PM transition. The saturation moment in the ordered state increases initially with Ni doping and reaches a maximum at $0.10< x < 0.15$, then decreases for $x=0.25$ followed by paramagnetic $M(H)$ behavior for $x \geq 0.30$.

The ordered moment $\mu\rm_{sat}$ per transition-metal atom in the AFM state obtained by extrapolating the high-field $M(H)$ data to zero field is small relative to the effective moment $\mu\rm_{eff}$ in the PM state, indicating that \scna\ is essentially an itinerant ferromagnetic system. In addition, the magnetic moment per formula unit versus applied field $H$ does not saturate at high fields up to 14~T\@. Instead, at the higher fields it increases approximately linearly with field as often observed in weak itinerant FMs. 
As noted above, the crystals are AFM in zero field with the ordered moments oriented within the $ab$~plane.  However, from the Weiss temperature in the Curie-Weiss law at $T>T_{\rm N}$ the dominant interactions are FM\@.

The maximum values of the $ab$-plane spin-flop-like field and the $ab$-plane critical field $H_{{\rm c}\,ab}$ at which the PM state is entered are both less than about 4~T for all $T < T_{\rm N}(x)$.  Therefore our isothermal $M_{ab} (H)$ data measured up to 14~T were analyzed in terms of Takahashi's spin-fluctuation theory for weak itinerant ferromagnets via $M^2$ vs $H/M$ and $M^4$ vs $H/M$ isotherms at different temperatures for fields above about 4~T\@. The linear behavior of the $M^4$ versus~$H/M$ isotherm at high fields for $T\approx T\rm_N$ (i.e., at the Curie temperature $T_{\rm C}$ within Takahashi's theory) observed for $x=0.10$, 0.15, and 0.20 indicates that the critical magnetic isotherm at $T_{\rm C}$ is strongly influenced by thermally-induced amplitude fluctuations.

Takahashi's spin-fluctuation spectral parameter $T_0$ obained from fitting the high-field $M_{ab}(H)$ measurents at $T=2$~K and $T=T_{\rm C}$ and the other spectral parameter $T_{\rm A}$ used as a fitting parameter yield an excellent fit of $\chi^{-1}(T>T_{\rm C})$ by Takahashi's theory for the normal-state inverse susceptibility, confirming that \scna\ acts like a weak itinerant ferromagnet for high fields $H_{ab}>4$~T at $T<T_{\rm  C}$ and for small fields at $T > T_{\rm C}$. 

The heat capacity $C{\rm_p}(T)$ data exhibit a Fermi-liquid behavior at low temperature for $x= 0$, 0.04, 0.06, 0.10, 0.15, and 0.50.  No clear feature in $C{\rm_p}(T)$ was observed near $T\rm_N$ for any of the crystals that exhibit AFM ordering, as similarly found~\cite{Anand2014} for the itinerant AFM \cca. This absence evidently arises from the small entropy change at $T_{\rm N}$ associated with the itinerant nature of the magnetism.  A low-temperature upturn in $C{\rm_p}(H=0,T)/T$ was observed for $x=0.25$.  This upturn was stronger for $0.30$ which is near the quantum-critical concentration separating the AFM and PM phases.  The upturns were fitted by  a $\ln T$ temperature dependence consistent with ferromagnetic quantum-critical fluctuations. These low-$T$ $C{\rm_p}(T)/T$ upturns and hence the FM fluctuations were suppressed in a field $H= 7$~T as expected for ferromagnetic fluctuations.  A scaling of the field-dependent heat capacity $\Delta C/T$ against $T/H^{0.25}$ was demonstrated, also suggestive of quantum-critical behavior.

Negative magnetoresistance was observed indicating a significant contribution of spin-disorder scattering. In most compositions the electrical resistivity at low temperatures follows a $\rho=\rho_0+AT^n$ dependence with $n=2$, as expected for a Fermi liquid. However,  deviations with $n<2$ are found for compositions $x=0.15$ and~0.30, the latter being near the boundary of the long-range magnetic-ordering composition  that also show a notable dependence on magnetic field. These observations are consistent with the non-Fermi-liquid behaviors observed in the  heat capacity measurements and suggest the occurrence of  a quantum critical point near $x=0.30$.

In conclusion, the Ni doping in \sca\ substantially affects the magnetic ground state and spin-spin correlations in \scna, causing a paramagnetic to antiferromagnetic phase transition as well as the spin-fluctuation-mediated non-Fermi-liquid state near $x=0.3$ where the long-range magnetic order disappears.  The long-range order thus terminates at an unusual quantum critical point at $x\approx 0.3$ at which the ferromagnetic-interaction-dominated antiferromagnetic phase undergoes a transition to the paramagnetic phase.



\section*{APPENDIX: TABLES OF FITTED PARAMETERS}


\begin{table*}
\caption{\label{CrystalData} Crystallographic data for \scna\ ($x=0 - 1$) single crystals at room temperature, including the fractional $c$-axis position parameter $z_{\rm As}$ of the As atoms, the tetragonal lattice parameters $a$ and~$c$, the unit cell volume $V_{\rm cell}$ containing two formula units of \scna, and the $c/a$ ratio. The compositions in the first column  were obtained from  EDS analyses.}
\begin{ruledtabular}
\begin{tabular}{ cccccc }
 Compound  & $z_{\rm As}$ & $a$ (\AA)  & $c$ (\AA) & $V_{\rm cell}$ (\AA$^3$) & $c/a$ \\
\hline
SrCo$_{2}$As$_2$                                              & 0.35770(11) 		&3.955(3)  		&  11.684(9) 	&   182.76(9) 		& 2.954(4) \\
Sr(Co$_{0.987(1)}$Ni$_{0.013(1)}$)$_2$As$_2$   & 0.3574(2)  		& 3.9601(16) 		&  11.653(5)  	&   182.74(17) 		& 2.942(2) \\
Sr(Co$_{0.96(1)}$Ni$_{0.04(1)}$)$_2$As$_2$      & 0.3579(3) 		&  3.958(3) 		& 11.667(12)	&   182.8(3) 		& 2.948(5) \\
Sr(Co$_{0.94(1)}$Ni$_{0.06(1)}$)$_2$As$_2$      & 0.35740(14)  	& 3.957(6)  		& 11.618(16) 	&  181.9(6) 		& 2.936(8)\\
Sr(Co$_{0.90(1)}$Ni$_{0.10(1)}$)$_2$As$_2$      &0.35759(14) 		& 3.970(2)  		& 11.566(7) 	& 182.3(2) 		& 2.913(3) \\
Sr(Co$_{0.85(2)}$Ni$_{0.15(2)}$)$_2$As$_2$      & 0.35769(11)  	& 3.976(5)  		&  11.530(16) 	& 182.3(5) 		&2.899(7)\\
Sr(Co$_{0.80(1)}$Ni$_{0.20(1)}$)$_2$As$_2$      & 0.35799(13) 		& 4.005(6)			&  11.482(16) 	& 184.2(6)			& 2.867(8)  \\
Sr(Co$_{0.75(1)}$Ni$_{0.25(1)}$)$_2$As$_2$      & 0.3581(4)		& 4.002(2) 		&  11.437(6) 	& 183.2(2) 		& 2.858(2)   \\
Sr(Co$_{0.70(1)}$Ni$_{0.30(1)}$)$_2$As$_2$      & 0.35799(10) 		& 4.005(5)  		& 11.420(15) 	& 183.2(5)			& 2.851(7) \\
Sr(Co$_{0.60(1)}$Ni$_{0.40(1)}$)$_2$As$_2$      & 0.35770(5) 		& 4.028(2)			& 11.181(6) 	& 181.4(2) 		& 2.776(2) \\
Sr(Co$_{0.50(1)}$Ni$_{0.50(1)}$)$_2$As$_2$      & 0.35775(7)		& 4.042(6)			& 11.030(18) 	& 180.2(6) 		& 2.729(8) \\
Sr(Co$_{0.40(1)}$Ni$_{0.60(1)}$)$_2$As$_2$      & 0.35840(9) 		& 4.074(5) 		&  10.850(14) 	&  180.1(5) 		& 2.663(6)\\
Sr(Co$_{0.30(1)}$Ni$_{0.70(1)}$)$_2$As$_2$      & 0.35886(5) 		& 4.0921(15)  		&  10.723(4) 	& 179.55(15) 		& 2.620(2) \\	
Sr(Co$_{0.20(1)}$Ni$_{0.80(1)}$)$_2$As$_2$      & 0.36010(4) 		& 4.1154(8)  		& 10.522(2) 	& 178.21(8) 		& 2.557(1) \\
Sr(Co$_{0.10(1)}$Ni$_{0.90(1)}$)$_2$As$_2$      & 0.36147(4) 		&  4.1366(9) 		&  10.355(2) 	& 177.19(9) 		& 2.503(1) \\
SrNi$_{2}$As$_2$\footnote{grown in Bi flux}        & 0.36244(8) 		& 4.1483(11)  		&  10.231(3) 	& 176.06(11) 		& 2.466(1) \\
\end{tabular}
\end{ruledtabular}
\end{table*}

\begin{table*}
\caption{\label{Tab.chidata1} Parameters obtained from Modified Curie-Weiss fits to $\chi(T)$ data between 100 and 300~K for \scna\ crystals using Eq.~(\ref{Eq:ChiFit}). Shown are  the $T$--independent contribution to the susceptibility $\chi_0$,  Curie constant per mole of formula units $C_\alpha$ with the applied field in the $\alpha = ab, c$ directions, the Weiss temperature $\theta\rm_{p\alpha}$, and the difference $\Delta\theta_{\rm p}$ between the $ab$~plane and $c$-axis Weiss temperatures $\Delta\theta_{\rm p} \equiv \theta_{{\rm p} ab} - \theta_{{\rm p}c}$. The effective Bohr magneton number per transition metal atom (Co and/or Ni) $p_{\rm_{eff\alpha}}$ was calculated from $C_\alpha$ using Eq.~(\ref{Eq:peffalpha}).}
\begin{ruledtabular}
\begin{tabular}{ccccccc}	
  										& 					 & $\chi_0$ 				& $C_{\alpha}$ 		    &  $p_{\rm eff\alpha}$ 	& $\theta_{\rm p\alpha}$ &	 $\Delta\theta_{\rm p}$\\
Compound 									&	Field Orientation	& $\rm{\left(10^{-4}~\frac{cm^3}{mol\,f.u.}\right)}$	 & $\rm{\left(\frac{cm^3 K}{mol\,f.u.}\right)}$    & & (K) & (K) \\
\hline
SrCo$_2$As$_2$                                 	& $H\parallel ab$ 		& $-4.2(2)$ 	& 0.92(1) 			&1.92(1)					& $-128(3)$	&$-39(4)$	\\
						                 	& $H\parallel c$ 		& $-6.1(2)$ 	& 1.12(1) 			&2.12(1) 					&$-167(3)$	&	\\
Sr(Co$_{0.987(1)}$Ni$_{0.013(1)}$)$_2$As$_2$    	& $H\parallel ab$ 		& 1.7(1)		& 0.442(4) 			& 1.329(6) 				&$-6.4(9)$	& $-0.5(9)$    		\\	
							               & $H\parallel c$ 		& 1.9(2)		& 0.438(5) 			&1.323(7) 				&$-6.9(1)$	& 	\\
Sr(Co$_{0.96(1)}$Ni$_{0.04(1)}$)$_2$As$_2$        & $H\parallel ab$ 		& 6.6(8) 		& 0.441(3) 			& 1.328(4) 				&10.4(6)		&4.5(7)		\\
							          	& $H\parallel c$ 		& 6.3(8)		& 0.466(2) 			&1.365(3) 				&5.9(4)		&	\\
Sr(Co$_{0.94(1)}$Ni$_{0.06(1)}$)$_2$As$_2$       	& $H\parallel ab$ 		& 5.4(2)		& 0.385(1) 			& 1.241(1) 				&23.1(2)		&6.0(5)		\\
										& $H\parallel c$ 		& 5.4(1)		& 0.419(3) 			&1.294(5) 				&17.1(4)		&	\\
Sr(Co$_{0.90(1)}$Ni$_{0.10(1)}$)$_2$As$_2$ 		& $H\parallel ab$ 		& 6.8(3) 		& 0.385(1) 			& 1.241(1) 				&32.9(2)		&1.7(4)		\\		
										& $H\parallel c$ 		& 5.6(1) 		& 0.411(2) 			&1.282(3) 				&31.2(3)		&	\\
Sr(Co$_{0.85(2)}$Ni$_{0.15(2)}$)$_2$As$_2$ 		& $H\parallel ab$ 		& 5.5(4) 		& 0.359(1) 			& 1.198(2) 				&35.2(3)		&4.5(4)		\\
										& $H\parallel c$ 		& 5.68(3)	 	& 0.393(1) 			&1.253(1) 				&30.7(2)		&	\\
Sr(Co$_{0.80(1)}$Ni$_{0.20(1)}$)$_2$As$_2$ 		& $H\parallel ab$		& 5.55(6) 	& 0.313(2) 			& 1.119(3) 				&25.6(5)		&4.2(3)		\\
										& $H\parallel c$ 		& 4.95(2) 	& 0.360(1) 			&1.200(1) 				&21.4(1)		&	\\
Sr(Co$_{0.75(1)}$Ni$_{0.25(1)}$)$_2$As$_2$ 		& $H\parallel ab$ 		& 14.4(3) 	& 0.245(1) 			& 0.989(2) 				&17.6(3)		&3.8(3)		\\
										& $H\parallel c$ 		& 12.7(3)		& 0.288(1) 			&1.073(2) 				&13.8(3)		&	\\
Sr(Co$_{0.70(1)}$Ni$_{0.30(1)}$)$_2$As$_2$ 		& $H\parallel ab$ 		& 5.14(3) 	& 0.199(1) 			& 0.892(2) 				&0.42(3)		&1.7(3)		\\
										& $H\parallel c$ 		& 1.18(2) 	& 0.219(1) 			&0.935(2) 				&$-1.3(3)$	&	\\
Sr(Co$_{0.50(1)}$Ni$_{0.50(1)}$)$_2$As$_2$ 		& $H\parallel ab$ 		& 7.8(2) 		& 0.164(4) 			& 0.81(1) 				&$-83(3)$		&$-3(4)$		\\
										& $H\parallel c$ 		& 7.7(5)		& 0.187(5) 			&0.86(1) 					&$-86(3)$		&	\\
\end{tabular}
\end{ruledtabular}
\end{table*}

\begin{table*}
\caption{\label{Tab.chidata2} Parameters obtained from modified Curie-Weiss fits to $\chi(T)$ data between 100 and 300~K for \scna\ using Eq.~(\ref{Eq:ChiFit}). Shown are the AFM transition temperature $T\rm_N$, the angle-averaged Curie constant per mol $C_{\rm {ave}}=(2 C_{ab} +C_c)/3$, the average Weiss temperature $\theta_{\rm p,ave} = (2 \theta_{{\rm p}ab} +\theta_{{\rm p}c})/3$, the angle-averaged effective Bohr magneton number per transition metal atom $p_{\rm_{eff,ave}}$ calculated from Eq.~(\ref{Eq:peffalpha}), and the ratio $f=\theta_{\rm p,ave}/T{\rm_N}$\@.  Also listed is the turn angle $kd$ between adjacent FM-aligned magnetic layers of a helix model obtained using Eq.~(\ref{Eq:kd}).}
\begin{ruledtabular}
\begin{tabular}{cccccccc}	
  			 							& $T\rm_N$		 & $C_{\rm ave}$ 	& $p_{\rm eff,ave}$  & $\theta_{\rm p,ave}$		 & $f=\theta_{\rm p,ave}/T_{\rm N}$ 	&$kd$ & $kd$ \\
Compound 									& (K) 			& $\rm{\left(\frac{cm^3 K}{mol}\right)}$	  &   & (K)  &	 		&  ($\pi$ rad)&	(degree) \\
\hline
SrCo$_2$As$_2$              					&		  		&0.99(1) 	  		&1.99(1)  		      	&$-141(2)$		  		&				&	&	\\						                            			
Sr(Co$_{0.987(1)}$Ni$_{0.013(1)}$)$_2$As$_2$    	&3.01(2)	  		&0.441(4) 			&1.33(1) 			&$-6.6(2)$				&$-2.2$			&	&	 \\								                   		 		
Sr(Co$_{0.96(1)}$Ni$_{0.04(1)}$)$_2$As$_2$        	&11.16(3)	  		&0.449(3)		 	&1.340(4) 			&8.9(5)					&0.8			&0.58 	&105 	\\							                  	 		
Sr(Co$_{0.94(1)}$Ni$_{0.06(1)}$)$_2$As$_2$       	&20.8(6)	  		&0.396(1)			&1.259(2) 			&21.1(3)				&1.01			&0.52	&93	\\													
Sr(Co$_{0.90(1)}$Ni$_{0.10(1)}$)$_2$As$_2$ 		&26.51(8)	  		&0.393(1)		 	&1.254(2)			&32.3(2)				&1.22			&0.44	&78	\\														
Sr(Co$_{0.85(2)}$Ni$_{0.15(2)}$)$_2$As$_2$ 		&25.1(3)	  		&0.370(1)		 	&1.217(1) 			&33.7(3)				&1.32			&0.39	&70	\\													
Sr(Co$_{0.80(1)}$Ni$_{0.20(1)}$)$_2$As$_2$ 		&10.79(6)	  		&0.328(2)		 	&1.146(3) 			&24.2(4)				&2.24			&0.41	&74	\\													
Sr(Co$_{0.75(1)}$Ni$_{0.25(1)}$)$_2$As$_2$ 		&3.08(1)	  		&0.259(1)		 	&1.018(2)  			&16.3(3)				&5.29			&0.60	&107	\\												
Sr(Co$_{0.70(1)}$Ni$_{0.30(1)}$)$_2$As$_2$ 		&		  			&0.206(1)		 	&0.907(2)  			&$-0.15(1)$				&				&&	\\												
Sr(Co$_{0.50(1)}$Ni$_{0.50(1)}$)$_2$As$_2$ 		&		  			&0.172(4)		 	&0.828(1)  			&$-84(3)$				&				&&	\\									
\end{tabular}
\end{ruledtabular}
\end{table*}

\begin{table*}
\caption{\label{Tab.MH} Parameters estimated from magnetization measurements of \scna\ at $T=2$~K\@.  Spin-flop  field $H_{\rm{SF}}$, critical field $H_{\rm{c}}$ at which $M_{ab}$ becomes approximately equal to $M_c$ in Fig.~\ref{Fig.MH} with increasing~$H$, saturation moment $\mu_{\rm sat}$ for $H||ab$ and $H||c$ obtained by extrapolating the high-field data in Fig.~\ref{Fig.MH} to $H=0$, angle-averaged saturation moments $\mu_{{\rm sat}, ab}$ and  $\mu_{{\rm sat}, c}$ , angle-averaged effective moment $\mu_{\rm eff, ave}$ from Table~\ref{Tab.chidata1}, and the Rhodes-Wohlfarth ratio $p_{\rm c}/p_{\rm sat}$ where $p_{\rm c}$ is given by Eq.~(\ref{Eq:pcDef}) and $p_{\rm sat} \equiv \mu_{\rm sat,ave}/\mu_{\rm B}$.}
\begin{ruledtabular}
\begin{tabular}{cccccccc}
  & & $H_{\rm{SF}}$ & $H_{\rm{c}}$   & $\mu_{\rm sat}$ & $\mu_{\rm sat,ave}$ & $\mu_{\rm eff,ave}$  &  $p_{\rm c}/p_{\rm sat}$ \\

Compound & Field Direction& (T) & (T) & $\rm{\left(\frac{\mu_B}{Co+Ni}\right)}$ & $\rm{\left(\frac{\mu_B}{Co+Ni}\right)}$ & $\rm{\left(\frac{\mu_B}{Co+Ni}\right)}$ \\
\hline
Sr(Co$_{0.987(1)}$Ni$_{0.013(1)}$)$_2$As$_2$		& $H\parallel ab$ 		&			&  			& 0.037(3) 		&0.035(3)		&1.33(1) 		&19(1)\\
										& $H\parallel c$ 		&			& 			& 0.0311(3)		&			&			&\\	
 Sr(Co$_{0.96(1)}$Ni$_{0.04(1)}$)$_2$As$_2$		&  $H\parallel ab$ 		&0.66(5)		&2.6(5)	 	& 0.067(4) 		&0.064(3)		&1.340(4) 	&10.4(5)\\
										& $H\parallel c$ 		&  			&3.4(3)		& 0.058(2) 		&			&			&\\	
 Sr(Co$_{0.94(1)}$Ni$_{0.06(1)}$)$_2$As$_2$		&  $H\parallel ab$ 		&0.91(4)		&2.7(1)  		& 0.118(1) 		&0.117(1)		& 1.259(2)	&5.18(6)\\
										& $H\parallel c$ 		&			&4.3(2) 		& 0.115(1)		&			&			&\\	
 Sr(Co$_{0.90(1)}$Ni$_{0.10(1)}$)$_2$As$_2$		&  $H\parallel ab$ 		&0.49(2)		&2.04(2)  	& 0.166(3) 		&0.167(3)		& 1.254(2)	&3.55(7)\\
										& $H\parallel c$ 		&			&3.37(5)	 	& 0.171(3) 		&			&			&\\	
 Sr(Co$_{0.85(2)}$Ni$_{0.15(2)}$)$_2$As$_2$		&  $H\parallel ab$ 		&0.41(2)		&1.62(4)  	& 0.165(1) 		&0.165(1)		&1.217(1) 	&3.48(2)\\
										& $H\parallel c$ 		&			&2.91(3)		& 0.165(2) 		&			&			&\\	
Sr(Co$_{0.80(1)}$Ni$_{0.20(1)}$)$_2$As$_2$		&  $H\parallel ab$ 		&0.23(3)		&0.64(4)  	& 0.128(2) 		&0.130(2)		&1.146(3) 	&3.99(7)\\
										& $H\parallel c$ 		&			&1.3(2)		& 0.135(2) 		&			&			&\\	
Sr(Co$_{0.75(1)}$Ni$_{0.25(1)}$)$_2$As$_2$		&  $H\parallel ab$ 		&			&0.60(6)  	& 0.092(6)		&0.094(5)		& 1.018(2)	&4.5(2)\\
										& $H\parallel c$ 		&			&0.65(4)		& 0.098(3) 		&			&			&\\	
Sr(Co$_{0.70(1)}$Ni$_{0.30(1)}$)$_2$As$_2$		&  $H\parallel ab$ 		&			& 0.61(7) 	& 0.048(4) 		&0.048(3)		& 0.907(2)	&7.2(4)\\
										& $H\parallel c$ 		&			& 0.58(8)		& 0.049(3) 		&			&			&\\		

\end{tabular}
\end{ruledtabular}
\end{table*}

\begin{table*}
\caption{\label{Tab.SFT}Spin-fluctuation parameters estimated with Takahashi's spin fluctuation theory in Eqs.~(\ref{Eq:M2}), (\ref{Eq:M0}) and (\ref{Eq:M4}) from $M^2$ vs $H/M$ and $M^4$ vs $H/M$ isotherm measurements for $x=0.04$, 0.06, 0.10, 0.15, 0.20, and 0.30. The fitted parameters are $T\rm_C$ (Curie--temperature), $T\rm_A$ (the width of the distribution of the dynamical susceptibility in the $q$ space), and $T\rm_0$ (the energy width of the dynamical spin fluctuation spectrum), and $F_1$ mode-mode coupling term. The error bars reflect systematic errors found from different field ranges of the fits. }

\begin{ruledtabular}
\begin{tabular}{c|c c c c c c c}
&\multicolumn{7}{c}{Data Obtained from $M^2$ versus $H/M$ Plots} \\

		&$M_0$						&slope($10^{-8}$)\footnotemark[1]	  				&$T\rm_C$	&$T\rm_{A1}$ 	&$T\rm_{01} $ 	&$F_1$	&$T\rm_C/T\rm_{01}$ 	 \\
$x$		&$\frac{\mu\rm_B}{\rm{Co+Ni}}$	&$\frac{[\mu\rm_B/(Co+Ni)]^3}{\rm Oe}$		& (K)			&(K)			&(K)	 		&(K)	 	&		  		\\
\hline
0.04 				&0.0417(3)	& 2.67(2)			& 5		&10874			&783			&40271		&0.006		\\			
0.06 				&0.1075(1)	&2.62(3)			& 20 	&10553  			&725			&40954 		&0.028		 \\			
0.10\footnotemark[2]	&0.1714(3)	&3.75(1)			& 		&5583			&290			&28662		&0.070 		 \\			
0.15\footnotemark[2]	&0.1694(2)	&3.52(2)			& 		&5734 		     &287			&30549		&0.070		\\
0.20\footnotemark[2]	&0.1293(3)	&2.38(2)			& 		&5886			&205			&45287		&0.061		 \\
\hline
&\multicolumn{7}{c}{Data Obtained from $M^4$ versus $H/M$ Plots at $T_{\rm C}$} \\
		&$M_0$						&slope($10^{-8}$)	  			&$T\rm_C$	 &$T\rm_{A2}$ 	&$T\rm_{02} $	&$F_1$	&$T\rm_C/T\rm_{02}$ 	 \\
$x$		&$\frac{\mu\rm_B}{(\rm{Co+Ni})}$	&$\frac{[\mu\rm_B/(Co+Ni)]^5}{\rm Oe}$	& (K)			&(K)			&( K)	 		&(K)	 	&		  		\\
\hline
0.10		&0.1666(2)	&0.19(1) 		& 20 		&3546		&1343			&2497 		&0.015 \\			
0.15		&0.1664(1)	&0.17(1)		& 20 		&3675  		&1214			&2967		& 0.016\\
0.20		&0.1255(2)	&0.08(2)		& 12.5 	&3442		&1226			&2579		&0.010 \\												
											
\end{tabular}
\end{ruledtabular}
\footnotetext[1]{slope $\frac{M^2}{H/M}$ from $M^2$ vs $H/M$ plots at $T=2$~K}
\footnotetext[2]{from $M(H)$ data at $T=2$~K, $T_{\rm C}$ is obtained from $M^4$ vs $H/M$ plots in the bottom part of the table}
\end{table*}

\begin{table*}
\caption{\label{Tab.RW} Magnetic and spin fluctuation parameters for \scna\  with $x=0.04$, 0.06, 0.10, 0.15, 0.20, and 0.30. The spontaneous (saturation) moment per Bohr magneton $p_{\rm sat}=M_{{\rm sat}, ab}/\mu_{\rm B}$ ($M(H=0-14~\rm T)$ at $T=2$~K extrapolated to $H=0$, from Fig~\ref{Fig:MH-14T}), the effective magnetic moment $\mu_{\rm eff}$ and Weiss temperature $\theta\rm_p$ are obtained from magnetization measurements. The error bars reflect systematic errors found from different field range of the fits. $p{\rm_c}/p_{\rm sat}$ is the Rhodes-Wholfarth ratio and $p{\rm_{eff}}/p_{\rm sat}$ is the Deuguchi-Takahashi ratio, where $p{\rm_{eff}}= \mu{\rm_{eff}}/\mu_{\rm B}$ from Table~\ref{Tab.chidata2}, and $p{\rm_c}= \sqrt{1+p_{\rm_{eff}}^2}-1$ from Eq.~(\ref{Eq:pcDef}). For the ratio $T_{\rm C}/T_0$, the Curie temperatures $T\rm_C$ are given in Table~\ref{Tab.SFT}. }

\begin{ruledtabular}
 \begin{tabular}{l c c c c c c c c|c}
&&&&&&&&\multicolumn{2}{c}{$T_{\rm C}/T_0$}\\
  			&  $p_{\rm sat}\mu_{\rm B}$ 							&$p\rm_{eff}\mu_{\rm B}$ 			&$p\rm_c\mu_{\rm B}$		& $p{\rm_{c}}/p_{\rm sat}$ 	&  $p{\rm_{eff}}/p_{\rm sat}$	& $\theta\rm_p$  	& $T\rm_C$	 &$M^2$ vs $H/M$ & $M^4$ vs $H/M$ \\
$x$			&$\rm{\left(\frac{\mu_B}{Co+Ni}\right)}$		& $\rm{\left(\frac{\mu_B}{Co+Ni}\right)}$	& $\rm{\left(\frac{\mu_B}{Co+Ni}\right)}$ 	&	 	&				& (K)			 	&(K)	 		& 			& \\
\hline
&&&&&&&&&\\
0.04 			&0.087(5)				&1.340(4)			&0.672(3)			&7.7(4)			&15.4(8)	&8.9			&5		&0.006 		&\\						
0.06 			&0.121(5) 				&1.259(2) 			&0.608(1)			&5.0(2)			&10.4(4)	&21.1			&20		&0.028  		 &\\						
0.10 			&0.174(1)				&1.245(2)			&0.604(1)			&3.43(2)		&7.20(5)	&32.3			&20		&0.070 		&0.015 \\					
0.15 			&0.173(2)				&1.217(2)  		     	&0.575(1)			&3.32(4)		&7.03(8)	&33.7			&20		& 0.070 		&0.016\\
0.20 			&0.135(2)				&1.146(3)  		      	&0.521(2)			&3.85(7)		&8.5(1)		&24.2			&12.5	& 0.061		&0.010 \\
0.25\footnotemark[1]			&0.094(5)				&1.018(2)			&0.427(1)			&4.5(2)			&10.8(5)	&16.3			&		&			&\\
0.30 			&0.057(3)				&0.907(2)  		      	&0.350(1)			&6.1(3)			&15.9(8)	&				&		&  			&\\
\end{tabular}
\end{ruledtabular}
\footnotetext[1]{$p_{\rm sat}$ from $M(H)$ at $T=2$~K in Fig.~\ref{Fig.MH}, extrapolated to $H=0$.}
\end{table*}

\begin{table*}
\caption{\label{Tab.HC} Parameters $\gamma$, $\beta$ and $\delta$ obtained by fitting the zero-field $C_{\rm p}(T)$ data of \scna\ by Eq.~(\ref{Eq.CpFit}). Also listed are the Debye temperature $\Theta \rm_D$ calculated from $\beta$ using Eq.~(\ref{Eq:thetaD}) and the density of states at the Fermi energy ${\cal D}(E\rm_F)$ calculated from $\gamma$ using Eq.~(\ref{Eq:DOS}). Another value of $\Theta\rm_D$ is obtained by fitting $C_{\rm p}(T)$ data by the Debye model according to Eq.~(\ref{Eq:Debye_Fit}).}

\begin{ruledtabular}
\begin{tabular}{ccccccc}
  											&  $\gamma$ 			&$\beta$ 			&$\delta$ 				& $\Theta\rm_D$  & $\Theta\rm_D$\footnotemark[1] & ${\cal D}_\gamma(E\rm_F)$   \\
Compound									&(mJ/mol K$^2$)		& (mJ/mol K$^4$)	 & ($10^{-3}$ mJ/mol K$^6$)	 	& (K)			& (K)   					 		&$\left(\frac{\rm states}{\rm eV~ f.u.}\right)$ \\
\hline
SrCo$_2$As$_2$       	 					& 49.2(1) 			& 0.291(5) 	& 0.66(4)		&322(1)		&288(1)		&20.86(4) \\														
Sr(Co$_{0.96(1)}$Ni$_{0.04(1)}$)$_2$As$_2$       	& 53.66(8) 			& 0.292(3) 	&1.52(3)		&321(1)		&299(2)		&22.74(3) \\														
Sr(Co$_{0.94(1)}$Ni$_{0.06(1)}$)$_2$As$_2$ 		& 55.87(6)			& 0.414(2) 	&0.91(2) 		&286.3(4)		&289(1		& 23.68(2)  \\													
Sr(Co$_{0.90(1)}$Ni$_{0.10(1)}$)$_2$As$_2$       	& 57.35(7) 			& 0.368(3) 	&0.732(3)		&298(1)		&300(1)		&24.33(3) \\														
Sr(Co$_{0.85(2)}$Ni$_{0.15(2)}$)$_2$As$_2$ 		& 59.14(2)			& 0.368(1)	&1.14(1) 		&297.8(2)		&297(2)		&25.07(1)  \\
Sr(Co$_{0.50(1)}$Ni$_{0.50(1)}$)$_2$As$_2$ 		& 42.82(9)			& 0.277(4)  	&4.25(5)		&327(1)		&285(4)		& 18.15(4) \\		

\end{tabular}
\end{ruledtabular}
\footnotetext[1]{$\Theta_{\rm D}$ calculated from Eq.~(\ref{Eq:Debye_Fit}).}
\end{table*}

\begin{table*}
\caption{\label{Tab.HC_SF} Parameters $\gamma$,  $\beta$,  $\delta$, $\kappa$, and $T_{\rm SF}$ obtained by fitting the zero-field $C_{\rm p}(T)/T$ data for \scna\ below 10~K by Eq.~(\ref{Eq:Cp_SF_Fit2}).  The values of $\Theta\rm_D$  and ${\cal D}_\gamma (E\rm_F)$ for both spin directions were obtained from the $\beta$ and $\gamma$ values using Eqs.~(\ref{Eq:thetaD}) and~(\ref{Eq:DOS}), respectively.}

\begin{ruledtabular}
\begin{tabular}{ccccccccc}
Compound 										&  $\gamma$ 		&$\beta$ 			&$\delta$ 		& $\kappa$		& $T_{\rm SF}$		& $\Theta\rm_D$  	& ${\cal D}_\gamma(E_F)$   \\
											& (mJ/mol\,K$^2$)	& (mJ/mol\,K$^4$)	& (mJ/mol\,K$^6$)	& (mJ/mol\,K$^2$)	&(K)				& (K)			&$\left(\frac{\rm states}{\rm eV\,f.u.}\right)$  \\
\hline
Sr(Co$_{0.80(1)}$Ni$_{0.20(1)}$)$_2$As$_2$ 			&58.9(2)			&0.53(1)			&0.00021(3)		&$-10.6(3)$ 		&10.5(2)			&263(1) 			&	25.0(1)\\
Sr(Co$_{0.75(1)}$Ni$_{0.25(1)}$)$_2$As$_2$ 			&60.5(1)  		&0.71(1) 			&0.00005(3)		&$-42.9(1)$ 		&5.39(4)			&239(1)   		&	25.7(1)\\
Sr(Co$_{0.70(1)}$Ni$_{0.30(1)}$)$_2$As$_2$ 			&56.4(1)			&0.81(2)			&0.0001(1)		&$-44.2(6)$		&4.89(5)			&229(2) 	 		&	23.9(1)\\
\end{tabular}
\end{ruledtabular}

\end{table*}

\clearpage

\acknowledgments

We are grateful to Prof.~Yoshinori Takahashi for helpful correspondence.  The research at Ames Laboratory was supported by the U.S. Department of Energy, Office of Basic Energy Sciences, Division of Materials Sciences and Engineering.  Ames Laboratory is operated for the U.S. Department of Energy by Iowa State University under Contract No.~DE-AC02-07CH11358.

\end{document}